\documentclass[a4paper,10pt]{article}

\usepackage{jheppub} 
\usepackage{xcolor}
\usepackage[T1]{fontenc} 
\usepackage{slashed}
\usepackage{comment}

\title{\boldmath Near-to-planar three-jet events
  at NNLL accuracy}

\author[a]{Luke Arpino,}
\author[a]{Andrea Banfi,}
\author[a,b]{Basem Kamal El-Menoufi}

\affiliation[a]{Department of Physics and Astronomy, University of Sussex,\\Sussex House, Brighton, BN1 9RH, UK}
\affiliation[b]{Consortium for Fundamental Physics, School of Physics and Astronomy, University of Manchester,\\Manchester M13 9PL, United Kingdom}

\emailAdd{l.arpino@sussex.ac.uk}
\emailAdd{a.banfi@sussex.ac.uk}
\emailAdd{basem.el-menoufi@manchester.ac.uk}

\abstract{We extend the ARES method for
  next-to-next-to-leading-logarithmic (NNLL) QCD resummations to
  three-jet event shapes in $\ee$ collisions in the near-to-planar
  limit. In particular, we define a NNLL radiator for three hard
  emitters, and discuss new features of NNLL corrections arising
  specifically in this case. As an example, we present predictions for
  the $D$-parameter, matched to exact NLO. After inclusion of
  hadronisation corrections in the dispersive approach, we compare our
  predictions with LEP1 data.}

\preprint{}

\newcommand{\as}{\alpha_s}

\newcommand{\cf}{C_F}
\newcommand{\ca}{C_A}
\newcommand{\tr}{T_R}
\newcommand{\nf}{n_F}

\newcommand{\ee}{e^+e^-}

\newcommand{\FNLL}{\mathcal{F}_{\rm NLL}}

\newcommand{\dFsc}{\delta \mathcal{F}_{\rm sc}}
\newcommand{\dFrec}{\delta \mathcal{F}_{\rm rec}}
\newcommand{\dFhc}{\delta \mathcal{F}_{\rm hc}}
\newcommand{\dFwa}{\delta \mathcal{F}_{\rm wa}}
\newcommand{\dFcor}{\delta \mathcal{F}_{\rm correl}}

\newcommand{\ycut}{y_{\rm cut}}

\newcommand{\pBorn}{\{\tilde{p}\}}
\newcommand{\Vsc}[1]{V_{\rm sc}(\pBorn,#1)}

\newcommand{\Vfull}[1]{V(\pBorn,#1)}

\newcommand{\dZ}{d{\cal Z}[R'_{\ell_i,\mathrm{NLL}}, \{k_i\}]}
\newcommand{\Rp}[1][]{R^\prime \ifx\\#1\\\else_{#1}\fi}
\newcommand{\Rs}[1][]{R^{\prime\prime}_{\mathrm{NNLL \ifx\\#1\\\else,#1\fi}}}
\newcommand{\RpNLL}[1][]{R^\prime_{\mathrm{NLL \ifx\\#1\\\else,#1\fi}}}
\newcommand{\RpNNLL}[1][]{\Delta R^\prime_{\mathrm{NNLL \ifx\\#1\\\else,#1\fi}}}

\newcommand{\muR}{\mu_R}

\begin{document} 
\maketitle
\flushbottom

\section{Introduction}

Jet dynamics, encapsulated in event-shape distributions and jet-rates, is one of
the most studied topics in QCD. These observables are designed to capture the
continuous energy-momentum flow in hadronic processes, and as such offer a
powerful probe of strong interactions. Jet observables probe disparate scales
across the energy spectrum, starting from high scales where fixed-order
perturbative calculations can be applied, and all the way down to $\Lambda_{\rm
  QCD}$ where the yet unexplained phenomenon of hadronisation dominates the
physics. Historically, and still up to this day, jet observables have been
utilised to accurately extract the strong coupling from data as well as to test
non-perturbative hadronisation models (see e.g.\ ref.~\cite{Patrignani:2016xqp}
and references therein).

The study of jet observables has been pivotal in understanding the all-orders
properties of QCD radiation, which subsequently lead to the discovery of
non-global logarithms \cite{Dasgupta:2001sh, Dasgupta:2002bw,Banfi:2002hw}.
Distributions in jet observables can be computed at fixed-order in perturbative
QCD \cite{Ellis:1980wv}, and such calculations have reached next-to-next-leading
order (NNLO) accuracy for a number of relevant QCD processes. In particular, for
$e^+ e^-$ annihilation, NNLO corrections to three-jet production have been
computed in
refs.~\cite{GehrmannDeRidder:2007hr,GehrmannDeRidder:2008ug,Weinzierl:2008iv,Weinzierl:2009ms}.

Fixed-order calculations are reliable when the value of the jet
observable is large. Nevertheless, the bulk of data lies in the region
of small observable value where the cross section is dominated by large
logarithms. The large logarithms emerge from the soft and/or collinear
regions of phase space, due to a miss-cancellation between real and
virtual corrections. Given a generic jet observable, the total
(normalised) cross section is denoted by $\Sigma(v)$, and represents
the fraction of events where the observable takes a value less than
$v$. In perturbation theory, $\Sigma(v)$ displays logarithmic terms,
$L \equiv\ln 1/v$, and the highest power that appears at each order
$\alpha_s^n$ of perturbation theory depends on the
observable. For double-logarithmic observables, $\Sigma(v)$ will
contain logarithms as high as $\alpha_s^n L^{2n}$. The fixed order
approximation of the cross section becomes unreliable in the regime
when $\alpha_s L\sim 1$, and resummation becomes mandatory for
theoretical consistency.

The primary concern of the resummation program is to reorganise the
perturbative series in such a way as to allow those large logarithms
to be isolated and resummed. Explicitly, the idea is to express
$\ln\Sigma(v)$ as a series of functions with successive logarithmic
accuracy. For double-logarithmic observables, we have
$\ln\Sigma(v) = L g_1(\alpha_s L)+g_2(\alpha_s L)+\alpha_s
g_3(\alpha_s L)+\dots$,
where $L g_1(\alpha_s L)$ resums the so-called leading logarithmic
(LL) terms, $\alpha_s^n L^{n+1}$, $g_2(\alpha_s L)$ the NLL ones,
$\alpha_s^n L^n$, $g_3(\alpha_s L)$ the NNLL ones,
$\alpha_s^n L^{n-1}$, and so on.

Next-to-leading logarithmic (NLL) resummations have been available for
many years for specific observables
\cite{Collins:1984kg,Catani:1991bd,Catani:1991kz,Catani:1992ua,Dokshitzer:1998kz,Bonciani:2003nt}. At
the present day NLL resummation is available for all (continuously)
{\em global} jet observables, which possess the property of recursive
infrared and collinear (rIRC) safety
\cite{Banfi:2001bz,Banfi:2003je,Banfi:2004yd}. The technique is based
on a semi-numerical approach developed in a series of works, and the
method is implemented in the computer program CAESAR
\cite{Banfi:2004yd}, which automatically verifies whether or not a
given observable is rIRC safe and continuously global. This paved the
way to a systematic study of event shapes in hadronic di-jet production
at NLL accuracy matched to next-to-leading order (NLO) results at
hadron colliders \cite{Banfi:2004nk,Banfi:2010xy}.

There is no doubt that the theoretical predictions of NLL resummation are
under a lot of tension due to various reasons. First, it remains true
that NLL resummed predictions have a sizeable theoretical uncertainty
which, when compared to current precision measurements, requires going
beyond NLL. Second, recent works have started to utilise resummation
results to test, and improve, the accuracy of parton shower
simulations \cite{Dasgupta:2018nvj}. Given the absolute importance of
parton showers for collider physics, it is mandatory that we push the
accuracy of resummation results and aim for the widest class of
observables. Third, event-shape distribution offer an important
testing ground for analytical models of non-perturbative hadronisation
corrections. Simultaneous fits of both the strong coupling and the
parameter controlling the leading hadronisation corrections have been performed
using NLL resummations for a variety of event-shapes
(see~\cite{Dokshitzer:1998qp,Salam:2001bd} for the most accurate
fits). Analogous studies using NNLL resummations exist only for a
limited number of
observables \cite{Gehrmann:2012sc,Abbate:2010xh,Hoang:2014wka,Hoang:2015hka},
it would be very interesting to have a new comprehensive picture of
leading hadronisation corrections using NNLL resummations.

In the past decade, much progress has taken place in NNLL
resummation. Nevertheless, most results available in the
literature are performed for two-jet observables, i.e. those which
vanish in the limit of two jets. Moreover, until very recently {\em
  most} NNLL resummations were observable specific, i.e.\ dependent on
whether a factorisation theorem holds for the observable.  Such
approaches made it possible to obtain full next-to-next-to-leading
logarithmic (NNLL) predictions for a number of global $e^+e^-$ event
shapes such as one minus the thrust $1-T$
\cite{Becher:2008cf,Abbate:2010xh,Monni:2011gb}, heavy jet mass
$\rho_{H}$ \cite{Chien:2010kc}, jet broadenings $B_T$ and $B_W$
\cite{Becher:2012qa}, $C$-parameter \cite{Hoang:2014wka}, energy-energy
correlation (EEC)
\cite{deFlorian:2004mp,Tulipant:2017ybb,Moult:2018jzp}, heavy
hemisphere groomed mass \cite{Frye:2016aiz}, and angularities
\cite{Procura:2018zpn,Bell:2018gce}.

Apart from $e^+ e^-$ annihilation, NNLL resummations have been
performed for a number of jet observables in other QCD processes. For
example, some results are available in deep inelastic
scattering~\cite{Kang:2013nha,Kang:2013wca,Kang:2013lga}, and for
hadronic collisions results were obtained when a colour singlet is
produced at Born level. For instance, the transverse momentum of a
colourless Boson in the final state \cite{Bozzi:2005wk,Becher:2010tm},
the variable $\phi^*$~\cite{Banfi:2011dx}, the beam
thrust~\cite{Stewart:2010pd,Berger:2010xi}, transverse
thrust~\cite{Becher:2015lmy}, and the leading jet's transverse
momentum
\cite{Becher:2012qa,Becher:2013xia,Banfi:2012jm,Stewart:2013faa}. We
have a limited number of NNLL resummations for processes with
more than two hard emitters, notably for heavy quark pair's transverse
momentum~\cite{Catani:2014qha,Zhu:2012ts} and
$N$-jettiness~\cite{Stewart:2010tn,Jouttenus:2011wh}.  Very recently,
resummations for the Boson's transverse momenta and related observables
has been pushed to N$^3$LL
accuracy~\cite{Bizon:2017rah,Bizon:2018foh}.

Despite the advances of these approaches, many observables do not exhibit the
factorisation requirements needed to carry out the resummation using, for
example, the SCET framework \cite{Bauer:2000yr}. This is particularly the case
for observables which cannot be expressed as a simple analytic function of
momenta, such as the thrust-major or the two-jet rate in the Durham algorithm.
Very recently, the ARES (Automated Resummation for Event Shapes) approach has
been completed and it is now possible to resum any rIRC safe di-jet observable,
in $e^+ e^-$ annihilation, at NNLL accuracy. The original development of ARES
focused on $e^+ e^-$ event shapes, but extensions thereof were presented for the
two-jet rate \cite{Banfi:2016zlc}. ARES performs the resummation in direct
space, i.e.\ without using any integral transforms, and only relies on the
factorisation properties of QCD matrix elements in the soft and/or collinear
limits. The fundamental ingredients of the method are as follows:
\begin{itemize}
\item The {\em analytic} cancellation of soft and collinear
  divergences, which relies on the exact exponentiation of infrared
  poles in QCD processes with coloured particles in the final
  state. This exponentiation is simple to implement in the case of
  two, as well as three, hard legs.
\item Unresolved emissions, owing to rIRC safety, yield a finite
  Sudakov radiator, which is analytically calculable in four
  dimensions. The radiator acts as a suppression factor for emissions
  contributing to the observable above $\delta v$, where $\delta$
  defines a resolution scale.
\item Resolved emissions do not generally exponentiate, but yield a
  class of functions starting at NLL accuracy. These functions are
  finite and directly calculable in four dimensions, thereby amenable
  to Monte Carlo integration. Each function has a distinct physical
  origin, and resums a given class of subleading logarithms that
  originate from various regions of phase space.
\end{itemize}
In this paper we extend the ARES method to resum three-jet
observables, which are global and rIRC-safe, up to NNLL accuracy. Our
work presents the first general NNLL resummation for three-jet
observables, and paves the way to future extensions of ARES to go
beyond three jets at NNLL accuracy. Given that the hard event is
comprised of three partons, we are able to follow the basic
constructions of the ARES approach outlined above. In particular, we
derive the Sudakov radiator suitable for three-jet
observables. Compared to previous two-jet results, the soft radiator
in our case exhibits a richer structure in that it depends explicitly
on the kinematics of the hard legs. The emission probability of soft
partons, which are emitted coherently from the all the hard legs,
takes the form of a sum over dipoles whose invariant masses end up
appearing in the Sudakov radiator. The radiator also receives a
contribution that is of pure collinear origin, and we report the full
extension of this contribution in the presence of three hard legs. The
dipole structure, in addition, leads to a new NNLL contribution
arising due to resolved emissions. The new function is called
$\Delta \mathcal{F}_{\rm wa}$ and presents a new addition to the
wide-angle NNLL function encountered in the ARES di-jet resummation
formula. Moreover, we take full account of spin-correlations that
become omnipresent due to the presence of a hard gluon in the Born
configuration. The latter introduce a new ingredient in the
resummation formula when the underlying gluon recoils against a hard
emission. This leads to a new NNLL function, which we call
$\Delta \mathcal{F}_{\rm rec}$, and we show how to compute it for a
generic observable.

The paper is organised as follows. In section~\ref{sec:observable} we
define our notation and the various Sudakov decompositions we employ
throughout the manuscript. We also give a lightening review of NLL
resummation previously obtained in \cite{Banfi:2000si}. Section
\ref{sec:nnll} contains the central piece of our work where we
explicitly derive the NNLL resummation master formula. The Sudakov
radiator is computed, before we move to formulate the various
correction functions. We make sure to elaborate on the new ingredients
that arise at NNLL, which manifest in two new correction functions. In
section \ref{sec:D-NNLL}, we apply our resummation formula to the
$D$-parameter, an example of an additive observable amenable to a
fully analytic treatment in ARES. Finally, in section
\ref{sec:checkandmatch} we validate our analytic resummation against
exact fixed-order calculations, and we present some simple
phenomenological studies.

\section{Kinematics and setup}
\label{sec:observable}
In this section we set up the kinematics and notation that will be used
throughout the paper. We explain the procedure we use to select three-jet events
and give a short review of how to perform NLL resummation for three-jet
observables in near-to-planar kinematics.

\paragraph{Three-jet Born kinematics.} 
At Born level, a three-jet event in $e^+e^-$ annihilation is made up
of a quark of momentum $p_1$, an antiquark $p_2$ and a gluon $p_3$: 
\begin{equation}
  \label{eq:Born-momenta}
  p_1 = E_1 (1,0,0,1)\,, \qquad p_2 = E_2 (1,0,\sin\theta_{12},\cos\theta_{12})\,,\qquad p_3=E_3(1,0,-\sin\theta_{13},\cos\theta_{13})\,,
\end{equation}
with $\theta_{12}$ and $\theta_{13}$ the angles between $p_1$ and
$p_2$, and $p_1$ and $p_3$ respectively. The relation between these
angles and energies and the variables that are typically used to
describe three-jet events is reported in
appendix~\ref{sec:kinematics}.

We consider event shapes that vanish in the three-jet limit,
i.e.\ $V(\{\tilde{p}\})=0$, where $\{\tilde p\}$ denotes the set
$\{\tilde p_1,\tilde p_2,\tilde p_3\}$, the actual final-state momenta,
which coincide with ($p_1,p_2,p_3$) in eq.~(\ref{eq:Born-momenta}) at
Born level. After many soft and/or collinear emissions $k_1,\dots,k_n$, the
three hard partons will recoil, and $\{\tilde p\}$ are the actual
final-state momenta after recoil.

\paragraph{Sudakov variables.} A single emission $k$ can be decomposed
along any pair of light-like momenta $(p_i,p_j)$, which constitute the
$(ij)$ dipole, as follows:
\begin{equation}
  \label{eq:Sudakov-k}
  k^{(ij)} = z^{(i)} \,p_i + z^{(j)} \,p_j + \kappa^{(ij)} \cos\phi^{(ij)} \,n^{(ij)}_{\rm in} + \kappa^{(ij)} \sin\phi^{(ij)} \, n^{(ij)}_{\rm out}\,,  
\end{equation}
where $n^{(ij)}_{\rm in}$ and $n^{(ij)}_{\rm out}$ are space-like
vectors such that $(n^{(ij)}_{\rm in})^2=(n^{(ij)}_{\rm out})^2=-1$,
given by
\begin{equation}
  \label{eq:nin-nout}
  n^{(ij)}_{\rm in} = \left(\cot\frac{\theta_{ij}}{2},\frac{\vec n_i+\vec n_j}{\sin\theta_{ij}}\right)\,,\qquad n^{(ij)}_{\rm out} = \left(0,\frac{\vec n_i\times \vec n_j}{\sin\theta_{ij}}\right)\,,\qquad \vec n_{\ell}\equiv\frac{\vec p_{\ell}}{E_\ell}\,,\quad \ell=i,j\,.
\end{equation}
We have also introduced the invariant transverse momentum with respect
to the $(ij)$ dipole
\begin{equation}
  \label{eq:ktdip}
  (\kappa^{(ij)})^2 = \frac{(2p_i k)(2k  p_j)}{(2p_i p_j)}\,,
\end{equation}
where $(p_i p_j)$ is a short-hand notation for the Lorentz-invariant
product $p_i \cdot p_j$. One can also choose to decompose the emission
$k$ along a {\em single} light-like momentum $p_\ell$. This can be
achieved by defining the light-like momentum
$\bar p_\ell=(E_\ell,-\vec p_\ell)$.  Explicitly,
\begin{align}
  \label{eq:Sudakov-leg}
k^{(\ell)} = x^{(\ell)} p_\ell + \bar{x}^{(\ell)} \bar{p}_\ell + k^{(\ell)}_\perp\,,
\end{align}
where $ k^{(\ell)}_\perp$ is a two-dimensional space-like vector lying
in the transverse plane to $\vec{p}_\ell$, and whose magnitude reads
\begin{align}
  - \left(k^{(\ell)2}_\perp\right) = \frac{(2 p_\ell k)(2 k  \bar{p}_\ell)}{(2 p_\ell \bar{p}_\ell)} \equiv \left(k_t^{(\ell)}\right)^2\,.
\end{align}
Note that, if $k$ is collinear to $\vec{p}_\ell$, we have
$\kappa^{(ij)}\to k_t^{(\ell)}$.

We now introduce the rapidity $\eta^{(ij)}$ with respect to a dipole
and its counterpart, $\eta^{(\ell)}$, with respect to leg $p_\ell$
\begin{equation}
  \label{eq:rapidity}
  \eta^{(ij)} \equiv \frac{1}{2} \ln \frac{z^{(i)}}{z^{(j)}} \,,\qquad\eta^{(\ell)} =\frac{1}{2} \ln \frac{x^{(\ell)}}{\bar x^{(\ell)}} \,.
\end{equation}
For an emission $k$ collinear to $p_i$ or $p_j$, the rapidities $\eta^{(ij)},\eta^{(i)},\eta^{(j)}$ are related as follows
\begin{equation}
  \label{eq:rapidity-links}
\eta^{(i)}\simeq \eta^{(ij)} + \ln\frac{2 E_i}{Q_{ij}}\,,\qquad\eta^{(j)} \simeq -\eta^{(ij)} + \ln\frac{2 E_j}{Q_{ij}}\,,\qquad Q^2_{ij}=2(p_i p_j)\,.
\end{equation}
From $z^{(i)},z^{(j)},x^{(\ell)},\bar x^{(\ell)}<1$, for a light-like
vector $k$ we obtain the rapidity bounds
\begin{equation}
  \label{eq:eta-boundaries}
  |\eta^{(ij)}| < \ln\frac{Q_{ij}}{\kappa^{(ij)}}\,,\qquad \eta^{(\ell)} < \ln \frac{2 E_\ell}{k_t^{(\ell)}}\,.  
\end{equation}
Last, we denote with $\phi^{(\ell)}$ the azimuthal angle of
$k^{(\ell)}_\perp$. We adopt the convention
$\phi^{(ij)}=\phi^{(\ell)}=0$ when an emission is in the plane formed
by $p_1,p_2,p_3$. Comparing the expressions of the component of $k$
outside the event plane in the Sudakov decompositions in
eqs.~(\ref{eq:Sudakov-k}) and (\ref{eq:Sudakov-leg}), we have that, for
an emission $k$ collinear to $\vec{p}_i$, $\phi^{(ij)}\simeq\phi^{(i)}$.

Notice that the light-like momenta we use in eqs.~\eqref{eq:Sudakov-k}
and \eqref{eq:Sudakov-leg} are not necessarily the Born momenta
$\{p_1,p_2,p_3\}$ in eq.~(\ref{eq:Born-momenta}), neither the actual
final state momenta, i.e.  $\{ \tilde{p} \}$. In fact, one can choose
a different Sudakov decomposition according to the kinematical limit
one is interested in, see for example~\cite{Catani:2011st}. The choice
we make will depend on the context and will be made explicit in the
subsequent derivations.

\paragraph{Lorentz-invariant phase space.} The particular Sudakov
variables we employ will depend on the form of singularities in the
squared matrix elements. Hence, we list here the Lorentz-invariant
phase-space measure expressed in various bases. For a massless
emission $k=(\omega,\vec k)$, we have:
\begin{align}
  \label{eq:PSm0dip}
  [dk]\equiv \frac{d^3 k}{(2\pi)^3 2\omega}& = \frac{ \kappa^{(ij)} d\kappa^{(ij)}}{(2\pi)^2} \frac{d\phi^{(ij)}}{2 \pi} \frac{d \eta^{(ij)}}{2} \Theta\left(\ln\frac{Q_{ij}}{\kappa^{(ij)}}-|\eta^{(ij)}| \right) \\ &= \frac{ k_t^{(\ell)} dk_t^{(\ell)}}{(2\pi)^2}  \frac{d\phi^{(\ell)}}{2 \pi} \frac{d \eta^{(\ell)}}{2}\Theta(\eta^{(\ell)})\Theta\left(\ln\frac{2 E_\ell}{k_t^{(\ell)}}-\eta^{(\ell)}\right)+\dots\,.
\end{align}
In the above equation, when we parameterise the phase-space in terms
of leg-variables $k_t^{(\ell)},\eta^{(\ell)},\phi^{(\ell)}$, we omit
the integration region corresponding to the anti-collinear direction
$\vec{\bar p}_\ell$, because this does not correspond to any collinear
singularity of QCD matrix elements. Also, the boundary of the region
collinear to $\vec{p}_\ell$ is conventionally chosen to be
$\eta^{(\ell)}=0$.

\paragraph{Selection of three-jet events.} In this paper we are
interested in studying event shapes in the near-to-planar limit. In
order to do this, we need a procedure to select hadronic events with
at least three jets. This could be, for instance, through a jet
algorithm that counts the number of well separated hard jets in the
event, or through a cut on some secondary, two-jet observable. This
constraint is represented by $\mathcal{H}(p_1,\dots,p_n)$, a function
of all hadron momenta $p_1,\dots,p_n$ that is 1 if an event passes the
cut and 0 otherwise. The function $\mathcal{H}(p_1,\dots,p_n)$ also
embodies exact energy-momentum conservation.  In our case, we use the
Durham algorithm~\cite{Catani:1991hj} and we select three-jet events
if the three-jet resolution variable $y_3(p_1,\dots,p_n)$ is greater
than $\ycut$.  Correspondingly, we have a total three-jet cross
section which, in $d$ dimensions, is given by
\begin{align}\label{eq:sigma3}
  \nonumber
  \sigma_{\mathcal{H}}&\equiv \sum_{n=3}^\infty\int d\Phi_n\frac{d\sigma_n}{d\Phi_n}\mathcal{H}(p_1,\dots,p_n)\\
  &=\sum_{n=3}^\infty \int d\Phi_n\frac{d\sigma_n}{d\Phi_n}\Theta\left(y_3(p_1,\dots,p_n)-\ycut\right)(2\pi)^d\delta^{(d)}(p_1+p_2+\dots+p_n)\,,
\end{align}
with $d\Phi_n$ the $n$-particle phase space. We now consider the
cumulative distribution of a three-jet event shape $V(p_1,\dots,p_n)$,
defined as: 
\begin{equation}
  \label{eq:SigmaH}
  \begin{split}
    \Sigma_{\mathcal{H}}(v)& \equiv \frac{1}{\sigma_{\cal H}}\sum_{n=3}^\infty\int d\Phi_n\frac{d\sigma_n}{d\Phi_n}\mathcal{H}(p_1,\dots,p_n)\Theta(v-V(p_1,\dots,p_n))
\\ &= \frac{1}{\sigma_{\cal H}}\sum_{n=3}^\infty\int d\Phi_n\frac{d\sigma_n}{d\Phi_n}\Theta\left(y_3(p_1,\dots,p_n)-\ycut\right)\Theta(v-V(p_1,\dots,p_n)) (2\pi)^d\delta^{(d)}(p_1+p_2+\dots+p_n)\,.
  \end{split}
\end{equation}
In near-to-planar kinematics, i.e.\ for $v\ll 1$, the cumulative
distribution $\Sigma_{\mathcal{H}}(v)$ assumes the factorised
form~(see e.g.~\cite{Banfi:2004yd})
\begin{equation}
  \label{eq:SigmaH-factorised}
  \Sigma_{\mathcal{H}}(v)\simeq \frac{1}{\sigma_{\mathcal{H}}}\int d\Phi_3\,\frac{d\sigma_3}{d\Phi_3}\,\Sigma(\{p_1,p_2,p_3\},v)\,\mathcal{H}(p_1,p_2,p_3)\,,
\end{equation}
where $p_1,p_2,p_3$ are now the three Born momenta in
eq.~(\ref{eq:Born-momenta}) and $d\sigma_3/d\Phi_3$ is given in
eq.~\eqref{eq:3jet-dx12}.  When $v\ll 1$, the function
$\Sigma(\{p_1,p_2,p_3\},v)$ develops large logarithms of
$v$, which we want to resum, up to a given logarithmic accuracy, to
all orders in the strong coupling. Notice that
eq.~\eqref{eq:SigmaH-factorised} we have implicitly mapped the final
state momenta, $\{ \tilde{p} \}$, to the Born level momenta,
$\{ p \}$. The details of this mapping are not important here. In
fact, with an IRC safe three-jet selection, in the presence of
infinitely soft and/or collinear emissions $k_1,\dots,k_n$, the
fineal-state momenta $\{ \tilde{p} \}$ always reduce to $\{ p \}$, and
the difference between $\mathcal{H}(\{\tilde p\},k_1,\dots,k_n)$ and
$\mathcal{H}(p_1,p_2,p_3)$ is suppressed by powers of $v$.

\paragraph{NLL resummation.}
We are interested in rIRC safe three-jet observables $V(\{p\},k_1,\dots k_n)$ in
the near-to-planar kinematics, in which $V(\{p\},k_1,\dots k_n)\ll 1$. We
require that, for a single soft emission collinear to leg $\ell$, our
observables behave as follows
\begin{equation}
  \label{eq:Vsc}
  V_{\rm sc}(\pBorn,k)\simeq d_\ell\left(\frac{k_t^{(\ell)}}{Q}\right)^a e^{-b_\ell \eta^{(\ell)}}g_\ell(\phi^{(\ell)})\,.
\end{equation}
In the above equation, $Q$ is a typical hard scale for the process
under consideration, in our case by default the centre-of-mass-energy
of the $e^+e^-$ collision. Note that IRC safety requires $a>0$ and
$b_\ell>-a$. The NLL resummation of near-to-planar three-jet
observables can be obtained from the general procedure of
refs.~\cite{Banfi:2003je,Banfi:2004yd}. At NLL accuracy, the
distribution $\Sigma(\{p_1,p_2,p_3\},v)$ introduced in
eq.~(\ref{eq:SigmaH-factorised}) reads
\begin{equation}
  \label{eq:Sigma-NLL}
  \Sigma_{\mathcal{B}}(\{p_1,p_2,p_3\},v) = e^{-R_{\rm NLL}(v)} \,\FNLL\left(\RpNLL(v)\right)\,,\qquad \RpNLL \simeq -v\frac{dR_{\rm NLL}}{dv}\,,
\end{equation}
where $R_{\rm NLL}(v)$ is the NLL radiator, encoding the probability
of observing no emissions $k_i$ with $V(\pBorn,k_i)>v$, and
$\RpNLL(v)$ is obtained from the logarithmic derivative of
$R_{\rm NLL}(v)$ neglecting all NNLL corrections. The NLL radiator,
obtained originally in ref.~\cite{Banfi:2004yd}, will be borne out as
a byproduct of our formalism in section 3. We recall here its
expression as a sum of contributions of the three dipoles
$q\bar q,qg,g\bar q$ that build up a three-jet configuration:
\begin{multline}
\label{eq:RNLL}
  R_{\rm NLL}(v) = \sum_{(ij)\in\{(q\bar q),(qg),(g\bar q)\}}\!\!\!\!\! C_{(ij)} \left(\sum_{\ell
      =i,j}\left[\right.  r_\ell(L) + r'_\ell(L)\left(\langle\ln(
      d_\ell\,g_\ell)\rangle-b_\ell \ln\frac{2E_\ell}{Q}\right) \right.\\ \left. \left.
    +B_\ell\, T\!\left(\frac{L}{a+b_\ell}\right) \right]+2\ln\frac{Q_{ij}}{Q}
  T\!\left(\frac{L}{a}\right) \right) \,,
\end{multline}
with $L\equiv \ln(1/v)$.  In the above equation, $C_{(ij)}$ is the
colour factor associated with dipole $(ij)$, namely $C_{(ij)}=C_A$
for $(ij)=(qg),(g\bar q)$ and $C_{(ij)}=2C_F\!-\!C_A$ for
$(ij)=(q\bar q)$. Then, we have the following integrals over selected
momentum regions:
\begin{equation}
  \label{eq:rell-T}
  \begin{split}
  r_\ell(L) & = \int_{Q e^{-\frac{L}{a+b_\ell}}}^{Q}\frac{dk_t}{k_t} \frac{\alpha^{\rm phys}_s(k_t)}{\pi} \ln\frac{Q}{k_t}+\frac{1}{b_\ell}
  \int^{Q e^{-\frac{L}{a+b_\ell}}}_{Qe^{-\frac{ L}{a}}}\frac{dk_t}{k_t}\frac{\alpha^{\rm phys}_s(k_t)}{\pi}\left(L+a\ln\frac{k_t}{Q}\right)\,,\\
  T(L)& = \int_{Q e^{-L}}^{Q}\frac{dk_t}{k_t} \frac{\alpha^{\rm phys}_s(k_t)}{\pi}\,,\qquad\qquad
  r'_\ell(L)=\frac{1}{b_\ell}\left[T\!\left(\frac{L}{a}\right)-T\!\left(\frac{L}{a+b_\ell}\right)\right]\,,
  \end{split}
\end{equation}
where $\alpha_s^{\rm phys}$ is the soft physical coupling defined in
ref.~\cite{Banfi:2018mcq}. To keep strict NLL accuracy, we calculate
the integrals above neglecting all subleading contributions. The
quantity $\langle\ln( d_\ell\,g_\ell)\rangle$ is an azimuthal average, which for any function $f(\phi)$ is defined as
\begin{equation}
  \label{eq:dg-average}
  \langle f\rangle\equiv\int_0^{2\pi}\frac{d\phi}{2\pi}f(\phi)\,.
\end{equation}
The terms proportional to the coefficient $B_\ell$ represent virtual
corrections of hard collinear origin down to the scale
$Q v^{\frac{1}{a+b_\ell}}$, the characteristic scale of hard collinear
radiation. The term proportional to $T(L/a)$ represents soft wide-angle
virtual corrections down to the scale $Q v^{\frac{1}{a}}$, the
characteristic scale of soft wide-angle radiation. In fact, at NLL
accuracy, when we perform the sum over dipoles, this is the only term
that does not appear as a sum of contributions of each individual leg,
but rather depends on the geometry of the underlying three-jet
event. Introducing $C_\ell$, the colour factor of leg $\ell$ ($C_F$
for a quark and $C_A$ for a gluon), $R_{\ell,{\rm NLL}}(v)\equiv 2C_\ell r_\ell(L)$, and $R'_{\ell,{\rm NLL}}(v)\equiv 2C_\ell r'_\ell(L)$, we can recast $R_{\rm NLL}(v)$ in the form: 
\begin{multline}
  \label{eq:RNLL-final}
R_{\rm NLL}(v) =  \sum_{\ell} \left[R_{\ell,{\rm NLL}}(v) + 
      R'_{\rm NLL,\ell}(v)\left(\langle\ln(
      d_\ell\,g_\ell)\rangle-b_\ell \ln\frac{2E_\ell}{Q}\right) 
    +\gamma^{(0)}_\ell\, T\!\left(\frac{L}{a+b_\ell}\right) \right]\\+2\,T\!\left(\frac{L}{a}\right)\left(\sum_{ij\in\{q\bar q,qg,g\bar q\}}\!\!\! C_{(ij)} \ln\frac{Q_{ij}}{Q}\right) \,.
\end{multline}
Here we have introduced $\gamma_\ell^{(0)}=2 C_\ell B_\ell$, which is
minus the coefficient of $\delta(1-x)$ in the splitting function
$P^{(0)}_{qq}(x)$ if $p_\ell$ is a quark, and of $P^{(0)}_{gg}(x)$ if
$p_\ell$ is a gluon, namely
\begin{equation}
  \label{eq:gamma_0}
  \gamma_\ell^{(0)}\equiv
      \begin{cases}
      & -\frac{3}{2} C_F \,,\qquad \qquad\quad\>\text{$p_\ell$ is a quark}\,,\\
      & -\frac{11 C_A-2 n_f}{12 C_A} \,,\qquad\quad \text{$p_\ell$ is a gluon}\,.
      \end{cases}
\end{equation}
At NLL accuracy, the main role played by real emissions is that of
cancelling the infrared singularities of virtual corrections. Only
soft and collinear emissions give a non-trivial contribution,
represented by the function $\FNLL\left(\RpNLL\right)$ in
eq.~\eqref{eq:Sigma-NLL}, where 
\begin{equation}
  \RpNLL \equiv \sum_\ell R'_{\ell,{\rm NLL}}\,.
\end{equation}
To define $\FNLL(\RpNLL)$ we have to parameterise the momentum of each
emission $k_i$ in terms of the leg $\ell_i$ to which the emission is
collinear, the azimuthal angle $\phi_i^{(\ell_i)}$, and the variable
$\zeta_i\equiv V(\{p\},k)/v$.  For event shapes only, we do not have
to specify the rapidity of emission $k_i$, since it can be suitably
integrated analytically. We then obtain the compact expression
\begin{equation}
  \label{eq:FNLL}
  \FNLL(\RpNLL) = \int \dZ \Theta\left(1-\frac{\Vsc{\{k_i\}}}{v}\right)\,,
\end{equation}
where we have used the short-hand notation
\begin{equation}
  \label{eq:dZ}
   \int \dZ G(\{k_i\}) \equiv \lim_{\delta\to 0} \delta^{\RpNLL} \sum_{n=0}^\infty\frac{1}{n!}\int\prod_{i=1}^n \left(\sum_{\ell_i} R'_{\ell_i,{\rm NLL}}\int_\delta^\infty \frac{d\zeta_i}{\zeta_i}\int_0^{2\pi}\frac{d\phi_i^{(\ell_i)}}{2\pi}\right) G(k_1,\dots,k_n)\,.
\end{equation}
In the above expression, $\delta$ is a cutoff. All emissions with
$\zeta_i<\delta$ are unresolved, and together with virtual corrections
build the factor $\delta^{\RpNLL}$.  Note that $\FNLL(\RpNLL)$ has to
be computed using $\Vsc{k_1,\dots,k_n}$, the approximate expression of
$\Vfull{k_1,\dots,k_n}$ when all emissions are soft and collinear, and
the quantity $\Vsc{k_1,\dots,k_n}/v$ is only a function of
$\{\zeta_i,\ell_i,\phi_i^{(\ell_i)}\}$. In the ARES formalism, some
formal manipulations have to be performed to obtain
$\Vsc{k_1,\dots,k_n}$ from $\Vfull{k_1,\dots,k_n}$. The CAESAR
program~\cite{Banfi:2004yd}, which automatically resums all rIRC
final-state observables at NLL accuracy, instead generates actual
momenta $k_i$, and with appropriate phase-space cuts forces them to be
soft and collinear. In that limit
\begin{equation}
  \label{eq:Vsc-Vfull}
  \frac{\Vsc{k_1,\dots,k_n}}{v} = \lim_{v\to 0}\frac{\Vfull{k_1,\dots,k_n}}{v}\,,
\end{equation}
where $\Vfull{k_1,\dots,k_n}$ is the actual observable evaluated on
the soft-collinear momenta $k_1,\dots,k_n$. The property of rIRC
safety ensures that the limit in eq.~\eqref{eq:Vsc-Vfull} exists, and
is only a function of $\{\zeta_i,\ell_i,\phi_i^{(\ell_i)}\}$. Given
the correspondence in eq.~\eqref{eq:Vsc-Vfull}, we have
$\Vsc{k_i}=v \zeta_i$.

A last remark is in order. When dealing with multiple soft and/or
collinear partons, an emission might be collinear to one of the final
state partons $\{\tilde p\}$ without being collinear to the
corresponding Born momentum that initiated the three-jet event. This
needs caution in the parametrisation of the soft and/or collinear
phase space.  This issue has been discussed in detail in
ref.~\cite{Banfi:2004yd} and recalled in
ref.~\cite{Banfi:2018mcq}. The outcome is that the light-like momenta
required to perform a Sudakov decomposition according to
eq.~(\ref{eq:Sudakov-leg}), which we will refer to as the ``emitters'',
might need to be redefined after each emission. In general, the
emitters do not coincide with the Born momenta in
eq.~(\ref{eq:Born-momenta}), but are related to those via a mapping,
whose details can be found in ref.~\cite{Banfi:2004yd}. Note that the
emitters coincide with the Born momenta in the limit where all
emissions are infinitely soft and/or collinear.

\section{NNLL resummation}
\label{sec:nnll}

In this section, we explain how to extend the ARES formalism to
three-jet observables. For the most part we rely on the previous
results obtained in ref.~\cite{Banfi:2018mcq}, while making sure to
stress the new features that arise in three-jet events. The quantity
of interest is the cumulative cross-section which reads
\begin{align}
  \label{eq:cumdef}
  \Sigma_{\cal H}(v) = \frac{1}{\sigma_{\cal H}} \sum_{n=0}^\infty \int d\Phi_{\tilde{3}+n}\, \frac{d\sigma\left(\{ \tilde{p} \},k_1, \dots, k_n\right)}{d\Phi_{\tilde{3}+n}} \Theta\left(v - V(\pBorn,k_1,\dots, k_n)\right)\, \mathcal{H}(\{\tilde p\},k_1,\dots,k_n) \,,
\end{align}
where $d\Phi_{\tilde{3}+n}$ is the phase spece for the final-state
momenta $\{\tilde{p}\}$ and the secondary emissions $k_1,\dots,k_n$.
Here we remind the reader that the jet-selection function $\mathcal{H}$
implicitly contains a delta function for conservation of total
four-momentum. In the near-to-planar limit, i.e. $v\ll1$, all
emissions $(k_1,..., k_n)$ are either soft and/or collinear. In this
region of phase space it is always possible to provide an explicit
mapping between the final state,
$(\tilde{p}_1,\tilde{p}_2,\tilde{p}_3)$, and the Born momenta
$(p_1,p_2,p_3)$ in eq.~(\ref{eq:Born-momenta}), see e.g.\
ref.~\cite{Banfi:2018mcq}. Once the Born momenta have been identified,
in the limit $v\to 0$ the cumulative cross-section
$\Sigma_{\cal H}(v)$ assumes the factorised form~\cite{Banfi:2004yd}
\begin{align}
  \label{eq:cumbasic}
  \nonumber
\Sigma_{\cal H}(v) &\simeq \frac{1}{\sigma_{\cal H}}\int d\Phi_3\,\frac{d\sigma_3}{d\Phi_3}\, \mathcal{H}(\{p\}) \\
&\times \mathcal{V}(\{p\}) \, \sum_{n=0}^\infty  S(n)\int \left(\prod_{i=1}^n
  [dk_i]\right)
   \,\mathcal{M}^2 \left(\{ \tilde{p} \},k_1, \dots, k_n\right) \Theta\left(v - V(\pBorn,k_1,\dots, k_n)\right) \,,
\end{align}
where $S(n)$ is a symmetry factor, e.g.~$1/n!$ for $n$ identical
gluons. In the above equation $\mathcal{V}(\{p\})$ represents virtual
corrections to the Born process, an explicit function of the {\em
  Born} momenta $\{p\}\equiv\{p_1,p_2,p_3\}$, normalised by
Born matrix element squared. Moreover, $\mathcal{M}^2$ represents the
real corrections.  Both $\mathcal{M}^2$ and $\mathcal{V}\, (\{p\})$
are divergent in four-dimensions, we therefore consider all our
expressions to be regularised in some way. Specifically, we adopt
dimensional regularisation, and all the quantities are computed in
$d=4-2\epsilon$ dimensions.

\subsection{Soft and collinear factorisation of matrix elements}

A basic ingredient of ARES is the factorisation properties of
QCD squared matrix elements, real and virtual. This makes the structure of
divergences manifest and allows for the sought after cancellation of infrared
poles. In this subsection, we recall the form of factorised amplitudes which are
needed to build the NNLL resumed cumulative distribution.
\paragraph{Soft limit of squared matrix elements.} The fundamental premise
of ARES is the {\em analytic} cancellation of infrared singularities. To
this aim, we start with the soft limit of real radiation, where we
first notice the following factorisation\footnote{Notice that for strictly soft radiation the mapping of the final state to Born momenta is trivial.}
\begin{align}
\label{eq:M2-soft}
\mathcal{M}^2 \left(\{ \tilde{p} \},k_1, \dots , k_n\right) \simeq \frac{d\sigma_3}{d\Phi_3}\,\mathcal{M}_s^2 \left(k_1, \dots , k_n\right) \,.
\end{align}
Since the final state contains only three coloured legs, it remains
true that the soft limit of the squared matrix elements is factorised
in terms of ``soft correlated blocks'', similar to the two-jet case
(see e.g.~\cite{Falcioni:2014pka})
\begin{align}
  \label{eq:clusters}
\nonumber
\mathcal{M}_s^2(k_1) &\equiv \tilde{M}_{\rm s}^2(k_1) \\\nonumber 
\mathcal{M}_s^2(k_1, k_2) &= \tilde{M}_{\rm s}^2(k_1) \tilde{M}_{\rm s}^2(k_2) + \tilde{M}_{\rm s}^2(k_1,k_2) \\ 
\mathcal{M}_s^2(k_1, k_2, k_3) &= \tilde{M}_{\rm s}^2(k_1) \tilde{M}_{\rm s}^2(k_2)
                     \tilde{M}_{\rm s}^2(k_3) + \left(\tilde{M}_{\rm s}^2(k_1)
                     \tilde{M}_{\rm s}^2(k_2,k_3) + \text{perm.}\right) +
                     \tilde{M}_{\rm s}^2(k_1,k_2,k_3)\ , \nonumber\\
\vdots 
\end{align}
As we shall explain
below, each $\tilde{M}^2_{\rm s} (k_1, \dots, k_n)$ then takes the form of a
sum over {\em dipoles}. The above decomposition in eq.~(\ref{eq:clusters}) is a
property of QCD. Nevertheless, one has to realise that beyond three
jets an equivalent expression is very difficult to obtain and will
involve complicated colour correlations~\cite{Falcioni:2014pka}.

\paragraph{Hard-collinear radiation.} The second ingredient is radiation in
the hard-collinear region of phase space. At a fixed logarithmic
accuracy, we only need to consider a fixed number of hard-collinear
emissions. The corresponding NNLL contributions are generated by a
single hard-collinear parton $k_{\rm hc}$ plus an ensemble of soft \emph{and}
collinear emissions $k_1,\dots,k_n$, all emitted independently. In
this region of phase space, we have the following factorisation
\begin{align}
\label{eq:M2-hc}
  \mathcal{M}^2 \left(\{ \tilde{p} \},k_{\rm hc},k_1, \dots , k_n\right) \simeq
  \frac{d\sigma_3}{d\Phi_3}\, {M}_{\rm hc}^2(\{ \tilde{p} \},k_{\rm hc})\,
  \prod_{i=1}^n \tilde{M}_{\rm s}^2(k_i)\,.
\end{align}
and the hard-collinear matrix element explicitly depends on the Born
momenta through the mapping mentioned above. We will write
\begin{align}
[dk_{\text{hc}}] {M}_{\rm hc}^2(\{ \tilde{p} \},k_{\rm hc}) =  \sum_{\ell} [dk_{\text{hc}}] {M}_{{\rm hc}, f_{\ell}}^2(\{ \tilde{p} \},k_{\rm hc})  \,,
\end{align}
where $f_\ell=q,g$ is the flavour of leg $\ell$. For a quark (or anti-quark) leg, we have
\begin{align}
  \label{eq:quarkhc}
  [dk_{\text{hc}}] {M}_{{\rm hc}, q}^2(\{
  \tilde{p} \},k_{\rm hc}) =
  \frac{dk_t^2}{k_t^2}\left(\frac{  4\pi \mu_R^2\, e^{-\gamma_E}}{k_t^2}\right)^\epsilon \frac{\alpha_s(k_t)}{2\pi}\, \frac{d\Omega_{2-2\epsilon}}{\Omega_{2-2\epsilon}}  dz\, \langle P_q(z;\epsilon)\rangle \,,
\end{align}
where $k_t$ is the emission's transverse momentum with respect to its
{\em emitter}, defined according to the procedure explained in
ref.~\cite{Banfi:2004yd} and recalled in ref.~\cite{Banfi:2018mcq}. We
have the angular measure in the $2-2\epsilon$-dimensional
transverse plane
\begin{align}
d\Omega_{2-2\epsilon} = d\Omega_{1-2\epsilon} (\sin\phi)^{-2\epsilon} d\phi, \quad \phi\in[0,\pi],  \quad \Omega_{2-2\epsilon} = \frac{2 \pi^{1-\epsilon}}{\Gamma(1 - \epsilon)} \ \ .
\end{align}
The function 
\begin{align}
  \label{eq:Pq-average}
\langle P_q(z;\epsilon)\rangle \equiv C_F \left( 
-2+(1-\epsilon)z\right)
\end{align}
is an azimuthally averaged splitting function where, to avoid
double-counting the soft-collinear region, we have appropriately
eliminated the divergent part of the full splitting function for
$z\to 0$. For a gluon we have
\begin{align}
  \label{eq:gluonhc}
  [dk_{\text{hc}}] {M}_{{\rm hc},g}^2(\{
  \tilde{p} \},k_{\rm hc}) =&
  \frac{dk_t^2}{k_t^2}\left(\frac{4 \pi \mu_R^2\, e^{-\gamma_E}}{k_t^2}\right)^\epsilon \frac{\alpha_s(k_t)}{2\pi} \frac{d\Omega_{2-2\epsilon}}{\Omega_{2-2\epsilon}} dz
   \left( \langle P_g(z;\epsilon)\rangle +
    \mathcal{T}(\{p\}) \Delta P_g(z,\phi;\epsilon) \right) \ \ ,
\end{align}
where the gluon averaged splitting function, with the soft divergence
subtracted, is given by
\begin{align}
  \langle  P_{g}(z,\epsilon) \rangle = C_A \left(z (1-z) - 2\right) + T_R n_f \left[1-\frac{2z(1-z)}{1-\epsilon}\right] \ \ .
\end{align}
For collinear splittings of a gluon, we need to keep track of spin
correlations with the hard event. These are accounted for by the
{\em un-averaged} splitting function
\begin{align}
  \Delta P_g(z,\phi;\epsilon)  = 4 z (1-z) \left( 2(1-\epsilon) \cos^2\phi - 1 \right) \left(\frac{C_A}{2} - \frac{T_R n_f}{1-\epsilon} \right) \ \ .
\end{align}
which has the property
\begin{equation}
  \label{eq:dPG-averaged}
  \int_0^{\pi} \, d\phi (\sin\phi)^{-2\epsilon }\,\Delta P_g(z,\phi;\epsilon)=0\,.
\end{equation}
Spin correlations do not simply factorise from the Born amplitude, therefore, we
have to introduce a new function $\mathcal{T}(\{p\})$ of the Born momenta. In
our case, we have~\cite{Ellis:1980wv}
\begin{align}
\mathcal{T}(\{p\}) = \frac{x_1+x_2-1}{x_1^2 + x_2^2} \,,
\end{align}
where $x_1,x_2$ are the invariants of the Born event defined in
appendix~\ref{sec:kinematics}.

\paragraph{Virtual corrections.} The last ingredient is the virtual corrections
\begin{align}
\label{eq:virtual}
\mathcal{V}(\{p\}) =  H(\{p\}, \alpha_s(Q)) \times e^{- \mathcal{S}(\{p\},\alpha_s(Q))} \times e^{- \mathcal{J}(\alpha_s(Q))}\,,
\end{align}
where $H(\{p\}, \alpha_s(Q))$ is a finite hard function,
$\mathcal{S}(\{p\},\alpha_s(Q))$ is a soft function~\cite{Falcioni:2014pka}
containing all soft singularities, while finally $\mathcal{J}(\alpha_s(Q))$
encapsulates all hard-collinear singularities. We choose to incorporate a
dependence on the Born momenta in the soft function, which we can always perform
provided we appropriately adjust the hard function at each fixed order in the
strong coupling. For the sake of clarity, let us pause and further discuss the
function $\mathcal{J}$. The latter admits the expression~\cite{Gardi:2009qi}
\begin{align}
  \label{eq:jet-function}
\mathcal{J}(\alpha_s(Q)) = \sum_{\ell =1}^3 \int^{Q^2}\frac{dk^2}{k^2} \sum_{n=1}^{\infty} \left( \frac{\alpha_s(k,\epsilon)}{2\pi}\right)^n  \gamma_{\ell}^{(n\!-\!1)} \ \ ,
\end{align}
where $\gamma_{\ell}^{(n\!-\!1)}$ comprise the coefficient of
$\delta(1\!-\!x)$ in the \emph{Altarelli-Parisi} splitting function
$P_{qq}^{(n\!-\!1)}(x)$ if leg $\ell$ is a quark or antiquark, and of
$P_{gg}^{(n\!-\!1)}(x)$ if leg $\ell$ is a gluon. The function
$\alpha_s(k,\epsilon)$ is the running coupling in $d=4\!-\!2\epsilon$
dimensions, defined as the solution of the $d$-dimensional
renormalisation group equation:
\begin{equation}
\mu_R^2\frac{d\alpha_s}{d\mu_R^2} = -\epsilon\, \alpha_s + \beta^{\rm (d=4)}(\alpha_s)\,,
\end{equation}
where $\beta^{\rm (d=4)}$ is the beta function in four dimensions,
given by the following expansion
\begin{equation}
  \label{eq:beta-function}
  \beta^{\rm (d=4)}(\alpha_s) = - \alpha_s^2 \sum_{n=0}^\infty \beta_n \alpha_s^n\,.
\end{equation}
In anticipation of our next steps, we separate the collinear jet
function as follows
\begin{align}
  \label{eq:jet-expanded}
e^{- \mathcal{J}(\alpha_s(Q))} \simeq e^{- R_{\rm hc}(v)}\,\left(  1 - \sum_{\ell=1}^{3} \int^{v^{\frac{2}{a+b_\ell} }Q^2}\!\! \frac{dk^2}{k^2} \left( \frac{\alpha_s(k,\epsilon)}{2\pi}\right)  \gamma_{\ell}^{(0)} \right)\,,
\end{align}
where the neglected terms, when combined with the corresponding real
corrections, give rise to N$^3$LL contributions. Last, the
hard-collinear radiator $R_{\rm hc}(v)$ can be defined at all
logarithmic orders as follows\footnote{The scale choice at which to partition the integral in eq.~\eqref{eq:Rhc} is motivated by considering the maximum emission's rapidity at a fixed observable value. Alternatively, one can motivate such choice by comparing the expansion of the resummation to fixed order results.}:
\begin{equation}
  \label{eq:Rhc}
  R_{\rm hc}(v) = \sum_{\ell =1}^3 \int^{Q^2}_{v^{\frac{2}{a+b_\ell} }Q^2}\frac{dk^2}{k^2} \sum_{n=1}^{\infty} \left(\frac{\alpha_s(k)}{2\pi}\right)^n \gamma_{\ell}^{(n-1)} \,.
\end{equation}
Last we have the hard function, $H(\{p\}, \alpha_s(Q))$, which is
quite involved as it captures all the finite terms in the one-loop
corrections to the Born event. Explicitly, in our case we have
\begin{align}
\nonumber
H(\{p\}, \alpha_s(Q))  = \sigma_0 &\bigg[ 1+ \frac{\alpha_s}{2\pi} \left( C_F \bigg( \frac{7\pi^2}{6} - 8 - \ln^2(1-x_3) \right) \\\nonumber
&+ C_A \left( \frac{7\pi^2}{12} + \frac12 (\ln^2(1-x_3) - \ln^2(1-x_1) - \ln^2(1-x_2) ) \right) \bigg) \\
& + \frac{\alpha_s}{2\pi} \frac{(1-x_1)(1-x_2)}{x_1^2 + x_2^2} F(x_1,x_2,x_3)\bigg] \,,
\end{align}
where $F(x_1,x_2,x_3)$ is given in ref.~\cite{Ellis:1980wv} and
$\sigma_0$ is the Born quark-antiquark total cross section.  Now that
we have the various ingredients of eq.~\eqref{eq:cumbasic}, we can
define two cumulants, a soft and a hard-collinear cumulant, each
encoding a separate non-overlapping portion of phase space.

\subsection{Soft cumulative distribution}
\label{eq:soft-dist}
The soft cumulant $\Sigma_{\rm soft}(v)$ is defined as
\begin{equation}
  \label{eq:softcumulant}
  \begin{split}
  \Sigma_{\rm soft}(v) = & e^{- R_{\rm hc}(v)} \int d\Phi_3\, \frac{d\sigma_3}{d\Phi_3} \mathcal{H}(\{p\})\, H(\{p\}, \alpha_s(Q))\, e^{- \mathcal{S}(\{p\},\alpha_s(Q))} \times \\
&  \times \sum_{n=1}^\infty S(n)\int \left(\prod_{i=1}^n
  [dk_i]\right)  \,\mathcal{M}_s^2 \left(k_1, \dots , k_n\right) \Theta\left(1 - \lim_{v\to 0} \frac{V(\pBorn,k_1, \dots, k_n)}{v} \right) \,,
  \end{split}
\end{equation}
which, as it stands, holds up to any logarithmic accuracy. 


\paragraph{Exponentiation.}
The cancellation of the soft divergences is performed in two steps. First, we {\em
  define} a resolution variable, and divide the soft emissions into
resolved and unresolved {\em clusters} according to the value of the
resolution variable. The clustering algorithm is explained in
ref.~\cite{Banfi:2018mcq}. The resolution variable is a fake
observable designed to cancel the soft divergences, with the only
condition being that it has to share the same leading logs as the
full observable. A natural choice is the soft-collinear limit of the
observable in the presence of a single emission given in
eq.~(\ref{eq:Vsc}).

The second step is to express the soft function, $\mathcal{S}$, in
terms of the soft blocks in eq.~(\ref{eq:clusters}) 
\begin{align}
\mathcal{S}(\{p\},\alpha_s(Q))  &= \int \frac{d^{d}k}{(2\pi)^{d}} \mathcal{W}(\{p\},\alpha_s(Q),k) \\
\mathcal{W}(\{p\},\alpha_s(Q),k) &\equiv \sum_{n=1}^\infty S(n)\int \left(\prod_{i=1}^n
  [dk_i]\right) \tilde{M}_{\rm s}^2(k_1,\dots,k_n) (2\pi)^{d}\delta^{(d)}(k-\sum_i
  k_i) \label{webdef} \ \ ,
\end{align}
where the function $\mathcal{W}$ is called a web, whose properties will
be discussed later on. Note that, the fact that we have written the
soft function in term of an integral over real emission matrix
elements implies definite kinematical boundaries for the
$k$-integration. Note that our representation is a choice. Other
representations, provided they correctly incorporate the soft singularities, lead to a
different hard function $H(\{p\},\alpha_s(Q))$.

The unresolved clusters drop from the theta function in
eq.~(\ref{eq:softcumulant}) and thus the unresolved soft blocks
exponentiate trivially\footnote{The property of rIRC safety guarantees that unresolved clusters contribute, at most, power corrections to the cross section.}. Hence, the soft cumulant becomes
\begin{align}\label{eq:softcumbasic}
\nonumber
\Sigma_{\rm soft}(v) = & e^{- R_{\rm hc}(v)} \,\frac{1}{\sigma_{\mathcal{H}}} \int d\Phi_{3}\, \frac{d\sigma_3}{d\Phi_3}  H(\{p\}, \alpha_s(Q)) \mathcal{H}(\{p\})  e^{- R_{\rm s}( v;\{p\})} \\\
&\times e^{- R_{\rm s}(\delta v;\{p\})} e^{ R_{\rm s}( v;\{p\})}  \sum_{n=1}^\infty S(n) \int_{\delta v} \left(\prod_{i=1}^n
  [dk_i]\right) \,\mathcal{M}_s^2 \left(k_1, \dots , k_n\right) \Theta\left(1 -\lim_{v\to 0} \frac{V(\pBorn,k_1, \dots, k_n)}{v} \right)  \ \ ,
\end{align}
where all soft divergences have been cancelled leaving a
manifestly finite Sudakov radiator
\begin{align}
  \label{eq:soft-radiator}
 R_{\rm s}(x;\{p\}) \equiv \int \frac{d^4k}{(2\pi)^4}\, \mathcal{W}(\alpha_s(Q),k)\, \Theta\left(V_{\rm sc}(\{\tilde{p}\}, k) - x \right) \ \ .
\end{align} 
Note that in eq.~(\ref{eq:softcumbasic}), the soft radiator is still a function
of the Born momenta and therefore still appears inside the integral over the
Born phase space. To simplify notation we will drop the explicit dependence on
the Born phase space in the remainder of the paper. In the above we have the
resolution parameter $\delta$, upon which the phase space is clustered. In
particular, the phase space of the resolved clusters is bounded from below by
making the corresponding resolution variable of each cluster bigger than $\delta
v$. All expressions in eq.~\eqref{eq:softcumbasic} are in four dimensions, and
the dependence on $\delta$ cancels out in all of the final expressions.

\subsubsection{Soft Radiator}
\label{sec:soft-radiator}
It is in fact quite straightforward to compute the soft radiator in
eq.~\eqref{eq:soft-radiator} by directly utilising the two-jet results
in ref.~\cite{Banfi:2018mcq}. First, we realise that the soft blocks
naturally take the form of a sum over dipoles. For example, the
single-emission soft block at tree level reads 
\begin{align}\label{eq:softamplitude}
\tilde{M}^2_s(k) = (4\pi \alpha_s)  \sum_{(ij)} C_{(ij)} \frac{(p_i p_j)}{(p_i k)(p_j k)}
\end{align}
where in our case
\begin{align}
C_{(q\bar{q})} = 2C_F - C_A, \quad C_{(gq)} = C_{(g\bar{q})} = C_A \ \ .
\end{align}
\noindent This allows us to express the web function as a sum over dipole webs as follows
\begin{align}
\mathcal{W}(\{\tilde p\},\alpha_s(Q),k) = \sum_{(ij)} C_{(ij)} w(\{\tilde p\},\alpha_s(Q),k^{(ij)})  \,.
\end{align}
The notation $k^{(ij)}$ stresses the fact that each
dipole web $w(\{\tilde p\},\alpha_s(Q),k^{(ij)})$ enjoys the property
\begin{align}
w(\{\tilde p\},\alpha_s(Q),k^{(ij)})  = w(\{\tilde p\},\alpha_s(Q),k^2,k^2+\kappa_{(ij)}^2) \,,
\end{align}
hence it is natural to express the momentum of the web in terms of the
dipole variables in eq.~(\ref{eq:Sudakov-k}).  In order to compute the
soft radiator we need to express also the resolution observable,
$V_{\rm sc}(k)$, in terms of same variables. This gives
\begin{equation}
  \label{eq:Vsc-ij}
  V_{\rm sc}(\{\tilde p\},k^{(ij)})= \sum_{\ell \in (ij)} d^{(ij)}_\ell\left(\frac{\kappa_{(ij)}}{Q_{ij}}\right)^a e^{- b_\ell \eta_\ell^{(ij)}} g_\ell(\phi^{(ij)}) \Theta(\eta_{\ell}^{(ij)})\,,
\end{equation}
where 
\begin{align}
\eta_i^{(ij)} = \eta^{(ij)} \ \ , \quad \eta_j^{(ij)} = - \eta^{(ij)} \ \,
\end{align}
and $d^{(ij)}_\ell$ is defined in such a way that $V_{\rm sc}^{(ij)}(k)$
reduces to the expression in eq.~(\ref{eq:Vsc}) when $k$ is collinear
to leg $\ell$. This gives the relation between $d^{(ij)}_\ell$ and the coefficient $d_\ell$ introduced in eq.~(\ref{eq:Vsc}):
\begin{equation}
  \label{eq:dell-ij}
  d^{(ij)}_\ell = d_\ell\,\left(\frac{Q_{ij}}{Q}\right)^a\,\left(\frac{Q_{ij}}{2 E_\ell}\right)^{b_\ell}\,.
\end{equation}
This allows us to define a soft radiator for each dipole as follows 
\begin{align}\label{eq:radijoriginal}
\mathcal{R}_{\ell}^{(ij)} (v) \equiv & \int \frac{d^4 k^{(ij)}}{(2\pi)^4}  w(\{\tilde p\},m^2,\kappa^2_{(ij)}+m^2)\, \Theta\left(d^{(ij)}_\ell \left(\frac{\kappa_{(ij)}}{Q_{ij}}\right)^a e^{- b_\ell \eta_\ell^{(ij)}} g_\ell(\phi^{(ij)})  -  v\right) \Theta(\eta_\ell^{(ij)}) \ \ ,
\end{align}
in terms of which the total soft radiator becomes
\begin{align}
R_{\text{s}} (v) = \sum_{(ij)} C_{(ij)} \sum_{\ell \in (ij)} \mathcal{R}_{\ell}^{(ij)} (v) \ \ .
\end{align}
Notice that in eq.~(\ref{eq:radijoriginal}) the sole dependence on the
Born kinematics is due to $d^{(ij)}_\ell$.  Furthermore, we
realise that the phase space measure in eq.~(\ref{eq:radijoriginal})
contains the rapidity of a massive web
momentum. Explicitly,
\begin{align}
d^4k^{(ij)} = \frac12 dy^{(ij)}\, dm^2 \, d\phi^{(ij)} \, \kappa_{(ij)} d\kappa_{(ij)}, \quad m^2 \equiv k^2 \ \ ,
\end{align}
where the rapidity is bounded as follows
\begin{align}
  \label{eq:rapidity-bound}
|y^{(ij)}| < \frac12 \ln \frac{Q_{(ij)}^2}{\kappa_{(ij)}^2 + m^2} \ \ .
\end{align}
Since the web is uniform is rapidity, ref.~\cite{Banfi:2018mcq} presented a simple
strategy to extract the mass dependence of the web. Hence, each dipole
contributes a radiator
\begin{align}\label{eq:radseparate}
\mathcal{R}_{\ell}^{(ij)} (v)  = \mathcal{R}_{\ell,\text{0}}^{(ij)} (v)  + \delta R_{\ell}^{(ij)} (v)  \ \ ,
\end{align}
where the subscript `$0$' means that eq.~(\ref{eq:radijoriginal}) is to
be evaluated with massless rapidity bounds, while
$\delta R_{\ell}^{(ij)} (v) $ is a mass correction that accounts for
the correct rapidity boundary of eq.~\eqref{eq:rapidity-bound}. At
NNLL accuracy, this is given by
\begin{multline}
  \label{eq:dR-mass}
   \delta R_{\ell}^{(ij)} (v)=\int \frac{d^4 k^{(ij)}}{(2\pi)^4}  w(\{\tilde p\},m^2,\kappa^2_{(ij)}+m^2)\,\Theta\left(\kappa^{(ij)}-v^{\frac{1}{a+b_\ell}} Q_{ij}\right)\times \\ \times\left[\Theta\left(\ln\sqrt{\frac{Q_{ij}^2}{\kappa^{(ij)}+m^2}}-\eta_\ell^{(ij)}\right)-\Theta\left(\ln\sqrt{\frac{Q_{ij}^2}{\kappa^{(ij)}}}-\eta_\ell^{(ij)}\right)\right]\,.
\end{multline}
To simplify the calculation of
$\mathcal{R}_{\ell,\text{0}}^{(ij)} (v)$ further we separate out the
dependence on the Born momenta and on the azimuthal angle by expanding
the step functions as follows
\begin{equation}
  \label{eq:thetaexpansion}
  \begin{split}
  &\Theta\left(d^{(ij)}_\ell \left(\frac{\kappa_{(ij)}}{Q_{ij}}\right)^a e^{- b_\ell \eta^{(ij)}} g_\ell(\phi^{(ij)})   -  v\right) =
    \Theta\left(\left(\frac{\kappa_{(ij)}}{Q_{ij}}\right)^a e^{- b_\ell \eta^{(ij)}}  -  v\right) \\ 
&+ \delta \left(\ln\left[ \left(\frac{\kappa_{(ij)}}{Q_{ij}}\right)^a e^{- b_\ell \eta^{(ij)}} \right] - \ln v \right) \ln\left(d_\ell g_\ell(\phi^{(ij)}) \right) \\ 
&+\frac12\,\delta^\prime \left( \ln\left[ \left(\frac{\kappa_{(ij)}}{Q_{ij}}\right)^a e^{- b_\ell \eta^{(ij)}} \right] -\ln v \right) \ln^2\left(d_\ell g_\ell(\phi^{(ij)}) \right) + \dots \,,
  \end{split}
\end{equation}
where we truncated appropriately for NNLL resummation. Using
eq.~(\ref{eq:thetaexpansion}), we write
\begin{align}
\label{eq:Rij}
\mathcal{R}_{\ell,\text{0}}^{(ij)} (v) = R_{\ell,\text{0}}^{(ij)} (v) + \left(R_{\ell,\text{0}}^{(ij)}\right)^{\prime} (v)\, \big\langle \ln\left(d^{(ij)}_\ell g_\ell  \right) \big\rangle \\ \nonumber
+ \frac12 \left(R_{\ell,\text{0}}^{(ij)}\right)^{\prime\prime} (v) \big\langle \ln^2\left(d^{(ij)}_\ell g_\ell \right) \big\rangle+\dots \ \ ,
\end{align}
where
\begin{equation}
  \label{eq:Rol}
  R_{\ell,\text{0}}^{(ij)} (v) =\int \frac{d^4 k^{(ij)}}{(2\pi)^4}  w(\{\tilde p\},m^2,\kappa^2_{(ij)}+m^2)\, \Theta\left(\left(\frac{\kappa_{(ij)}}{Q_{ij}}\right)^a e^{- b_\ell \eta_\ell^{(ij)}}  -  v\right) \Theta(\eta_\ell^{(ij)}) \ \ ,
\end{equation}
and
\begin{align}
\left(R_{\ell,\text{0}}^{(ij)}\right)^{\prime} = - v \frac{dR_{\ell,\text{0}}^{(ij)} (v)}{dv}, \quad  \left(R_{\ell,\text{0}}^{(ij)}\right)^{\prime\prime} = - v \frac{d\left(R_{\ell,\text{0}}^{(ij)}\right)^{\prime} (v)}{dv}\,.
\end{align}
With this setup, it is straightforward to see that the dipole radiator in
Eq.~(\ref{eq:Rol}) is identical to the soft radiator of two-jet observables,
which has been obtained in ref.~\cite{Banfi:2018mcq}. The only difference is
that now $\alpha_s$ is a function of the invariant mass of each dipole, instead
of the hard scale $Q$ as in \cite{Banfi:2018mcq}. Nevertheless, we can re-expand
the coupling to cast our results as a function of the resummation variable
\begin{align}
\lambda \equiv \alpha_s(Q) \beta_0 \ln\left(\frac{1}{v}\right) , \quad \quad \beta_0 = \frac{11 C_A - 2 n_f}{12 \pi} \ \ .
\end{align}
Implementing these steps we obtain
\begin{align}
R_{\ell,\text{0}}^{(ij)} (v) &= - \frac{\lambda}{\alpha_s(Q) \beta_0} g^{(\ell)}_{1}(\lambda) - g^{(\ell)}_{2} (\lambda) - \mathfrak{g}^{(\ell)}_{2} (\lambda,Q_{ij})  - \frac{\alpha_s(Q)}{\pi} \left(g^{(\ell)}_{3} (\lambda) + \mathfrak{g}^{(\ell)}_{3} (\lambda, Q_{ij}) \right) \label{masslessrad}  \ \ , \\
\delta R_{\ell}^{(ij)} (v)  &= -  \frac{\alpha_s(Q)}{\pi} \delta g^{(\ell)}_{3} (\lambda) \label{masscor} \ \ ,
\end{align}
where
\begin{align}
&g_1^{(\ell)} (\lambda) =   \frac{ (a+b_\ell-2\lambda) \ln \left(1 - \frac{2\lambda}{a+b_\ell}\right) - (a - 2 \lambda ) \ln \left(1 - \frac{ 2 \lambda }{a}\right)}{4\pi  b_\ell \beta _0 \lambda} \label{eq:g1} \ \ , \\\nonumber
&g_2^{(\ell)} (\lambda) = \bigg[ \frac{ K^{(1)} \left(a \ln \left( 1 - \frac{2 \lambda
                           }{a}  \right)- (a+b_\ell) \ln \left( 1 - \frac{2 \lambda
                           }{a+b_\ell}\right)\right)}{8 \pi ^2 b_\ell \beta _0^2}  \\\nonumber
                         &+  \frac{\beta _1 (a+b_\ell) \ln ^2\left( 1 -\frac{2\lambda}{a+b_\ell}  \right)}{8 \pi b_\ell \beta_0^3} + \frac{\beta_1 (a+b_\ell)
                           \ln \left(1 - \frac{2 \lambda }{a+b_\ell}\right)}{4\pi b_\ell \beta_0^3} \\
                         & - \beta_1\frac{a \ln \left( 1 - \frac{2 \lambda }{a}\right) \left(\ln \left(1- \frac{2 \lambda
                           }{a}\right)+2\right)}{8 \pi b_\ell \beta_0^3} \bigg] \label{eq:g2} \ \ , \\
&\mathfrak{g}_2^{(\ell)}(\lambda,Q_{ij}) =  - \lambda^2 \frac{dg_1^{(\ell)}}{d\lambda} \ln\left(\frac{Q_{ij}^2}{Q^2}\right) \label{eq:g2frak} \ \ ,  \\\nonumber                          
&g_3^{(\ell)} (\lambda) =  \bigg[ K^{(1)}  \frac{
                           \beta _1 \left( a^2 (a+b_\ell-2 \lambda )
                           \ln \left( 1 - \frac{2 \lambda }{a}\right)
                           - (a+b_\ell)^2 (a-2\lambda) \ln \left( 1 -
                           \frac{2 \lambda }{a+b_\ell}\right) + 6
                           b_\ell \lambda^2 \right)}{8 \pi b_\ell
                           \beta_0^3 (a-2\lambda) (a+b_\ell-2\lambda)}
  \\\nonumber
&+\frac{\left(\text{$\beta_1 $}^2 (a+b_\ell)^2 (a-2 \lambda ) \ln ^2\left(1-\frac{2 \lambda }{a+b_\ell}\right)-4 b_\ell \lambda ^2 \left(\text{$\beta_0 $}
   \text{$\beta_2 $}+\text{$\beta_1 $}^2\right)\right)}{8 b_\ell
\text{$\beta_0 $}^4 (a-2 \lambda ) (a+b_\ell-2 \lambda )}\\\nonumber
&-\frac{a \ln \left(1-\frac{2 \lambda }{a}\right) \left(2 \text{$\beta_0 $} \text{$\beta_2 $} (a-2 \lambda )+a \text{$\beta_1 $}^2 \ln
   \left(1-\frac{2 \lambda }{a}\right)+4 \text{$\beta_1 $}^2 \lambda
 \right)}{8 b_\ell \text{$\beta_0 $}^4 (a-2 \lambda )}\\\nonumber
&+\frac{(a+b_\ell) \ln \left(1-\frac{2 \lambda }{a+b_\ell}\right) \left(\text{$\beta_0 $} \text{$\beta_2 $} (a+b_\ell-2 \lambda )+2 \text{$\beta_1 $}^2 \lambda 
   \right)}{4 b_\ell \text{$\beta_0 $}^4 (a+b_\ell-2 \lambda )}\\
                         &- K^{(2)} \frac{ 2 \lambda^2 }{16 \pi^2 (a-2\lambda) (a+ b_\ell - 2\lambda) \beta_0^2} \bigg] \label{eq:g3}  \ \ , \\ 
&\mathfrak{g}_3^{(\ell)}(\lambda, Q_{ij}) = \pi \beta_0 \lambda^2 \left( \frac{g_1^{(\ell)}(\lambda)}{d\lambda}  + \frac{\lambda}{2} \frac{d^2g_1^{(\ell)}(\lambda)}{d\lambda^2}\right)  \ln^2\left(\frac{Q_{ij}^2}{Q^2}\right)  \\
&- \pi \beta_0 \lambda \left(\frac{dg_2^{(\ell)}(\lambda)}{d\lambda}   + \frac{\beta_1}{\beta_0^2} \lambda \frac{d g_1^{(\ell)}(\lambda)}{d\lambda} \right)   \ln\left(\frac{Q_{ij}^2}{Q^2}\right)  \label{eq:g3frak}            \ \ ,                     
\end{align}
and $K^{(1)}$ and $K^{(2)}$ are the coefficients of the soft physical coupling
defined in ref.~\cite{Banfi:2018mcq}. Finally, the mass correction reads
\begin{align}
 \delta g_3^{(\ell)}(\lambda) = - \zeta(2) \frac{\lambda}{2\left(a + b_\ell - 2\lambda\right)}  \ \ ,
 \end{align}
which, up to NNLL accuracy, has no dependence on the dipole kinematics.

\subsubsection{Hard-collinear radiator}
\label{sec:unres-hc}

In ARES, the Sudakov radiator receives a contribution from the virtual
hard-collinear region truncated at the collinear scale\footnote{Although the jet
  function, eq.~\eqref{eq:jet-function}, only captures the hard-collinear poles,
  it remains true that the transcendental terms in eqs.~\eqref{B2q} and
  \eqref{B2g} come from soft-regular terms.}
\begin{align}
v_{\text{hc}} = v^{\frac{1}{a+b_\ell}} Q
\end{align}
which is completely independent of the dipole structure, i.e. it only knows
about the emitting leg, and hence the dependence on $b_\ell$. Following
ref.~\cite{Banfi:2018mcq}, we write
\begin{align}
R_{\text{hc}} (v) = \sum_{\ell} R_{\text{hc},\ell} (v) 
\end{align}
where
\begin{align}
R_{\text{hc},\ell} (v)  = - h_2^{(\ell)}(\lambda) - \frac{\alpha_s}{\pi} h_3^{(\ell)}(\lambda) \ \ .
\end{align}
The various functions are expressed in terms of the coefficient of $\delta(1-x)$
in the regularised Altarelli-Parisi as follows
\begin{align}
h_2^{(\ell)}(\lambda) &= \frac{\gamma_\ell^{(0)}}{2 \pi \beta_0} \ln \left( 1 - \frac{2\lambda}{a+b_\ell} \right) \label{eq:h2} \ \ , \\
h_3^{(\ell)}(\lambda) &= \gamma^{(0)}_\ell \frac{\beta_1 \left((a+b_\ell) \left(\ln \left( 1- \frac{2 \lambda}{a+b_\ell}\right)\right)+ 2\lambda \right)}{2\beta^2_0 \left(a + b_\ell - 2 \lambda\right)} -\gamma^{(1)}_\ell \frac{\lambda}{2\pi \beta_0 (a+b_\ell-2\lambda)} \label{eq:h3} \ \ ,
\end{align}
where
\begin{align}
\gamma^{(0)}_{q} &=\gamma^{(0)}_{\bar{q}}  = -\frac{3}{2} C_F, \quad   \gamma^{(0)}_{g} = - 2 \pi \beta_0          \ \ , \\ 
\gamma^{(1)}_{q,\bar{q}} & = - \frac{C_F}{2} \left(C_F \left(\frac34 - \pi^2 + 12 \zeta_3\right) + C_A \left( \frac{17}{12} + \frac{11 \pi^2}{9} - 6 \zeta_3 \right) - n_f \left(\frac{1}{6} + \frac{2\pi^2}{9}\right) \right) \label{B2q} \ \ , \\
\gamma^{(1)}_{g} &= \frac{n_f}{2} C_F + \frac23 n_f C_A - C_A^2 \left(\frac83 + 3 \zeta_3 \right)\label{B2g} \ \ .
\end{align}
This concludes the analytic construction of the Sudakov radiator. 

\subsubsection{Soft resolved clusters and correction functions}

The contribution of resolved clusters to the cumulant, the second line
of eq.~(\ref{eq:softcumulant}), arranges itself in the form of a
correction function, where contributions with successive logarithmic
accuracy can be systematically extracted. This correction function is
\begin{align}
\label{eq:Fsoft}
  \mathcal{F}_{\text{s}}(v) = e^{- R_{\rm s}(\delta v; \{p\})} e^{ R_{\rm s}( v; \{p\})}  \sum_{n=1}^\infty S(n) \int_{\delta v} \left(\prod_{i=1}^n [dk_i]\right) \,\mathcal{M}_s^2 \left(k_1, \dots , k_n\right) \Theta\left(1 - \lim_{v\to 0}\frac{V(\pBorn,k_1, \dots, k_n)}{v} \right) \ \ .
\end{align}
The above expression starts at NLL accuracy. The next-to-leading
logarithms emerge only from soft and collinear emissions widely
separated in angle, which build the function $\FNLL$. The remaining
contributions emerge from relaxing, one at a time, the approximations
performed to obtain $\FNLL$. This systematic procedure gives rise to
various NNLL functions, which we analyse in the following. To avoid
clutter of notation, we will drop the dependence on $\{p\}$ in the
soft radiator $R_{\rm s}$.

The first thing to notice is that the observable in eq.~\eqref{eq:Fsoft} is
\emph{not} constrained to be evaluated in the soft limit. In fact, one of the
gluons $k_1, \dots, k_n$ can be hard and collinear, although emitted with the
soft matrix element. This introduces an overlap between the soft and hard
collinear regions, which we need to isolate. Let us denote this hard-collinear
gluon as $k_{\rm hc}$. With this gluon, we have
\begin{equation}
\label{eq:Vhc}
\lim_{v\to 0}\frac{V(\pBorn,k_{\rm hc},k_1, \dots, k_n)}{v}\equiv \frac{V_{\rm hc}(\pBorn,k_{\rm hc},k_1, \dots, k_n)}{v}\,.
\end{equation}
Therefore, subtracting the double counting with the soft-collinear region, we
obtain the following NNLL contribution
\begin{equation}
  \label{eq:Fs/hc}
\begin{split}
  \mathcal{F}_{\rm s/hc} &\equiv e^{- R_{\rm s}(\delta v)} e^{ R_{\rm s}( v)} \int [dk_{\rm hc}]  \sum_{n=1}^\infty S(n) \int_{\delta v} \left(\prod_{i=1}^n [dk_i]\right) \,\mathcal{M}_s^2 \left(k_{\rm hc}, k_1, \dots , k_n\right) \\
                           &\times \left[ \Theta\left(1 - \frac{V_{\text{hc}}(\pBorn, k_{\rm hc},k_1, \dots, k_n)}{v} \right) - \Theta\left(1 - \frac{V_{\text{sc}}(\pBorn, k_{\rm hc}, k_1, \dots, k_n)}{v} \right) \right]\,.
\end{split}
\end{equation}
This contribution will be incorporated in the function $\dFrec$, to be discussed
later on in section~\ref{sec:hard-collinear} along with the other NNLL functions
of hard-collinear origin.

Given this logic, another contribution naturally arises when one of the
emissions is soft, but emitted at large angle. This gives the
following correction
\begin{align}
\nonumber
\mathcal{F}_{\text{wa}}(v) &\equiv \ e^{- R_{\rm s}(\delta v)} e^{ R_{\rm s}( v)} \int [dk]  \sum_{n=1}^\infty S(n) \int_{\delta v} \left(\prod_{i=1}^n [dk_i]\right) \,\mathcal{M}_s^2 \left(k, k_1, \dots , k_n\right) \\
                           &\times \left[ \Theta\left(1 - \frac{V_{\text{wa}}(\pBorn, k,k_1, \dots, k_n)}{v} \right) - \Theta\left(1 - \frac{V_{\text{sc}}(\pBorn, k, k_1, \dots, k_n)}{v} \right) \right]
\label{eq:wideangle} \ \ ,
\end{align}
where, in eq.~(\ref{eq:wideangle}), $V_{\text{wa}}$ means that we need
to probe the observable in the limit when a single soft gluon $k$ is
emitted at large angles. In the second step function, the observable
is evaluated as if $k$ were soft and collinear. Without this subtraction,
$\mathcal{F}_{\rm wa}$ would contain NLL terms that are part of the function
$\FNLL$ defined in eq.~(\ref{eq:FNLL}). Notice that we dropped the
resolution parameter, $\delta$, from the integral over the extra
emission, $k$, since the difference of theta functions renders the
result finite in the limit $\delta \to 0$. Notice importantly that
$\mathcal{F}_{\rm wa}$, as it stands, contain contributions beyond
NNLL accuracy. We will show below how to isolate the NNLL
contributions. When this is done we get
$\mathcal{F}_{\rm wa}\simeq(\alpha_s/\pi) \dFwa(\lambda)$. The explicit expression of $\dFwa(\lambda)$ will be discussed later.

After extracting $\mathcal{F}_{\rm wa}$ from eq.~(\ref{eq:Fsoft}), we observe
that we still have logarithms of arbitrary accuracy. At NLL, it suffices to
treat all soft emissions as independent, while starting at NNLL we need to take
into account the correlated portion of the double-soft squared matrix element,
as follows\footnote{The factor of 1/2 in the phase space of correlated partons
  is strictly for identical gluons, therefore, one has to multiply the
  $q\bar{q}$ portion by 2.}:
\begin{align}\label{eq:scbasic}
\nonumber
  \mathcal{F}_{\text{s}}(v)  &= \mathcal{F}_{\rm wa}(v) \\\nonumber
  &+ e^{- R_{\rm s}(\delta v)} e^{ R_{\rm s}( v)}  \sum_{n=1}^\infty \frac{1}{n!} \int_{\delta v} \left(\prod_{i=1}^n [dk_i]\right) \, \tilde{M}_s^2 \left(k_i \right) \Theta\left(v - V_{\text{sc}}(\pBorn,k_1, \dots, k_n) \right) \\\nonumber
  &+ e^{- R_{\rm s}(\delta v)} e^{ R_{\rm s}( v)} \sum_{n=1}^\infty \frac{1}{n!} \int_{\delta v} \left(\prod_{i=1}^n [dk_i]\right) \, \tilde{M}_s^2 \left(k_i \right) \times \\ 
  &\times \frac{1}{2!} \int_{\delta v} [dk_a] [dk_b] \tilde{M}^2_{\text{s}}(k_a,k_b) \Theta\left(v - V_{\text{sc}}(\pBorn,k_a, k_b, k_1, \dots, k_n) \right) \ \ .
\end{align}
The above form is valid up to NNLL accuracy.  We now aim to re-arrange
it for manifestly finite integrals that produce exact logarithmic
accuracy. We concentrate first on the part that contains the
correlated matrix element $\tilde{M}^2_{\text{s}}(k_a,k_b)$. The 
dependence on $\delta$ in the integration over $(k_a,k_b)$ can be
eliminated in the most elegant way by replacing the strong coupling
for each soft emission $\tilde{M}_s^2 \left(k_i \right)$, eq.~(\ref{eq:softamplitude}),
by the physical coupling that defines the soft radiator
\cite{Banfi:2018mcq}
\begin{align}
  \label{eq:physcoupreplace}
\tilde{M}^2_{\text{s}} (k) \to 8\pi \sum_{(ij)} C_{(ij)} \frac{\alpha_s^{\text{phys}}(\kappa_{(ij)}) }{\kappa_{(ij)}^2} \ \ .
\end{align}
With this replacement, we can isolate the function
$\mathcal{F}_{\rm correl}(v)$ that starts at NNLL accuracy, and is given by
\begin{align}
  \label{eq:fcorrelbasic}
\nonumber
\mathcal{F}_{\rm correl}(v) &=e^{- R_{\rm s}(\delta v)}e^{ R_{\rm s}( v)} \sum_{n=1}^\infty \frac{1}{n!} \int_{\delta v} \left(\prod_{i=1}^n [dk_i]\right) \,\tilde{M}_{\text{s}}^2 \left(k_i \right) \\\nonumber
& \times  \frac{1}{2!} \int [dk_a] [dk_b] \tilde{M}^2_{\text{s}}(k_a,k_b)\bigg[ \Theta \left(v - V_{\text{sc}}(\pBorn, k_a, k_b, k_1, \dots, k_n)\right)  \\
&-\Theta \left(v-  \lim_{m^2 \to 0} V_{\text{sc}}(\pBorn, k_a+ k_b, k_1, \dots, k_n)\right)   \bigg] \ \ .
\end{align}
In eq.~(\ref{eq:fcorrelbasic}), $m^2$ is the invariant mass of the correlated
pair, i.e.  $m^2 \equiv (k_a + k_b)^2$. The second step function
precisely encapsulates the inclusive limit of the double emission that
allowed us to introduce the physical coupling in the soft-collinear
matrix element squared. Notice in particular that we dropped $\delta$
from the double emission phase space because, once again, the integral
is manifestly finite in the limit $\delta \to 0$.

The previous steps leave us with the following expression
\begin{align}
  \label{eq:fscstart}
  \nonumber
  \mathcal{F}_{\text{s}}(v)  &= \mathcal{F}_{\rm wa}(v)+\mathcal{F}_{\rm correl}(v)\\ 
  & +e^{- R_{\rm s}(\delta v)} e^{ R_{\rm s}( v)}  \sum_{n=1}^\infty \frac{1}{n!} \int_{\delta v} \left(\prod_{i=1}^n [dk_i]\right) \, \tilde{M}_s^2 \left(k_i \right) \Theta\left(v - V_{\text{sc}}(\pBorn,k_1, \dots, k_n) \right) \,. 
\end{align}
The last step is to extract $\FNLL$ from the second line of the above
equation, and isolate the remaining NNLL soft contributions.  First,
we expand the exponential prefactor in eq.~(\ref{eq:fscstart}), up to
NNLL, as follows
\begin{align}\label{eq:expandfsc}
\begin{split}
e^{- R_{\rm s}(\delta v)} e^{ R_{\rm s}( v)} \simeq \delta^{R^{\prime}_{\rm s, \text{NLL}}} \left( 1 -  \Delta R^{\prime}_{\rm s, \text{NNLL}}  \ln\frac{1}{\delta} - \frac12 R^{\prime\prime}_{\rm s, \text{NNLL}}\ln^2 \frac{1}{\delta} \right)\,,
\end{split}
\end{align}
where $\Delta R^\prime_{\rm s, \text{NNLL}}$ denotes NNLL contributions
to the first derivative of the full soft radiator, $R_{\rm s}(v)$.
Now, using eq.~(\ref{eq:physcoupreplace}), for each leg in a
certain dipole we implement the following transformation of variables
\begin{align}\label{eq:varchange}
\ln (v \zeta)  =  a \ln \frac{\kappa_{ij}}{Q_{ij}} - b_\ell \eta_\ell^{(ij)} + \ln \left(d_\ell^{(ij)} g_\ell \right), \quad \xi^{(\ell)} = \frac{a+b_\ell}{b_\ell \eta_\ell^{(ij)} - a \ln (\kappa_{ij}/Q_{ij})} \eta_\ell^{(ij)} \ \ .
\end{align}
In particular, $\xi^{(\ell)}$ represents the rapidity fraction of the
emission, i.e. the ratio of the emission's rapidity with respect to
leg $\ell$ to the maximum available rapidity at fixed observable value
$v \zeta$. The essence of the above
transformation is that the observables we are interested in are event
shapes and therefore do not depend on the rapidity fraction
$\xi^{(\ell)}$. This allows us to integrate out this variable for each
emission, and we reconstruct the logarithmic derivative of the
massless radiator. In terms of these new variable, for each emission,
we can write
\begin{equation}
  \label{eq:M2sc-newvariables}
  \int_{\delta v} [dk]\tilde{M}_{\rm s}^2 \left(k \right) = \sum_{(ij)} C_{(ij)}\sum_{\ell\in(ij)} \int_{\delta }^\infty \frac{d\zeta}{\zeta} \int_0^{2\pi}\frac{d\phi^{(ij)}}{2\pi} (R_{\ell,\text{0}}^{(ij)})^\prime\left (\frac{v\zeta}{d_\ell^{(ij)} g_\ell(\phi^{(ij)})}\right) \ \ .
\end{equation}
Now it is straightforward to extract various contributions to
$\mathcal{F}_{\rm s}(v)$ by expanding $(R_{\ell,\text{0}}^{(ij)})'(\zeta v)$
around $(R_{\ell,\text{0}}^{(ij)})'(v)$ as follows:
\begin{equation}
  \label{eq:Rpdip-expanded}
  (R_{\ell,\text{0}}^{(ij)})^\prime\left (\frac{v\zeta}{d_\ell^{(ij)} g_\ell(\phi^{(ij)})}\right) = (R_{\ell,\text{0}}^{(ij)})'(v) +\Delta (R_{\ell,\text{0},\text{NNLL}}^{(ij)})' (v) + (R_{\ell,\text{0},\text{NNLL}}^{(ij)})'' (v) \ln \frac{ \left(d^{(ij)}_\ell g_\ell(\phi^{(ij)}) \right) }{\zeta}\,.
\end{equation}
We can then write
\begin{align}
\mathcal{F}_{\text{s}}(v) = \mathcal{F}_{\text{NLL}}(\lambda)  + \mathcal{F}_{\text{wa}}(v)+ \mathcal{F}_{\text{clust}}(v) + \Delta \mathcal{F}_{\text{s}}(v) \ \ .
\end{align}
In the above equation, $\FNLL(\lambda)$ is given by 
 \begin{align}
   \label{eq:fnllbasic}
   \nonumber
   \mathcal{F}_{\text{NLL}}(v) &=  \delta^{R'_{\rm s, \text{NLL}}}\\
                               &\times \sum_{n=1}^\infty \frac{1}{n!} \prod_{i=1}^n \sum_{(ij)} C_{(ij)}\sum_{\ell_i \in(ij)} \int_{\delta }^\infty \frac{d \zeta_i}{d \zeta_i}\, \int_0^{2\pi}\frac{d\phi_i^{(ij)}}{2\pi} \, (R_{\ell_i,\text{0},\text{NLL}}^{(ij)})' (v)  \,   \Theta\left(1 -  \frac{V_{\text{sc}}(\pBorn,k_1, \dots, k_n)}{v} \right) \ \ .
 \end{align}
 To avoid confusion, we introduced the notation $\ell_i$ in the above equation
 to denote the legs in a fixed dipole to which the $i^{\text{th}}$ soft emission
 belongs. This expression can be further simplified by observing that, when an
 emission $k$ is soft and collinear to leg $\ell$, the azimuthal angle
 $\phi^{(ij)}$ reduces to the azimuthal angle with respect to the leg, i.e.
 $\phi^{(\ell)}$. Then, we can also eliminate the sum over dipoles by defining a
 leg-dependent quantity
 \begin{equation}
   \label{eq:Rpleg}
   R'_{\ell,{\rm NLL}}(v)=\sum_{\{(ij)|\ell\in(ij)\}} C_{(ij)} (R_{\ell,\text{0},\text{NLL}}^{(ij)})' (v) \  \  .
 \end{equation}
 Using this substitution, we can recast $\mathcal{F}_{\text{NLL}}$
 into the standard form of eq.~\eqref{eq:FNLL}. We are left with the
 task of extracting the NNLL corrections contained in the leftover
 $\Delta\mathcal{F}_s$. Using eqs.~(\ref{eq:expandfsc})
 and~(\ref{eq:Rpdip-expanded}), this function can be recast in the
 form
 \begin{align}
   \label{eq:dfsbasic}
  \nonumber
  \Delta \mathcal{F}_{\text{s}}(v) &= \delta^{R'_{\rm NLL}} \sum_{n=0}^\infty \frac{1}{n!} \prod_{i=1}^n  \left(\sum_{\ell_i = 1}^3 \int_{\delta }^\infty \frac{d\zeta_i}{\zeta_i}\int_0^{2\pi}\frac{d\phi_i^{(\ell_i)}}{2\pi} \, R'_{\ell_i,\mathrm{NLL}}(v) \right)     \\ \nonumber
                                                        &\times \left( \sum_{(ij)} \sum_{\ell\in(ij)} \int_{\delta }^\infty \frac{d\zeta}{\zeta} \int_0^{2\pi}\frac{d\phi^{(ij)}}{2\pi}\left[ \Delta (R_{\ell,\text{0},\text{NNLL}}^{(ij)})' (v) + (R_{\ell,\text{0},\text{NNLL}}^{(ij)})'' (v) \ln \frac{ \left(d^{(ij)}_\ell g_\ell(\phi^{(ij)}) \right) }{\zeta}\right] \right.\times  \\ \nonumber
                                    &  \times\Theta\left(1 - \frac{V_{\text{sc}}(\pBorn, k, k_1, \dots, k_n)}{v} \right)
  \\ 
                                    &\left .
                                      - \left( \Delta R^{'}_{\rm s, \text{NNLL}}  \ln\frac{1}{\delta} + \frac12 R^{''}_{\rm s, \text{NNLL}}\ln^2 \frac{1}{\delta} \right) \times \Theta\left(1 - \frac{V_{\text{sc}}(\pBorn,k_1, \dots, k_n)}{v}\right) \right) \ \ .
\end{align}
We first eliminate as much as possible the dependence on the cutoff
$\delta$. The procedure, introduced in ref.~\cite{Banfi:2014sua},
consists in writing logarithms of $\delta$ as integrals over an
auxiliary variable $\zeta$. Using
\begin{align}
\Delta R'_{\rm s, \text{NNLL}} &= \sum_{(ij)} \sum_{\ell\in(ij)}\left( \Delta (R_{\ell,\text{0},\text{NNLL}}^{(ij)})' + (R_{\ell,\text{0},\text{NNLL}}^{(ij)})''\, \big\langle \ln\left(d^{(ij)}_\ell g_\ell  \right) \big\rangle  \right) \ \ , \\
R''_{\rm s, \text{NNLL}} &=  \sum_{(ij)} \sum_{\ell\in(ij)} (R_{\ell,\text{0},\text{NNLL}}^{(ij)})'' \,,
\end{align}
we obtain
\begin{equation}
  \label{eq:dfs-simplified1}
  \begin{split}
    \Delta \mathcal{F}_{\text{s}}(v) &= \delta^{R'_{\rm NLL}} \sum_{n=0}^\infty \frac{1}{n!} \prod_{i=1}^n  \left(\sum_{\ell_i=1}^3 \int_{\delta }^\infty \frac{d\zeta_i}{\zeta_i}\int_0^{2\pi}\frac{d\phi_i^{(\ell_i)}}{2\pi} \, R'_{\ell_i,\mathrm{NLL}}(v) \right)  \times \\ & \times
   \sum_{(ij)} \sum_{\ell\in(ij)} \int_{\delta }^\infty \frac{d\zeta}{\zeta} \int_0^{2\pi}\frac{d\phi^{(ij)}}{2\pi}\left[ \Delta (R_{\ell,\text{0},\text{NNLL}}^{(ij)})' (v) + (R_{\ell,\text{0},\text{NNLL}}^{(ij)})'' (v) \ln \frac{ \left(d^{(ij)}_\ell g_\ell(\phi^{(ij)}) \right) }{\zeta}\right] \times  \\ 
    &  \times \left[\Theta\left(1 -  \frac{V_{\text{sc}}(\pBorn, k, k_1, \dots, k_n)}{v} \right)-\Theta(1-\zeta)\Theta\left(1 -  \frac{V_{\text{sc}}(\pBorn, k_1, \dots, k_n)}{v} \right)\right] \ \ .
  \end{split}
\end{equation}
We can further simplify the above expression by extracting a piece
that is purely soft and collinear, and that can be seen as the
generalisation of $\dFsc$ introduced for two-jet
observables~\cite{Banfi:2014sua}. However, we anticipate that, in the
current case, $\Delta \mathcal{F}_{\rm s}$ contains a term that is
manifestly of wide-angle origin, and therefore is more naturally
associated with $\mathcal{F}_{\rm wa}$. First, similar to
eq.~(\ref{eq:Rpleg}) we can define another leg-dependent function
 \begin{equation}
   \label{eq:Rsleg}
   R^{''}_{\ell,{\rm NNLL}}(v)=\sum_{\{(ij)|\ell\in(ij)\}} C_{(ij)} (R_{\ell,\text{0},\text{NNLL}}^{(ij)})''(v)\,.
 \end{equation}
 Then, using the expression of $d_\ell^{(ij)}$ in
 eq.~(\ref{eq:dell-ij}), rewrite eq.~\eqref{eq:dfs-simplified1} in the
 form
\begin{equation}
  \label{eq:dfs-simplified}
  \begin{split}
    \Delta \mathcal{F}_{\text{s}}(v) &= \delta^{R^\prime_{\rm NLL}} \sum_{n=0}^\infty \frac{1}{n!} \prod_{i=1}^n  \left(\sum_{\ell_i=1}^3 \int_{\delta }^\infty \frac{d\zeta_i}{\zeta_i}\int_0^{2\pi}\frac{d\phi_i^{(\ell_i)}}{2\pi} \, R'_{\ell_i,\mathrm{NLL}}(v) \right)   \sum_{\ell=1}^3  \int_{0 }^\infty \frac{d\zeta}{\zeta} \int_0^{2\pi}\frac{d\phi^{(\ell)}}{2\pi} \times \\ & \times
  \bigg[R_{\ell,0,\text{NNLL}}'' (v) \left(\ln\frac{d_\ell g_\ell(\phi^{(\ell)})}{\zeta}-b_\ell\ln\frac{Q}{2 E_\ell} \right)\\
  &+\sum_{\{(ij)|\ell\in(ij)\}}C_{(ij)}\left(\Delta (R_{\ell,\text{0},\text{NNLL}}^{(ij)})' (v)+(a+b_{\ell})(R_{\ell,\text{0},\text{NNLL}}^{(ij)})'' (v)\ln\frac{Q_{ij}}{Q}\right)\bigg] \times  \\ 
  &  \times \left[\Theta\left(1 -\frac{V_{\text{sc}}(\pBorn, k, k_1, \dots, k_n)}{v} \right)-\Theta(1-\zeta)\Theta\left(1 -  \frac{V_{\text{sc}}(\pBorn, k_1, \dots, k_n)}{v} \right)\right]
  \end{split}
\end{equation} 
Using the explicit expression of the full radiator given in
section~\ref{sec:soft-radiator}, we have, to NNLL accuracy,
\begin{equation}
\Delta (R_{\ell,\text{0},\text{NNLL}}^{(ij)})' (v) =-\alpha_s\beta_0 \frac{d}{d\lambda}g^{(\ell)}_2(\lambda) - 2\lambda (R^{(ij)}_{\ell,0,\mathrm{NNLL}})'' \ln\frac{Q_{ij}}{Q}
\end{equation}
We stress that, at NNLL accuracy, $(R^{(ij)}_{\ell,0,\mathrm{NNLL}})''$
does not depend on the dipole kinematics, but only on the leg contained in the
dipole $(ij)$. Combining all terms that depend on $Q_{ij}$ we obtain, to NNLL
accuracy
\begin{equation}
  \label{eq:Rs-Qij}
  (a+b_\ell-2\lambda) (R^{(ij)}_{\ell,0,\mathrm{NNLL}})'' \ln\frac{Q_{ij}}{Q} =\frac{\alpha_s(v^{1/a}Q)}{a \pi} \ln\frac{Q_{ij}}{Q}\,, 
\end{equation}
which corresponds clearly to a term of soft wide-angle origin. Last, we define
\begin{equation}
  \label{eq:dRpleg}
  \Delta R'_{\ell, {\rm NNLL}}(v)\equiv\sum_{\{(ij)|\ell\in(ij)\}} C_{(ij)} \left(-\alpha_s\beta_0 \frac{dg_2^{(\ell)}}{d\lambda}\right) \,.
\end{equation}
Introducing everywhere the soft-collinear measure $\dZ$ defined in
eq.~(\ref{eq:dZ}), we can write, at NNLL accuracy,
\begin{equation}
  \label{eq:dfs-NNLL}
  \begin{split}
    & \mathcal{F}_{\rm wa}(v)  =
    \frac{\alpha_s(Q)}{\pi}\dFwa(\lambda)\,,\qquad \mathcal{F}_{\rm
      correl}(v)=\frac{\alpha_s(Q)}{\pi}\dFcor(\lambda)\,, \\ & \Delta
    \mathcal{F}_{\rm s} =
    \frac{\alpha_s(Q)}{\pi}\left(\dFsc(\lambda)+\Delta\mathcal{F}_{\rm
        wa}(\lambda)\right)\,.
  \end{split}
\end{equation}
Collecting all these functions together we obtain our final expression:
\begin{equation}
  \label{eq:Fs-final}
  \mathcal{F}_{\rm s}(v)=\FNLL(\lambda)+\frac{\alpha_s(Q)}{\pi}\left(\dFsc(\lambda)+\dFwa(\lambda)+\Delta\mathcal{F}_{\rm wa}(\lambda)+\dFcor(\lambda)\right)\,.
\end{equation}
We now briefly derive the form of each NNLL correction that is suitable for numerical integration.

\paragraph{Soft-collinear NNLL correction.}

We collect from eq.~(\ref{eq:dfs-simplified}) all the terms that
depend explicitly on each leg, and not on the event geometry, and we
make use of the new function $\Delta R^\prime_{\ell,\mathrm{NNLL}}$
  defined in eq.~(\ref{eq:dRpleg}). This gives the generalisation of
  the soft-collinear function $\dFsc$ introduce for the two-jet
  case in ref.~\cite{Banfi:2014sua}: 
\begin{multline}
  \label{eq:dFsc}
    \dFsc(\lambda) = \frac{\pi}{\as(Q)}\int_0^\infty\frac{d\zeta}{\zeta} \int_0^{2\pi}\frac{d\phi}{2\pi}\sum_{\ell=1}^3 \left[ \Delta R'_{\ell,\mathrm{NNLL}} + R''_{\ell,\mathrm{NNLL}} \left(\ln \frac{d_\ell g_\ell(\phi)}{\zeta}-b_\ell\ln\frac{2E_\ell}{Q}\right)\right] \times \\ \times \int \dZ
    \left[\Theta\left(1-\frac{\Vsc{k,\{k_i\}}}{v}\right)-\Theta(1-\zeta)
\Theta\left(1-\frac{\Vsc{\{k_i\}}}{v}\right)\right]\,,
  \end{multline}

\paragraph{Soft wide-angle NNLL correction.}
Let us move to eq.~\eqref{eq:wideangle} and
extract the NNLL contribution. Since the emission $k$ is at largest
angle with respect to all the others, to this aim all soft
emissions are independent, hence
\begin{align}
\mathcal{M}_s^2 \left(k, k_1, \dots , k_n\right) \simeq \tilde{M}^2_{\text{s}} (k) \prod_{i=1}^n \tilde{M}^2_{\text{s}} (k_i) \ \ .
\end{align}
where, once again, the single-emission soft block is defined with the
physical coupling. For soft and collinear emissions, we can introduce
the soft-collinear measure $\dZ$ following the same steps as for
$\FNLL$. Furthermore, for the soft wide-angle emission $k$, we perform
a change of variables that reflects the dependence of
$\mathcal{F}_{\rm wa}$ on the dipole kinematics. We then use the
Sudakov variables of eq.~(\ref{eq:Sudakov-k}), and for each dipole
$(ij)$ we further introduce
\begin{align}\label{eq:transwideangle}
\zeta \equiv \frac{1}{v} \left( \frac{\kappa^{(ij)}}{Q_{ij}} \right)^a  \ \ .
\end{align}
Following the same steps as in ref.~\cite{Banfi:2014sua}, at NNLL
accuracy, we obtain
$\mathcal{F}_{\rm wa}(v)=(\alpha_s(Q)/\pi)\, \dFwa(\lambda)$, where
\begin{align}\label{eq:wideanglezeta}
\nonumber
\delta\mathcal{F}_{\text{wa}}(\lambda) &=  \sum_{(ij)} C_{(ij)} \frac{\alpha_{\text{s}}(v^{1/a} Q)}{a\,\alpha_s(Q)} \int_0^\infty \frac{d\zeta}{\zeta} \int _{-\infty}^{\infty} d\eta^{(ij)} \int_0^{2\pi} \frac{d\phi^{(ij)}}{2\pi}\times  \\
&\times \int \dZ \left[ \Theta\left(1 -\frac{V_{\rm wa}(\pBorn, k^{(ij)},\{k_i\})}{v} \right) - \Theta\left(1 - \frac{V_{\text{sc}}(\pBorn,k^{(ij)}, \{k_i\})}{v} \right) \right] \ \ .
\end{align}
We then collect from eq.~(\ref{eq:dfs-simplified}) all terms that
contain the ratios $Q_{ij}/Q$. This gives the new NNLL function
\begin{multline}
  \label{eq:DeltaF-wa}
  \Delta \mathcal{F}_{\rm wa}(\lambda) = \sum_{(ij)} C_{(ij)} \frac{\alpha_{\text{s}}(v^{1/a} Q)}{a\,\alpha_s(Q)} \ln\frac{Q_{ij}}{Q} \int_0^\infty \frac{d\zeta}{\zeta}\int_0^{2\pi} \frac{d\phi}{2\pi}\times  \\
\times  \int \dZ\times\left[\Theta\left(1 - \lim_{v\to0}\frac{ V_{\text{sc}}(\pBorn, k, \{k_i\})}{v} \right)-\Theta\left(1 - \lim_{v\to0}\frac{ V_{\text{sc}}(\pBorn, \{k_i\}}{v} \right)\Theta(1-\zeta)\right] \ \ .
\end{multline}

\paragraph{Soft correlated NNLL correction.}
Eq.~(\ref{eq:fcorrelbasic}) can be simplified further to extract the NNLL
contributions. First, we write the correlated portion of the double-emission
tree-level matrix element in terms of the variables introduced in
appendix~\ref{sec:correl-kinematics}:
\begin{align}
  \label{eq:doublesoft}
  \frac{1}{2!} \int [dk_a] [dk_b] \tilde{M}^2_{\text{s},0}(k_a,k_b) &= \sum_{(ij)} C_{(ij)} \sum_{\ell\in(ij)} \int \frac{d\kappa_{ij}}{\kappa_{ij}} \frac{d\phi^{(ij)}}{2\pi} d\eta_\ell^{(ij)} \frac{\alpha_s(\kappa_{ij})}{\pi} \times \\
                                                                    & \times \frac{\alpha_s(\kappa_{ij})}{2\pi}  \int_0^\infty \frac{d\mu^2}{\mu^2(1+\mu^2)} \int_0^1 dz \int_0^{2\pi} \frac{d\phi}{2\pi} \frac{1}{2!}\, \mathcal{A}^2\left(z, \mu,\phi \right) \ \ ,
\end{align}
where
\begin{align}
\mathcal{A}^2 \equiv C_A (2\mathcal{S} + \mathcal{H}_g) + n_f \mathcal{H}_q \ \ ,
\end{align}
and $\mathcal{S},\mathcal{H}_{g},\mathcal{H}_{q}$ can be found in
appendix~\ref{sec:correl-kinematics}. Note that the variables
$\kappa^{(ij)},\eta_\ell^{(ij)},\phi^{(ij)}$ refer to the Sudakov decomposition
of the parent momentum $k = k_a + k_b$, and the construction is explained in
Appendix~\ref{sec:correl-kinematics}. Now in eq.~(\ref{eq:fcorrelbasic}) we
change variables in a similar fashion to eq.~(\ref{eq:varchange})
\begin{align}\label{eq:zetaFcorrel}
\zeta = \lim_{\mu^2\to 0}\frac{V_{\rm sc}(k_a + k_b)}{v}, \quad \xi^{(\ell)} = \frac{a+b_\ell}{b_\ell \eta_\ell^{(ij)} - a \ln (\kappa_{ij}/Q_{ij})} \eta_\ell^{(ij)} \ \ .
\end{align}
Owing to the fact that the observable does not depend on $\xi^{(\ell)}$, we can
integrate it out analytically and find
\begin{align}
\nonumber
\frac{1}{2!} \int [dk_a] [dk_b] \tilde{M}^2_{\text{s},0}(k_a,k_b) &= \sum_{(ij)} C_{(ij)} \sum_{\ell\in(ij)} \frac{\lambda (R_{\ell,0,\rm NNLL}^{(ij)})''(v)}{2 \pi a \beta_0}  \times \\
&\times \int_0^\infty \frac{d\zeta }{\zeta} \int_0^{2\pi}\frac{d\phi^{(ij)}}{2\pi}\int_0^\infty \frac{d\mu^2}{\mu^2(1+\mu^2)} \int_0^1 dz \int_0^{2\pi} \frac{d\phi}{2\pi} \frac{1}{2!}\, \mathcal{A}^2\left(z, \mu,\phi \right)  \ \ .
\end{align}
Using eq.~(\ref{eq:Rsleg}), and the fact that $k_a,k_b$ are soft and collinear
to the same leg $\ell$, we can approximate $\phi^{(ij)}\simeq \phi^{(\ell)}$,
and finally obtain $\mathcal{F}_{\rm correl}(v)=(\alpha_s(Q))/\pi)
\dFcor(\lambda)$, where
\begin{align}
\label{eq:dFcorrel-final}
  \nonumber
  \delta\mathcal{F}_{\rm correl}(v) &=\sum_{\ell} \frac{\lambda R^{''}_{\ell,\rm NNLL}}{2 a \beta_0\alpha_s(Q)} \int_0^\infty \frac{d\zeta }{\zeta}\int_0^{2\pi} \frac{d\phi^{(\ell)}}{2\pi}\int_0^\infty \frac{d\mu^2}{\mu^2(1+\mu^2)} \int_0^1 dz \int_0^{2\pi} \frac{d\phi}{2\pi} \frac{1}{2!}\, \mathcal{A}^2\left(z, \mu,\phi \right)
                                      \times \\
                                    & \times \int \dZ
                                      \bigg[ \Theta \left(1 -\frac{V_{\text{sc}}(\pBorn, k_a, k_b, \{k_i\})}{v}\right) -\Theta \left(1 -\lim_{\mu^2 \to 0} \frac{V_{\text{sc}}(\pBorn, k_a+ k_b,\{k_i\})}{v}\right)   \bigg] \ \ .
\end{align}

\paragraph{The master NNLL formula for the soft cumulative distribution.} We now
put together all the ingredients to write down our master formula for the soft
cumulant, valid up to NNLL accuracy
\begin{align}\label{eq:softmaster}
  \begin{split}
  \Sigma^{\rm NNLL}_{\rm soft}(v) &=  e^{- R_{\rm hc}(v)} \int d\Phi_3\, \frac{d\sigma_3}{d\Phi_3} \mathcal{H}(\{p\})\, H(\{p\}, \alpha_s(Q))\, e^{- R_{\rm s}( v;\{p\})} \times \\
& \times \left[\mathcal{F}_{\rm NLL}(\lambda) + \frac{\alpha_s}{\pi} \left(\delta \mathcal{F}_{\rm sc}(\lambda) + \delta \mathcal{F}_{\rm wa} (\lambda)+ \delta \mathcal{F}_{\rm correl}(\lambda) + \Delta \mathcal{F}_{\rm wa} (\lambda)\right)\right] \ \ .
  \end{split}
\end{align}

\subsection{Hard-collinear cumulative distribution}
\label{sec:hard-collinear}

Up to NNLL, the hard-collinear cumulant reads
\begin{align}
\nonumber
\Sigma^{\rm NNLL}_{\rm hc}(v) &=  e^{- R_{\rm hc}(v)} \,\frac{1}{\sigma_{\mathcal{H}}} \int d\Phi_{3}\, \frac{d\sigma_3}{d\Phi_3}  H(\{p\}, \alpha_s(Q)) \mathcal{H}(\{p\})  e^{- R_{\rm s}( v; \{p\})}  \\\ \nonumber
& \times \delta^{R_{\rm NLL}^\prime}\sum_{n=0}^\infty \frac{1}{n!} \int_{\delta} \left(\prod_{i=1}^n
  [dk_i]\right) \, \tilde{M}^2_{\rm s} \left(k_i\right) \bigg[ \int [dk_{\rm hc}] M^2_{\rm hc}(k_{\rm hc})  \Theta\left(v - V(\pBorn, k_{\rm hc}, k_1, \dots, k_n) \right) \\
 &- \sum_{\ell=1}^3  \int_0^{Qv^{1/a+b_\ell}}  \frac{dk}{k} \frac{\alpha_s(k,\epsilon)}{\pi} \gamma^{(0)}_\ell  \Theta\left(v - V(\pBorn, k_1, \dots, k_n)\right) \bigg]\ \ .
\end{align}
Our first task is to cancel the collinear divergence in the above expression. To
this aim we notice that the singularity is encoded solely in the portion of the
hard-collinear matrix element proportional to the Born amplitude, namely the
pieces having the averaged splitting functions in eqs.~\eqref{eq:quarkhc} and
\eqref{eq:gluonhc}. We will show below that the extra piece in
eq.~(\ref{eq:gluonhc}), proportional to the un-averaged splitting function,
produces a finite term due to the vanishing of the azimuthal average,
eq.~\eqref{eq:dPG-averaged}, as $k_t \to 0$.

We can partition the above expression into various pieces in order to
arrange for manifestly finite expressions that could then be evaluated
in 4 dimensions. The steps follow ref.~\cite{Banfi:2014sua}, albeit
with a new contribution arising from the spin-correlations of the
gluons. We have
\begin{align}\label{hcmaster}
\nonumber
\Sigma^{\rm NNLL}_{\rm hc}(v) &=  e^{- R_{\rm hc}(v)} \,\frac{1}{\sigma_{\mathcal{H}}} \int d\Phi_{3}\, \frac{d\sigma_3}{d\Phi_3}  H(\{p\}, \alpha_s(Q)) \mathcal{H}(\{p\}) \, e^{- R_{\rm s}(v;\{p\})}  \\
&\times \left( \mathcal{F}_{\rm NLL}(\lambda) \sum_{\ell=1}^3  C^{(1)}_{\mathrm{hc},\ell} + \mathcal{F}_{\rm rec} + \mathcal{F}_{\rm hc} + \mathcal{T}(\{p\}) \frac{\alpha_s(Q)}{\pi}  \Delta \mathcal{F}_{\rm rec}(\lambda) \right)
\end{align}
where
\begin{align}\label{eq:hcconstant}
\nonumber
C^{(1)}_{\mathrm{hc},\ell} &= \left(4\pi\mu_R^2\, e^{-\gamma_E}\right)^\epsilon \sum_{\ell=1}^3 \int_0^Q\frac{dk_t}{k_t^{1+2\epsilon}} \frac{\alpha_s(k_t)}{\pi} \int \frac{d\Omega_{2-2\epsilon}}{\Omega_{2-2\epsilon}}\int_0^1 dz\, \langle P_{f_\ell}  (z,\epsilon) \rangle\, \Theta\left(v - V_{\rm sc}(k_{\rm hc}) \right) \\
&-\sum_{\ell=1}^3  \int_0^{Qv^{1/a+b_\ell}}  \frac{dk}{k} \frac{\alpha_s(k,\epsilon)}{\pi} \gamma^{(0)}_\ell \ \ ,
\end{align}
comprises a constant leftover after cancelling the collinear divergence. We can
easily evaluate eq.~(\ref{eq:hcconstant}) using the splitting functions given in
eqs.~\eqref{eq:quarkhc} and \eqref{eq:gluonhc}. For a (anti)-quark we
get\footnote{To obtain eq.~\eqref{eq:Chc-q}, one first expands the step function
  similar to eq.~\eqref{eq:thetaexpansion}.}
\begin{equation}
  \label{eq:Chc-q}
  C^{(1)}_{\mathrm{hc},\ell}= \frac{\alpha_s\left(Q v^{\frac{1}{a+b_\ell}}\right)}{2\pi} C_F \left(
  \frac{7}{2} \frac{b_\ell}{a+b_\ell}+\frac{3}{a+b_\ell}\left(\langle\ln d_\ell g_\ell \rangle-b_\ell\ln\frac{2 E_\ell}{Q}\right)+\frac{1}{2}\right)  \,.
\end{equation}
This result coincides with that of ref.~\cite{Banfi:2018mcq} for two hard legs.
For a gluon we have
\begin{multline}
  \label{eq:Chc-g}
  C^{(1)}_{\mathrm{hc},\ell}=   \frac{\alpha_s\left(Q v^{\frac{1}{a+b_\ell}}\right)}{2\pi} \bigg[
    \left(\frac{67}{18} C_A-\frac{13}{9} T_R n_f \right)\frac{b_\ell}{a+b_\ell} \\ + \frac{1}{a+b_\ell}\left(\frac{11}{3} C_A-\frac{4}{3} T_R n_f \right)\left(\langle\ln d_\ell g_\ell\rangle-b_\ell\ln\frac{2 E_\ell}{Q}\right)+\frac{1}{3}T_R n_f  \bigg] \,.
\end{multline}
Moreover, we have two correction functions. The first arises solely due to our
choice of regularisation in eqs.~(\ref{eq:hcconstant}) and (\ref{eq:frecbasic}),
and reads
\begin{align}\label{eq:fhcbasic}
\nonumber
\mathcal{F}_{\rm hc} &= \delta^{R_{\rm NLL}^\prime} \bigg[ \sum_{\ell=1}^3 \int \frac{dk_t}{k_t} \frac{\alpha_s(k_t)}{\pi} \int_0^\pi \frac{d\phi}{\pi} \int_0^1 dz \, \langle P_{f_\ell}(z,0) \rangle \sum_{n=0}^\infty \frac{1}{n!} \int_{\delta} \left(\prod_{i=1}^n
  [dk_i]\right) \, \tilde{M}^2_{\rm s} \left(k_i\right) \\ 
  &\times  \left( \Theta\left(v - V_{\rm sc}(\pBorn, k, k_1, \dots, k_n) \right) - \Theta\left(v - V_{\rm sc}(\pBorn, k_1, \dots, k_n) \right) \, \Theta\left( v- V_{\rm sc}(k) \right) \right) \bigg]   \ \ .
\end{align}
We can further simplify the above expressions by introducing the phase-space
measure over soft-collinear emissions. We first introduce the observable
fraction of the hard emission
\begin{align}
\label{eq:zeta-original}
\zeta \equiv \frac{1}{v} \frac{d_\ell g_\ell(\phi)}{(z^{(\ell)})^{b_\ell}}\left(\frac{k_t}{Q}\right)^{a+b_\ell} \ \ ,
\end{align}
and using eq.~\eqref{eq:varchange} for the soft-collinear emissions we find
$\mathcal{F}_{\rm hc}(v)=(\alpha_s/\pi)\dFhc(\lambda)$
\begin{multline}
  \label{eq:dFhc}
    \dFhc(\lambda) = \sum_{\ell=1}^3 \frac{\as(Qv^{1/(a+b_\ell)})}{\as(Q)(a+b_\ell)} \int_0^\infty\frac{d\zeta}{\zeta} \int_0^{\pi}\frac{d\phi^{(\ell)}}{\pi} \int_0^1 dz^{(\ell)} \langle P_{f_\ell}(z^{(\ell)},0) \rangle  \\ \times
    \int \dZ\times\left[\Theta\left(1 - \frac{ V_{\text{sc}}(\pBorn, k, \{k_i\})}{v} \right)-\Theta\left(1 - \frac{ V_{\text{sc}}(\pBorn, \{k_i\}}{v} \right)\Theta(1-\zeta)\right] \ \ .
\end{multline}
Notice that in the above expression we can send the upper limit of the $\zeta$
integral to infinity, with corrections suppressed by powers of $v$. The
second correction incorporates the recoil of the event shape due to the
hard-collinear emission
\begin{align}\label{eq:frecbasic}
\nonumber
\mathcal{F}_{\rm rec} &= \delta^{R_{\rm NLL}^\prime} \bigg[ \sum_{\ell=1}^3 \int \frac{dk_t}{k_t} \frac{\alpha_s(k_t)}{\pi} \int_0^{2\pi} \frac{d\phi}{2\pi} \int_0^1 dz \, \langle P_{f_\ell}(z,0) \rangle \sum_{n=0}^\infty \frac{1}{n!} \int_{\delta} \left(\prod_{i=1}^n
  [dk_i]\right) \, \tilde{M}^2_{\rm s} \left(k_i\right) \\ 
  &\times  \left( \Theta\left(1 - \frac{V_{\rm hc}(\pBorn, k, k_1, \dots, k_n)}{v} \right) - \Theta\left(1 - \frac{V_{\rm sc}(\pBorn, k, k_1, \dots, k_n)}{v} \right)\right) \bigg]   \ \ ,
\end{align}
where the extra emission, $k$, is treated as soft and collinear in the second
step function. This correction can be conveniently combined with the overlap
function $\mathcal{F}_{\rm s/hc}$ introduced in eq.~(\ref{eq:Fs/hc}). Performing
the same formal manipulations that lead to eq.~\eqref{eq:dFhc} (see also
ref.~\cite{Banfi:2014sua} for details) we obtain that, at NNLL accuracy,
$\mathcal{F}_{\rm rec}+ \mathcal{F}_{\rm s/hc}=(\alpha_s/\pi)\dFrec$, where
\begin{multline}
  \label{eq:dFrec}
    \dFrec(\lambda) = \sum_{\ell=1}^3 \frac{\as(Qv^{1/(a+b_\ell)})}{\as(Q)(a+b_\ell)} \int_0^\infty\frac{d\zeta}{\zeta} \int_0^{\pi}\frac{d\phi^{(\ell)}}{\pi} \int_0^1 dz^{(\ell)} \left(\frac{2C_\ell}{z^{(\ell)}}+\langle P_{f_\ell}(z^{(\ell)},0) \rangle\right)\\ \times
    \int \dZ\times  \left[\Theta\left(1 - \frac{V_{\text{hc}}(\pBorn, k, \{k_i\})}{v} \right)-\Theta\left(1 -  \frac{V_{\text{sc}}(\pBorn, k, \{k_i\})}{v}\right)\right] \ \ ,
\end{multline}
where $C_q = C_{\bar{q}} = C_F$ and $C_g = C_A$. Note that, in
eqs.~\eqref{eq:fhcbasic} and \eqref{eq:dFrec}, we obtain identical results if we
use the following alternative definition for $\zeta$
\begin{equation}
  \label{eq:zeta-alt}
  \zeta \equiv \frac{1}{v} \left(\frac{k_t}{Q}\right)^{a+b_\ell}\,.
\end{equation}
In fact, what really matters is only the scaling of the hard-collinear
emission $k$ with respect to $k_t$. It is crucial that we pause here
to address an intricate point in the derivation of
eq.~\eqref{eq:dFrec}. The squared matrix elements in
eqs.~\eqref{eq:quarkhc} and \eqref{eq:gluonhc} are expressed in terms
of the transverse momentum with respect to the emitter. Consequently,
we have to express the observable in terms of the same set of
variables utilised in the matrix elements, which might turn out to be
non-trivial depending on the observable. Any specific event shape will
either use an axis in its definition, e.g. the thrust axis, or will
depend on the relative transverse momentum between the particles of
the final state. For soft emissions, the situation is simple because
the direction of the emitter is the same as the direction of the final
state hard momenta, up to terms that vanish as $k_t^2 \to 0$. For a
hard-collinear emission, extra care must be taken. The transverse
momentum appearing in the observable defines the integration variable
$\zeta$, and its precise relation to $k_t$ in the emission
probability, i.e.\ eqs.~\eqref{eq:quarkhc} and \eqref{eq:gluonhc},
must be explicitly worked out.

Finally we have a new correction which is absent in the case of di-jet
observables, and is due to the spin-correlation in the final state. Explicitly,
we have
\begin{align}
\nonumber
\Delta \mathcal{F}_{\rm rec}(\lambda) = \delta^{R_{\rm NLL}^\prime} \frac{\pi}{\alpha_s(Q)}& \left(4 \pi \mu_R^2\, e^{-\gamma_E}  \right)^\epsilon\bigg[ \int_0^Q \frac{dk_t}{k^{1+2\epsilon}_t} \frac{\alpha_s(k_t)}{\pi}\int \frac{d\Omega_{2-2\epsilon}}{\Omega_{2-2\epsilon}}  \int_0^1 dz \, \Delta P_{g} (z,\phi;\epsilon) \\
&\times \sum_{n=0}^\infty \frac{1}{n!} \int_{\delta v} \left(\prod_{i=1}^n
  [dk_i]\right) \, \tilde{M}^2_{\rm s} \left(k_i\right)
  \, \Theta\left(v - V_{\rm hc}(\pBorn, k_{\rm hc}, k_1, \dots, k_n) \right)\biggr] \ \ .
\end{align}
The above integral is indeed finite, but requires extra care. The collinear
divergence is regulated because as $k_t \to 0$, the azimuthal average of $\Delta
P_{g} (z,\phi;\epsilon)$ vanishes identically in $d=4-2\epsilon$. The finite
contribution that arises can be isolated. Let us change variables according to
eq.~\eqref{eq:zeta-alt} and extract the NNLL contribution\footnote{This equation
  is valid in the limit $\epsilon \to 0$, as long as $a+b_g>0$ which is
  guaranteed by IRC safety of the observable.} 
\begin{align}
\label{eq:DFreceps}
\nonumber
\Delta \mathcal{F}_{\rm rec}(\lambda) &= \delta^{R_{\rm NLL}^\prime}\left(\frac{4 \pi \mu_R^2\, e^{-\gamma_E}}{Q^2}  \right)^\epsilon \frac{\as(Qv^{1/(a+b_g)})}{\as(Q)(a+b_g)} v^{-2\epsilon} \bigg[ \int_0^{1/v} \frac{d\zeta}{\zeta^{1+2\epsilon}}\int \frac{d\Omega_{2-2\epsilon}}{\Omega_{2-2\epsilon}}  \int_0^1 dz \, \Delta P_{g} (z,\phi;\epsilon) \\
&\times \sum_{n=0}^\infty \frac{1}{n!} \int_{\delta v} \left(\prod_{i=1}^n
  [dk_i]\right) \, \tilde{M}^2_{\rm s} \left(k_i\right) \Theta\left(v - V_{\rm hc}(\pBorn, k_{\rm hc}, k_1, \dots, k_n) \right)\biggr] \ \ .
\end{align}
Now we can use (see for example \cite{deFlorian:2001zd})
\begin{align}
\frac{1}{\zeta^{1+2\epsilon}} = -\frac{1}{2\epsilon} \delta(\zeta) \left[ 1 - 2\epsilon \ln\left(\frac{1}{v}\right) + \mathcal{O}(\epsilon^2) \right] + \frac{1}{\zeta_+} \ \ ,
\end{align}
where the plus distribution is defined as follows
\begin{align}
\int_0^{1/v} d\zeta \frac{f(\zeta)}{\zeta_+} \equiv \int_0^{1/v} d\zeta \frac{f(\zeta)- f(0)}{\zeta} \ \ .
\end{align}
Hence, as promised the pole term disappears because the azimuthal average
vanishes in $4-2\epsilon$ dimensions, {\em cf.} eq.~\eqref{eq:dPG-averaged}.
Applying the plus prescription yields a finite result
\begin{multline}
\Delta \mathcal{F}_{\rm rec}(\lambda) = \frac{\alpha_s(Qv^{1/(a+b_g)})}{(a+b_g)\alpha_s(Q)} \int_0^{1/v}\frac{d\zeta}{\zeta} \int_0^{\pi} \frac{d\phi^{(\ell)}}{\pi}\int_0^1 dz^{(\ell)} \, \Delta P_{g} (z^{(\ell)},\phi^{(\ell)}) \times \\
  \times \int \dZ \left[\Theta\left(1 - \frac{ V_{\text{hc}}(\pBorn, k, \{k_i\})}{v} \right)-\Theta\left(1 - \frac{ V_{\text{sc}}(\pBorn, \{k_i\}}{v} \right)\right] \ \ .
\end{multline}
Although the second step function vanishes because of the azimuthal
average, it is still quite important to keep it in order for numerical
integration to be feasible. The goal is to utilise the second step
function as a regulator in a Monte Carlo integration. To simplify the
implementation one ideally wants to push the limit of the $\zeta$
integral to infinity. For most observables, the first integral is
effectively cut off by the observables constraint, so we can push the
limit of the $\zeta$ integration to infinity. There are however a
number of observables, especially those who are affected by
cancellations of the contribution of emissions with comparable values
of $\zeta$, for which the integral is damped by the result of the
integration over the soft-collinear measure $\dZ$ (see e.g.\ appendix
H ref.~\cite{Banfi:2004yd}). For those observables, the integral in
$\zeta$ converges for $R'_{\rm NLL}$ lower that a certain critical
value, which is the region in which our resummation is valid. Note
that this consideration applies to $\mathcal{F}_{\rm NLL}$ and to
all NNLL corrections, and we recall it here for completeness.  We can
also split the second integral at $\zeta=1$, and the contribution from
$1$ to $1/v$ vanishes identically upon azimthal integration. Hence,
one can identically recast the above expression as follows:
\begin{multline}
\label{eq:finalDfrecMC}
\Delta \mathcal{F}_{\rm rec}(\lambda) = \frac{\alpha_s(Qv^{1/(a+b_g)})}{(a+b_g)\alpha_s(Q)} \int_0^\infty\frac{d\zeta}{\zeta} \int_0^{\pi} \frac{d\phi^{(\ell)}}{\pi}\int_0^1 dz^{(\ell)} \, \Delta P_{g} (z^{(\ell)},\phi^{(\ell)}) \times \\
  \times \int \dZ \left[\Theta\left(1- \frac{V_{\text{hc}}(\pBorn, k, \{k_i\})}{v} \right)- \Theta(1-\zeta) \Theta\left(1 - \frac{ V_{\text{sc}}(\pBorn, \{k_i\})}{v}\right)\right] \,,
\end{multline}
Eq.~\eqref{eq:finalDfrecMC} is one of the main results of this paper,
and is suitably defined for numerical evaluation.

\subsection{Additive observables}
\label{sec:additive}

For additive observables, such as the $D$-parameter, we can obtain closed form
expressions for all NNLL functions. Additivity implies that the observable can
be written as the sum of contributions of individual emissions. For soft and
collinear emissions $k_1,\dots,k_n$, this means
\begin{align}
  \label{eq:additive}
\Vsc{k_1, \dots, k_n}  = \sum_{i=1}^n V_{\text{sc}}(\{\tilde{p}\}, k_i) \ \ .
\end{align}
The evaluation of eq.~(\ref{eq:fnllbasic}) becomes simple and yields
the well-known result (see e.g.~\cite{Banfi:2004yd})
\begin{equation}
  \label{eq:FNLL-additive}
  \FNLL=\frac{e^{-\gamma_E R'_{\rm NLL}}}{\Gamma(1 + R'_{\rm NLL})},
\end{equation}
where $R'_{{\rm NLL}} \equiv R'_{{\rm s, NLL}}$. Using eq.~\eqref{eq:additive} we can compute $\dFsc$,
$\Delta\mathcal{F}_{\rm wa}$ and $\dFcor$ using the procedure
described in appendix C of ref.~\cite{Banfi:2014sua}, and we get
\begin{align}
\nonumber
  \delta \mathcal{F}_{\text{sc}}(\lambda) = & - \mathcal{F}_{\text{NLL}}(\lambda) \sum_{\ell =1}^3 \left\{ \left[ \Delta R'_{\ell, \rm NNLL}+ R''_{\ell,\rm NNLL} \,\left(\langle\ln\left(d_\ell g_\ell \right) \rangle-b_\ell\ln\frac{2E_\ell}{Q}\right)  \right] \left(\psi^{(0)}(1+\RpNLL)) + \gamma_E \right)\right. \\
  & \left.+\frac12 R''_{\ell,\rm NNLL} \left( \left(\psi^{(0)}(1 + R'_{\rm NLL}) + \gamma_E \right)^2 - \psi^{(1)}(1+ R'_{\rm NLL}) + \frac{\pi^2}{6} \right)\right\} \label{eq:scollfinal} \ \  ,\\
  \Delta\mathcal{F}_{\rm wa}(\lambda) &= - \mathcal{F}_{\rm NLL}(\lambda) \left( \psi^{(0)}(1 + R'_{\rm NLL}) + \gamma_E \right)  \frac{\alpha_s(v^{1/a}Q)}{a \,\alpha_s(Q)} \sum_{(ij)} C_{(ij)}  \ln\left(\frac{Q_{ij}}{Q} \right) \label{eq:finalDwa} \ \  .
\end{align}
Proceeding to the wide-angle correction, all what we really need is to probe the
observable, when a single soft emission $k$ is emitted at wide angle. If we
parametrise $k$ using the Sudakov decomposition in eq.~(\ref{eq:Sudakov-k}), and
for an additive observable, we obtain
\begin{align}
  \label{eq:Vsc-additive}
  \frac{V_{\rm sc}(\pBorn, k^{(ij)}, \{ k_i \})}{v} &= \zeta f^{(ij)}_{\text{sc}}(\eta^{(ij)},\phi^{(ij)})
+ \sum_i \zeta_i 
\end{align}
where $\zeta$ is defined in eq.~(\ref{eq:transwideangle}), and 
\begin{equation}
\label{eq:fwascoll}
  f^{(ij)}_{\text{sc}}(\eta^{(ij)},\phi^{(ij)})  =\sum_{\ell\in(ij)} d_\ell^{(ij)} e^{- b_\ell \eta_\ell^{(ij)}} g_\ell(\phi^{(ij)}) \, \Theta(\eta_{\ell}^{(ij)}) \ \ .
\end{equation}
Also,
\begin{align}
  \label{eq:Vwa-additive}
  \frac{V_{\rm wa}(\pBorn, k^{(ij)}, \{ k_i \})}{v} &= \zeta f^{(ij)}_{\text{wa}}(\eta^{(ij)},\phi^{(ij)})
+ \sum_i \zeta_i \ \ .
\end{align}
Using the above relations in eq.~\eqref{eq:wideanglezeta}, we find 
\begin{align}\label{eq:dFwa-additive}
\delta\mathcal{F}_{\text{wa}}(\lambda)  = \mathcal{F}_{\text{NLL}}(\lambda) \sum_{(ij)} \frac{C_{(ij)}}{a} \frac{\alpha_{\text{s}}(v^{1/a} Q)}{\alpha_s(Q)} \int _{-\infty}^{\infty} d\eta^{(ij)} \int_0^{2\pi} \frac{d\phi^{(ij)}}{2\pi} \ln \frac{f^{(ij)}_{\text{sc}}(\eta^{(ij)},\phi^{(ij)})}{f^{(ij)}_{\text{wa}}(\eta^{(ij)},\phi^{(ij)})} \ \ .
\end{align}
Here, it is important to notice that the wide-angle correction is sensitive to
the invariant mass of the dipole in contrast to the soft-collinear correction in
eq.~\eqref{eq:scollfinal}. For $\dFcor$, we follow the procedure of
ref.~\cite{Banfi:2018mcq}. In particular, for $k_a,k_b$ collinear to leg $\ell$,
we can write
\begin{align}
  \label{eq:Vcorrel-additive}
  \frac{V_{\rm sc}(\pBorn, k_a,k_b,k_1, \dots, k_n)}{v} &= \zeta \,f_{\rm correl}(z,\mu,\phi,\phi^{(\ell)}) 
                                                + \sum_{i=1}^n\zeta_i  \ \ ,
\end{align}
where $\zeta$ is defined in eq.~\eqref{eq:zetaFcorrel}. After some formal
manipulations, we obtain
\begin{multline}
  \label{eq:dFcor-additive}
  \dFcor(\lambda)=-\FNLL(\lambda)\sum_{\ell=1}^3 \frac{ \lambda R''_{\ell,\rm NNLL}}{2 a \beta_0\alpha_s(Q)} \int_0^{2\pi} \frac{d\phi^{(\ell)}}{2\pi}\times \\ \times \int_0^\infty \frac{d\mu^2}{\mu^2(1+\mu^2)} \int_0^1 dz \int_0^{2\pi} \frac{d\phi}{2\pi} \frac{1}{2!}\, \mathcal{A}^2\left(z, \mu,\phi \right)\ln f^{(\ell)}_{\rm correl}(z,\mu,\phi,\phi^{(\ell)})  \ \ .
\end{multline}
Now we discuss NNLL contributions induced by hard-collinear radiation. For
additive observables, following appendix C of~\cite{Banfi:2014sua}, for the
hard-collinear correction $\dFhc$, we find
\begin{equation}
  \label{eq:dFhc-additive}
  \dFhc(\lambda) = -\FNLL(\lambda) \sum_{\ell=1}^3 \frac{\as(Qv^{1/(a+b_\ell)})}{\as(Q)(a+b_\ell)} \gamma_\ell^{(0)} \left( \psi^{(0)}(1 + \RpNLL) + \gamma_E \right) \ \ ,
\end{equation}
where $\gamma_\ell^{(0)}$ arises due to the integral over the splitting
function. Now we move to computing the function $\dFrec$. First, we write
\begin{align}
  \frac{V_{\rm hc}(\pBorn, k, \{ k_i \})}{v} &= \zeta f^{(\ell)}_{\rm hc}(z^{(\ell)},\phi^{(\ell)}) + \sum_i \zeta_i \label{additivehc} \ \ , \\
  \frac{V_{\rm sc}(\pBorn, k, \{ k_i \}) }{v} &= \zeta f^{(\ell)}_{\rm sc}(z^{(\ell)},\phi^{(\ell)})  +  \sum_i \zeta_i \ \ ,
\end{align}
where $\zeta$ is now given by eq.~\eqref{eq:zeta-alt}, and thus we have
\begin{align}
f^{(\ell)}_{\rm sc}(z^{(\ell)},\phi^{(\ell)})  =\frac{d_\ell g_\ell (\phi)}{z^{(\ell)}} \left(\frac{Q}{2E_\ell}\right)^{b_\ell} \ \ .
\end{align}
Now we follow almost identical steps to the treatment of the soft
wide-angle correction and we get 
\begin{align}
\nonumber
  \delta\mathcal{F}_{\rm rec}(\lambda) = &\mathcal{F}_{\rm NLL}(\lambda) \sum_{\ell =1}^3 \frac{\alpha_s\left(Q v^{1/(a+b_\ell)}\right)}{(a+b_\ell) \alpha_s(Q)}  \int_0^{\pi} \frac{d\phi^{(\ell)}}{\pi} \times \\
  &\times \int_0^1 dz^{(\ell)}\,\left(\frac{2 C_\ell}{z^{(\ell)}}+ \langle P_{f_\ell}(z^{(\ell)},0) \rangle \right) \ln\frac{f^{(\ell)}_{\rm sc}(z^{(\ell)},\phi^{(\ell)})}{f^{(\ell)}_{\rm hc}(z^{(\ell)},\phi^{(\ell)})} \ \ .
\end{align}
Finally, we have the new function $\Delta \mathcal{F}_{\rm
  rec}$.
Instead of starting from eq.~\eqref{eq:finalDfrecMC}, we show here
that this function can be computed directly using eq.~\eqref{eq:DFreceps} and
use additivity, as follows:
\begin{align}
\nonumber
\Delta \mathcal{F}_{\rm rec}(\lambda) &= \left(\frac{4 \pi \mu_R^2\, e^{-\gamma_E}}{Q^2}  \right)^\epsilon \frac{\as(Qv^{1/(a+b_g)})}{\as(Q)(a+b_g)} v^{-2\epsilon} \bigg[ \int_0^{1/v} \frac{d\zeta}{\zeta^{1+2\epsilon}}\int \frac{d\Omega_{2-2\epsilon}}{\Omega_{2-2\epsilon}}  \int_0^1 dz \, \Delta P_{g} (z,\phi;\epsilon) \\
&\times \int \dZ \Theta\left(1 - \sum_i \zeta_i - \zeta f^{(f_g)}_{\rm hc}(z^{(\ell)},\phi^{(\ell)}) \right)\biggr] \ \ .
\end{align}
Owing to rIRC safety, we can rescale the $\zeta_i$'s, {\em cf.}
ref.~\cite{Banfi:2014sua}, to construct $\mathcal{F}_{\rm NLL}$
\begin{align}\label{eq:finalDfrec}
\nonumber
\Delta\mathcal{F}_{\rm rec}(\lambda)  &= \mathcal{F}_{\rm NLL}(\lambda)  \left(\frac{4 \pi \mu_R^2\, e^{-\gamma_E}}{Q^2}  \right)^\epsilon \frac{\as(Qv^{1/(a+b_g)})}{\as(Q)(a+b_g)} v^{-2\epsilon}\times \\
\times&\bigg[ \int_0^{1/f^{(f_g)}_{\rm hc}} \frac{d\zeta}{\zeta^{1+2\epsilon}}\int \frac{d\Omega_{2-2\epsilon}}{\Omega_{2-2\epsilon}}
  \int_0^1 dz \, \Delta P_{g} (z,\phi;\epsilon) \left( 1- f^{(f_g)}_{\rm hc}(z^{(\ell)},\phi^{(\ell)})\zeta \right)^{R'_{\rm NLL}}\biggr] \ \ .
\end{align}
Now this integral is well defined in $4-2\epsilon$ dimensions, and therefore we
can rescale $\zeta \to \zeta/f^{(f_g)}_{\rm hc}$
\begin{align}
\nonumber
\Delta\mathcal{F}_{\rm rec}(\lambda)  &= \mathcal{F}_{\rm NLL}(\lambda)  \left(\frac{4 \pi \mu_R^2\, e^{-\gamma_E}}{Q^2}  \right)^\epsilon \frac{\as(Qv^{1/(a+b_g)})}{\as(Q)(a+b_g)} v^{-2\epsilon}\times \\
\times&\bigg[  \int_0^{1} \frac{d\zeta}{\zeta^{1+2\epsilon}} \left( 1- \zeta \right)^{R'_{\rm NLL}}\int \frac{d\Omega_{2-2\epsilon}}{\Omega_{2-2\epsilon}}
  \int_0^1 dz \, \Delta P_{g} (z,\phi;\epsilon)  \left(f^{(f_g)}_{\rm hc}(z^{(\ell)},\phi^{(\ell)}) \right)^{2\epsilon}\biggr] \ \ ,
\end{align}
where now the $\zeta$ integral can be trivially performed and yields
\begin{align}
\nonumber
\Delta\mathcal{F}_{\rm rec}(\lambda)  &= \mathcal{F}_{\rm NLL}(\lambda)  \left(\frac{4 \pi \mu_R^2\, e^{-\gamma_E}}{Q^2}  \right)^\epsilon \frac{\as(Qv^{1/(a+b_g)})}{\as(Q)(a+b_g)} v^{-2\epsilon}\times \\
\times&\bigg[\frac{\Gamma(-2\epsilon) \Gamma(1+R'_{\rm NLL})}{ \Gamma(1+R'_{\rm NLL} - 2\epsilon)}  \int \frac{d\Omega_{2-2\epsilon}}{\Omega_{2-2\epsilon}}
  \int_0^1 dz \, \Delta P_{g} (z,\phi;\epsilon)  \left(f^{(f_g)}_{\rm hc}(z^{(\ell)},\phi^{(\ell)}) \right)^{2\epsilon}\biggr] \ \ .
\end{align}
Finally, we recall eq.~\eqref{eq:dPG-averaged} and expand the above equation
around $\epsilon = 0$ to find our final expression
\begin{align}
\nonumber
\Delta\mathcal{F}_{\rm rec}(\lambda)  = - \mathcal{F}_{\rm NLL}(\lambda) \frac{\as(Qv^{1/(a+b_g)})}{\as(Q)(a+b_g)}\bigg[ \int_0^\pi \frac{d\phi^{(\ell)}}{\pi}
  \int_0^1 dz^{(\ell)} \, \Delta P_{g} (z^{(\ell)},\phi^{(\ell)};0)  \ln f^{(f_g)}_{\rm hc}(z^{(\ell)},\phi^{(\ell)}) \biggr] \ \ .
\end{align}

\section{NNLL resummation of the $\mathbf{D}$-parameter in
  the near-to-planar limits}
\label{sec:D-NNLL}

As a proof of concept, in this article we concentrate on a specific three-jet event shape,
the $D$-parameter. This is defined in terms of the
determinant of the spherocity tensor \cite{Ellis:1980wv}
\begin{equation}
  \label{eq:spherocity-tensor}
  \Theta_{\alpha\beta}=\frac{1}{Q}\sum_i \frac{p_{i\alpha} p_{i\beta}}{E_i}\,,
\end{equation}
where the sum runs over all hadron momenta $p_i$ and $Q$ is the centre-of-mass energy of $e^+e^-$ annihilation. The spherocity tensor has three eigenvalues $\lambda_1,\lambda_2,\lambda_3$ satisfying
$\lambda_1+\lambda_2+\lambda_3=\mathrm{Tr}\,\Theta=1$. Using these eigenvalues we construct the $C$-parameter
\begin{equation}
  \label{eq:Cpar-def}
  C=3(\lambda_1\lambda_2+\lambda_1\lambda_3+\lambda_2\lambda_3)\,,
\end{equation}
and the $D$-parameter
\begin{equation}
  \label{eq:Dpar-def}
  D = 27\det \Theta = 27\lambda_1\lambda_2\lambda_3\,.
\end{equation}
For an isotropic
event all eigenvalues are equal to $1/3$, and hence both the $C$- and the $D$-parameter
are equal to 1.
Another useful form of the $D$-parameter is~\cite{Larkoski:2018cke}
\begin{equation}
  \label{eq:Dpar-new}
  D=\frac{27}{Q^3}\sum_{i<j<k} \frac{[\vec p_i\cdot(\vec p_j \times \vec p_k)]^2}{E_i E_j E_k} \ \ ,
\end{equation}
which is very convenient to obtain analytic expressions for the
$D$-parameter in the soft and collinear limits, as needed to compute the various components of our resummation master formula.
In particular, in the presence of multiple soft emissions $k_1,\dots,k_n$, eq.~\eqref{eq:Dpar-new} can be approximated as follows: 
\begin{equation}
  \label{eq:Dpar-soft-start}
  D(\{\tilde p\},k_1,\dots,k_n)\simeq\frac{27}{Q^3} \sum_{j<k=2}^3 \sum_i\frac{[\vec k_i\cdot(\vec{\tilde{p}}_j \times \vec{\tilde{p}}_k)]^2}{\omega_i \tilde E_j \tilde E_k}\,,
\end{equation}
where $k_i=(\omega_i,\vec k)$. Note that, in the presence of soft emissions, the
final-state hard momenta $\tilde p_1,\tilde p_2,\tilde p_3$ can be
approximated by their Born counterparts $p_1,p_2,p_3$. Therefore 
\begin{equation}
  \label{eq:Dpar-soft}
  D(\{\tilde p\},k_1,\dots,k_n)\simeq\frac{27}{Q^3}\sum_{j<k=2}^3 E_j E_k \sin^2\theta_{jk} \sum_i \frac{k_{ix}^2}{\omega_i}\,,
\end{equation}
where $k_{ix}$ is the component of $\vec k_i$ in the direction of
$\vec p_j\times \vec p_k$, i.e.\ out of the plane formed by the Born
momenta $p_1,p_2,p_3$.  Using the fact that, for three particles~(see
e.g.\ \cite{Larkoski:2018cke})
\begin{equation}
  \label{eq:Cpar-3part}
  C=3 \lambda_1\lambda_2=\frac{3}{Q^2}\sum_{j<k=2}^3 E_j E_k \sin^2\theta_{jk}\,,
\end{equation}
we obtain the final expression for the $D$-parameter in the presence of soft emissions:
\begin{equation}
  \label{eq:Dpar-soft-final}
  D(\{\tilde p\},k_1,\dots,k_n)\simeq 27 \lambda_1\lambda_2 \sum_i \frac{k_{ix}^2}{Q \omega_i}\,,
\end{equation}
where $\lambda_1 \lambda_2$ has to be computed using \emph{Born} momenta. 

\paragraph{NLL resummation.}
To compute the NLL resummation of the $D$-parameter we consider its
behaviour after a single soft emission, collinear to leg $\ell$. Using
eq.~\eqref{eq:Dpar-soft-final} and the Sudakov parametrisation in
eq.~(\ref{eq:Sudakov-leg}) we obtain
\begin{equation}
  \label{eq:Dsc}
  D(\{\tilde p\},k) \simeq 54 \lambda_1 \lambda_2 
    \frac{k_t^{(\ell)}}{Q}  e^{-\eta^{(\ell)}} \sin^2 \phi^{(\ell)}\,.
\end{equation}
Comparing the above expression with eq.~(\ref{eq:Vsc}) we get:
\begin{equation}
  \label{eq:D-coeffs}
  a=1\,,\quad b_\ell = 1\,, \quad d_\ell = 54 \lambda_1\lambda_2 \,,\quad g_\ell(\phi) = \sin^2\phi\,,\qquad \ell=1,2,3 \ \ .
\end{equation}
This information is enough to compute the resummed cumulant at NLL. We first
note that the $D$-parameter is additive, i.e. obeys eq.~\eqref{eq:additive},
which is clear from eq.~\eqref{eq:Dpar-soft-final}. The parameters in
eq.~\eqref{eq:D-coeffs} allows us to directly compute the NLL radiator using the
following relation
\begin{equation}
  \label{eq:lnd-average}
  \begin{split}
    \langle\ln(d_\ell g_\ell)\rangle = \ln\frac{d_\ell}{4}\ \ ,
  \end{split}
\end{equation}
and plugging $a = b_\ell =1$ in eq.~\eqref{eq:RNLL}. Finally,
$\mathcal{F}_{\rm NLL}$ is given by eq.~\eqref{eq:FNLL-additive}, where one
merely computes the logarithmic derivative $R'_{\rm NLL}$.

\paragraph{NNLL resummation.}
In order to use our prescription for the NNLL radiator, we first construct
$d_{\ell}^{(ij)}$ for each dipole by combining eq.~\eqref{eq:D-coeffs} with
eq.~(\ref{eq:dell-ij}). Once we have $d_{\ell}^{(ij)}$, we can compute the soft
NNLL radiator using the formulae of section \ref{sec:soft-radiator}.
In particular, we utilise the following relation
\begin{equation}
  \label{eq:lnd2-average}
  \begin{split}
    \langle\ln^2(d_\ell g_\ell)\rangle = \ln^2\frac{d_\ell}{4}+\frac{\pi^2}{3}\,.
  \end{split}
\end{equation}
The hard-collinear coefficients $C_{\mathrm{hc},\ell }^{(1)}$ can be
computed by replacing $\langle\ln(d_\ell g_\ell)\rangle$ in
eqs.~\eqref{eq:Chc-q} and~\eqref{eq:Chc-g} with the appropriate
expression in eq.~\eqref{eq:lnd-average}.

We now consider the various real-emission NNLL corrections. The
function $\dFsc$ is the one for additive observables, and is given by eq.~\eqref{eq:scollfinal}. Furthermore, due to additivity both the wide angle, $\Delta\mathcal{F}_{\rm wa}$, and the hard-collinear, $\dFhc$, functions are given by eqs.~\eqref{eq:finalDwa} and \eqref{eq:dFhc-additive}.

\begin{figure}[htbp]
\begin{center}
      \includegraphics[width=0.5\textwidth]{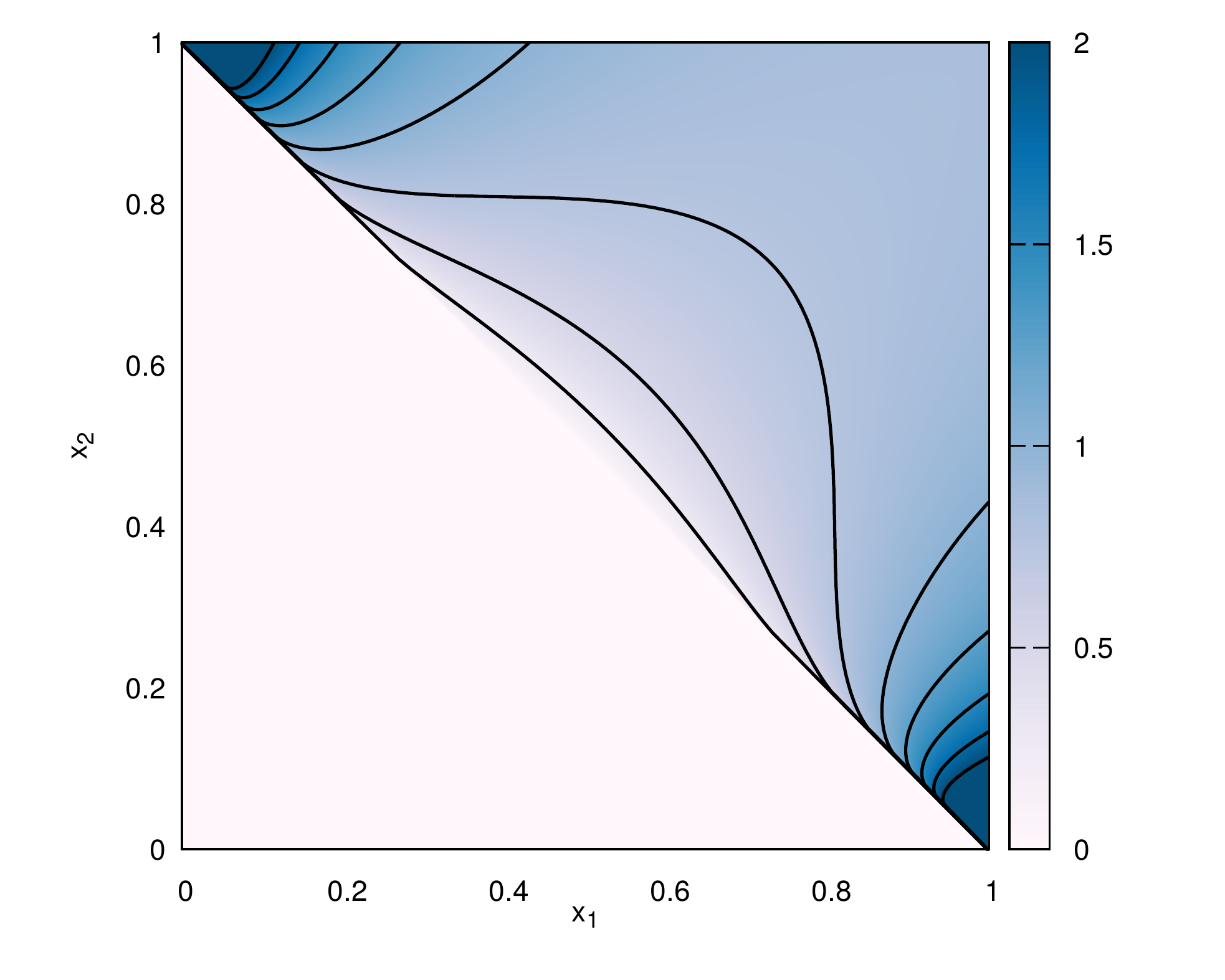}
\end{center}
\caption{A contour plot that displays our numerical results for the integral $I_{\rm wa}(\theta_{12})$ defined in eq.~\eqref{eq:Iwaij}.}
\label{Fwaplot}
\end{figure}

To compute recoil corrections, $\dFrec$ and $\Delta\mathcal{F}_{\rm rec}$, we need to obtain the expression for
the $D$-parameter after a single hard splitting of leg $\ell$. This
produces an emission $k$ with a fraction $z^{(\ell)}$ of the energy
$E_\ell$ of the parent momentum $p_\ell$, and a final-state momentum
$\tilde p_\ell$ carrying the remaining energy fraction $1\!-\!
z^{(\ell)}$. The important point to notice is that both $k$ and $\tilde{p}_\ell$ carry equal and opposite out of plane
momenta, $\tilde p_{\ell, x}= - k_x$.  From eq.~(\ref{eq:Dpar-new}), labelling
the remaining two hard partons with the indexes
$\ell_1,\ell_2\ne \ell$, we have to consider four terms
\begin{equation}
  \label{eq:Dpar-hc}
\begin{split}
  D_{\rm hc}(\{p\},k)& =\frac{27}{Q^3} \left\{\frac{\left[\vec k\cdot(\vec p_{\ell_1}\times\vec p_{\ell_2})\right]^2}{z^{(\ell)} E_\ell E_{\ell_1} E_{\ell_2}}
+ \frac{\left[\vec{\tilde p}_\ell\cdot(\vec p_{\ell_1}\times\vec p_{\ell_2})\right]^2}{(1-z^{(\ell)}) E_\ell E_{\ell_1} E_{\ell_2}}+
\frac{\left[\vec k\cdot(\vec{\tilde p}_\ell\times\vec p_{\ell_1})\right]^2}{z^{(\ell)}(1-z^{(\ell)}) E_\ell^2 E_{\ell_1}}+
\frac{\left[\vec k\cdot(\vec{\tilde p}_\ell\times\vec p_{\ell_2})\right]^2}{z^{(\ell)}(1-z^{(\ell)}) E_\ell^2 E_{\ell_2}}
 \right\}\\ & 
=\frac{27}{Q^3}\sum_{j<k=2}^3 E_j E_k \sin^2\theta_{jk}\frac{k_x^2}{z^{(\ell)}(1\!-\!z^{(\ell)}) E_{\ell}}= 27\lambda_1\lambda_2 \frac{k_x^2}{z^{(\ell)}(1\!-\!z^{(\ell)})E_\ell Q}\,.
\end{split}
\end{equation}
If we add an arbitrary number of soft and collinear emissions
$k_1,\dots,k_n$, their transverse momenta are much smaller than that
of the hard collinear emission, which is the only one that effectively
recoils against the hard parton $\tilde p_{\ell}$. In particular, the
soft emissions do not change the direction of the emitter, up to
non-singular corrections. Therefore, $k_x$ is the out-of-event-plane
component of the transverse momentum with respect to the emitter
$p_\ell$. Therefore, $k_x$ coincides with the emission's transverse
momentum with respect to the emitter $p_\ell$, and we get
\begin{equation}
  \label{eq:Dpar-hc-final}
  D_{\rm hc}(\{p\},k,k_1,\dots,k_n)= \frac{k_t^2}{Q^2} f_{\rm hc}^{(\ell)}(z^{(\ell)},\phi^{(\ell)}) +  D_{\rm sc}(\{p\},k_1,\dots,k_n)\,
\end{equation}
with
\begin{equation}
  \label{eq:D-fhc}
f_{\rm hc}^{(\ell)}(z^{(\ell)},\phi^{(\ell)}) =  \frac{27 \lambda_1\lambda_2 Q}{ z^{(\ell)}(1-z^{(\ell)}) E_\ell} \sin^2\phi^{(\ell)} \ \ .
\end{equation}
This means that the $D$-parameter is additive also in the presence of
an extra hard and collinear emission. For $z\to 0$ we have
\begin{equation}
  \label{eq:D-fsc}
f_{\rm sc}^{(\ell)}(z^{(\ell)},\phi^{(\ell)}) = \frac{27\lambda_1\lambda_2\,Q}{z^{(\ell)} E_\ell} \sin^2\phi^{(\ell)}  \ \ .
\end{equation}
Using eqs.~\eqref{eq:D-fhc} and~\eqref{eq:D-fsc}, as well as the
additivity of the $D$-parameter, we can compute $\dFrec$ using
eq.~\eqref{eq:dFrec} as follows:
\begin{equation}
  \label{eq:dFrec-Dpar}
  \begin{split}
    \dFrec(\lambda) &= \FNLL(\lambda) \sum_{\ell=1}^{3} \frac{\as(\sqrt{D} Q)}{2\, \as(Q)} \int_0^{\pi} \frac{d\phi^{(\ell)}}{\pi} \int_0^1 dz^{(\ell)} \left(\frac{2 C_\ell}{z^{(\ell)}}+ \langle P_{f_\ell}(z^{(\ell)},0) \rangle \right)  \ln \left(1-z^{(\ell)}\right) \\
    &= \FNLL(\lambda) \frac{\as(\sqrt{D} Q)}{2\, \as(Q)} \left( 2 C_F \left( \frac{5}{4} - \frac{\pi^2}{3} \right) + C_A \left( \frac{67}{36} - \frac{\pi^2}{3} \right)- T_R n_f \frac{13}{18} \right) \,.
  \end{split}
\end{equation}
Using eq.~\eqref{eq:D-fhc} we can also compute $\Delta \mathcal{F}_{\rm rec}$
from eq.~\eqref{eq:finalDfrec}, as follows:
\begin{equation}
  \label{eq:DeltaFrec-Dpar}
  \begin{split}
    \Delta\mathcal{F}_{\rm rec}(\lambda) & = -\mathcal{F}_{\rm NLL}(\lambda) \frac{\alpha_s\left(\sqrt{D}\,Q \right)}{2\,\alpha_s(Q)}
    \left(\frac{C_A}{2}-T_R n_f\right)\int_0^1 \!\! dz\, 4z(1-z) \int_0^\pi
    \frac{d\phi}{\pi}\, (2\cos^2\phi-1)\,\ln \left(\sin^2\phi\right) \\ &=
    \mathcal{F}_{\rm NLL}(\lambda) \frac{\alpha_s\left(\sqrt{D}\,Q \right)}{3\,\alpha_s(Q)}
    \left(\frac{C_A}{2}-T_R n_f\right) \ \ .
  \end{split}
\end{equation}
The next NNLL correction we need to compute is $\dFwa$. According to
eq.~(\ref{eq:Dpar-soft-final}), for soft emissions the $D$-parameter
is additive, so we can make use of the general expression in
eq.~\eqref{eq:dFwa-additive}. To achieve this we need to recast the
expression of the $D$-parameter, with a single soft wide-angle
emission, in the form of eq.~\eqref{eq:Vwa-additive}. Using the
Sudakov decomposition in eq.~(\ref{eq:Sudakov-k}), for the dipole
$(ij)$ we obtain
\begin{equation}
  \label{eq:Dwa-ij}
\begin{split}
  D_{\rm wa}(\{\tilde p\},k)& =27 \lambda_1\lambda_2 \frac{k_x^2}{\omega Q}\\  & = 27 \lambda_1\lambda_2\frac{\kappa^{(ij)} }{Q}
\frac{\sin (\theta_{ij}/2)\sin^2 \phi^{(ij)}}{\cosh (\eta^{(ij)} + \eta_0^{(ij)}) + \cos (\theta_{ij}/2) \cos \phi^{(ij)}}= \frac{\kappa^{(ij)} }{Q_{ij}} f_{\rm wa}^{(ij)}(\eta^{(ij)}, \phi^{(ij)})\,,
\end{split}
\end{equation}
where
\begin{equation}
  \label{eq:fwa-ij}
 f_{\rm wa}^{(ij)}(\eta^{(ij)}, \phi^{(ij)}) = 27 \lambda_1\lambda_2\frac{Q_{ij}}{Q} \frac{\sin (\theta_{ij}/2) \sin^2 \phi^{(ij)}}{(\cosh (\eta^{(ij)} + \eta_0^{(ij)}) + \cos (\theta_{ij}/2) \cos \phi^{(ij)})}\,.
\end{equation}
Using eq.~(\ref{eq:fwascoll}) and eq.~(\ref{eq:D-coeffs}), we find
\begin{equation}
  \label{eq:Dwa-fsc}
  f_{\rm sc}^{(ij)}(\eta^{(ij)}, \phi^{(ij)}) = 54 \lambda_1 \lambda_2 \frac{Q_{ij}}{Q}\sin\frac{\theta_{ij}}{2}\sin^2 \phi^{(ij)}\left[e^{-(\eta^{(ij)} + \eta_0^{(ij)})}\,\Theta(\eta^{(ij)}) + e^{\eta^{(ij)} + \eta_0^{(ij)}}\,\Theta(-\eta^{(ij)})\right] \,.
\end{equation}
Inserting the above expressions in eq.~\eqref{eq:dFwa-additive} we
obtain
\begin{equation}
  \label{eq:dFwa-Dpar}
  \begin{split}
    \dFwa(\lambda) 
    &= \FNLL(\lambda) \sum_{(ij)} C_{(ij)} \frac{\as(DQ)}{\as(Q)} \int_0^{2\pi} \frac{d\phi}{2\pi} \int_{-\infty}^\infty d\eta \times \\
    &\times \left(
      \ln \left[
        2 e^{-(\eta+\eta_0^{(ij)})}
        \left( \cosh(\eta + \eta_0^{(ij)}) +
          \cos\frac{\theta_{ij}}{2} \cos \phi \right)
      \right]
      \Theta(\eta) \right.
    \\
    &+
    \left. \ln \left[ 2 e^{\eta+\eta_0^{(ij)}} \left( \cosh(\eta + \eta_0^{(ij)}) + \cos\frac{\theta_{ij}}{2} \cos \phi \right) \right] \Theta(-\eta) \right) \\
& = \FNLL(\lambda) \sum_{(ij)} C_{(ij)} \frac{\as(DQ)}{\as(Q)}  \left((\eta_0^{(ij)})^2+2 I_{\rm wa}(\theta_{ij})\right)\,,
  \end{split}
\end{equation}
where 
\begin{equation}
  \label{eq:Iwaij}
  I_{\rm wa}(\theta_{ij}) \equiv \int_0^\infty \!\! d\eta\,\ln \left[ e^{-\eta} \left( \cosh\eta + \sqrt{\cosh^2\eta - \cos^2 \frac{\theta_{ij}}{2}} \right) \right]
\end{equation}
In fig.~\ref{Fwaplot} we provide a plot of the integral in
eq.~\eqref{eq:Iwaij} as a function of the three-parton variables
$(x_1,x_2)$ defined in Appendix~\ref{sec:kinematics}. The plot shows the explicit result only for the $q\bar{q}$ dipole, and we choose not to explicitly show the similar plots for either the $qg$ or the $\bar{q}g$ dipoles. The only difference is simply that the contours rotate in the $(x_1,x_2)$ plane.


The last contribution we need to compute is $\dFcor$. Since the
$D$-parameter is additive, we can again use the general formula for additive
observables in eq.~\eqref{eq:dFcor-additive}. We recast the $D$-parameter, with
two soft-collinear emissions, in the form of eq.~\eqref{eq:Vcorrel-additive}.
This gives
\begin{equation}
  \label{eq:Dcor}
  f_{\rm correl}(\mu,z,\phi,\phi^{(\ell)})=1+\mu^2 \frac{\sin^2(\phi+\phi^{(\ell)})}{\sin^2\phi^{(\ell)}} \ \ ,
\end{equation}
which is the same for all three legs. This gives
\begin{equation}
  \label{eq:dFcor-Dpar}
  \dFcor(\lambda)=-\FNLL(\lambda) \sum_{\ell =1}^3 \frac{\lambda R^{''}_{\ell,0,\rm NNLL}}{2\beta_0\alpha_s(Q)}\left(C_A\langle\ln f_{\rm correl}\rangle_{C_A}+
  n_f\langle \ln f_{\rm correl}\rangle_{n_f}\right)\ \ ,
\end{equation}
and the various integrals are computed via a Monte Carlo routine
\begin{equation}
  \label{eq:lnf-average}
  \begin{split}
    \langle\ln f_{\rm correl}\rangle_{C_A}&=\frac{1}{2!}\int_0^{2\pi}\frac{d\phi^{(\ell)}}{2\pi}
  \int_0^\infty\frac{d\mu^2}{\mu^2(1+\mu^2)}\int_0^1 dz\int_0^{2\pi}\frac{d\phi}{2\pi}(2\mathcal{S}+\mathcal{H}_g)\ln f_{\rm correl}(\mu,z,\phi,\phi^{(\ell)}) = 1.8139 \ \ , \\
    \langle\ln f_{\rm correl}\rangle_{n_f}&=T_R\int_0^{2\pi}\frac{d\phi^{(\ell)}}{2\pi}
  \int_0^\infty\frac{d\mu^2}{\mu^2(1+\mu^2)}\int_0^1 dz\int_0^{2\pi}\frac{d\phi}{2\pi} \mathcal{H}_q \ln f_{\rm correl}(\mu,z,\phi,\phi^{(\ell)})= 1.1562 \ \ .
  \end{split}
\end{equation}

\section{Validation and phenomenology}
\label{sec:checkandmatch}

In this section we validate the analytic results of
section~\ref{sec:D-NNLL}, match them to fixed order at NLO and finally
compare our matched distribution to LEP1 data. We do so in two
steps. First, we use the Monte Carlo event generator EVENT2 to check
most, but not all, of the pieces in the expansion of the
resummation. For the $D$-parameter, EVENT2 provides results at LO,
i.e.\ $\mathcal{O}(\alpha_s^2)$, and thus we will validate all terms
in the expansion at $\mathcal{O}(\alpha_s^2)$ up to NNLL. Second, we
use NLOJet++ to match the resummation at NLO, i.e.\
$\mathcal{O}(\alpha_s^3)$. Below we explain our choice of the matching
scheme and point out interesting aspects of the resulting
phenomenology.

\subsection{Partial validation using EVENT2}
\label{sec:event2}

In this subsection we start with EVENT2 to validate various ingredients in the
resummed cumulative distribution of section~\ref{sec:D-NNLL}. Given that the
Born event is already at $\mathcal{O}(\alpha_s)$, all the pieces in the
expansion of the resummation that starts at $\mathcal{O}(\alpha_s)$ can then be
checked against EVENT2. Table~\ref{tab:table1} lists these various terms, which
contribute at successive logarithmic accuracy. Moreover, the results of EVENT2
are sufficient to validate the geometry dependence in the radiator at NLL, i.e.\
the $d_\ell$-dependent term in eq.~\eqref{eq:Rol}.
\begin{table}[h!]
\begin{center}
\begin{tabular}{ |c| c | } 
 \hline
 LL & $g_1^{(\ell)}(\lambda)$   \\ 
 \hline
 NLL & $h_2^{(\ell)}(\lambda)$ \\ 
 \hline
 NNLL & $H(\{p\}, \alpha_s(Q))$ \,  $C^{(1)}_{{
\rm hc}, \ell}$ \, $\delta\mathcal{F}_{\rm wa}(\lambda)$ \, $\delta\mathcal{F}_{\rm rec}(\lambda)$ \, $\Delta\mathcal{F}_{\rm rec}(\lambda)$ \\ 
 \hline
\end{tabular}
\caption{The various contributions in the expansion of the resummation that are
  amenable to validation against EVENT2.}
\label{tab:table1}
\end{center}
\end{table}

\begin{figure}[htbp]
\begin{center}
      \includegraphics[width=0.5\textwidth]{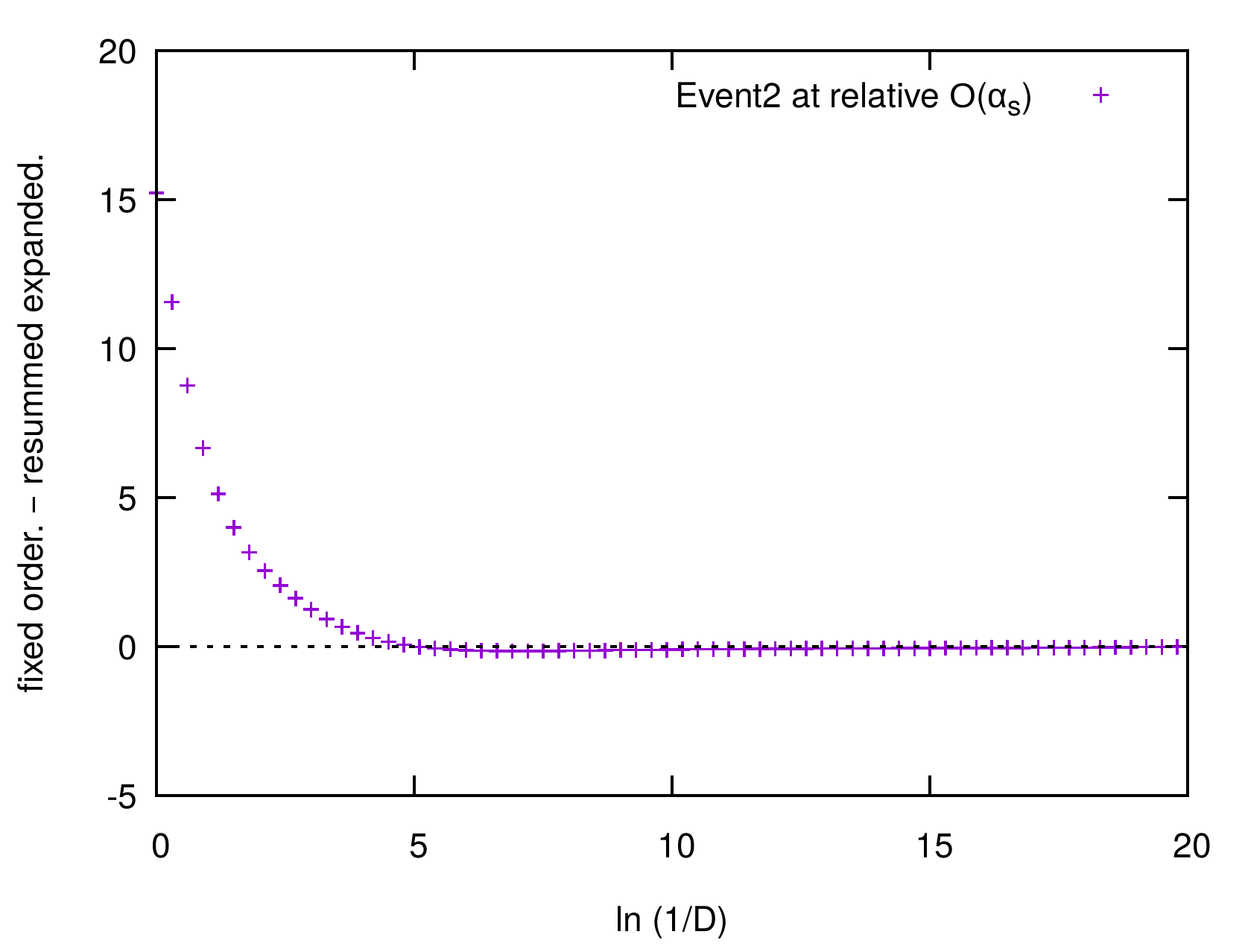}
\end{center}
\caption{The plot shows the difference between the fixed-order result
  for the D parameter at LO, i.e. $\mathcal{O}(\alpha_s^2)$, using
  EVENT2 and the expansion of the resummation from
  section~\ref{sec:D-NNLL}.}
\label{fig:a2const}
\end{figure}

In fig.~\ref{fig:a2const} we expand the resummation of the cumulative
cross section and subtract the result from EVENT2. We see indeed that
the difference is consistent with zero, and displays an asymptotic
behaviour for the entire resummation region.

\begin{figure}[htbp]
\begin{center}
      \includegraphics[width=0.5\textwidth]{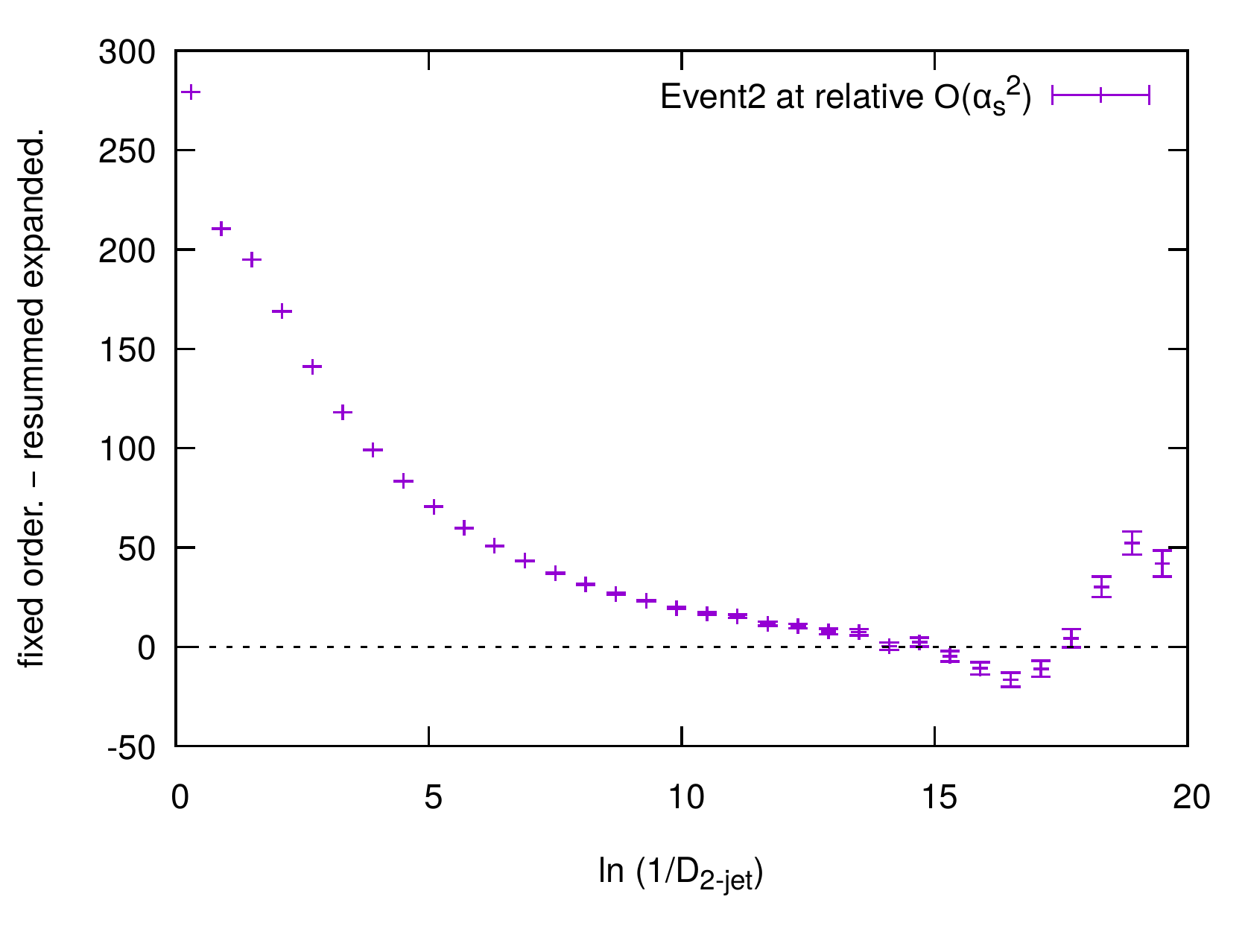}
\end{center}
\caption{The plot shows the difference between the EVENT2 result for $D_{2-\rm jet}$ at NLO, i.e. $\mathcal{O}(\alpha_s^2)$, and the expansion of the resummation.}
\label{fig:fakeD}
\end{figure}

Moreover, we can isolate and validate an extra NNLL function using
EVENT2 results, i.e. $\delta\mathcal{F}_{\rm
  correl}(\lambda)$.
Indeed, we can not achieve this directly because the fixed-order
expansion of $\delta\mathcal{F}_{\rm correl}(\lambda)$ starts at
$\mathcal{O}(\alpha_s^2)$. Nevertheless, we can look for an event
shape whose behaviour in the presence of correlated emissions is
identical to that of the $D$ parameter. We shall call this observable
$D_{2-\rm jet}$, and it reads
\begin{align}
D_{2-\rm jet} = \frac{1}{Q^2}  \sum_{i < j} \frac{[\hat{n}_{\rm beam} \cdot(\vec p_i \times \vec p_j)]^2}{E_i E_j}  \ \ ,
\end{align}
where $\hat{n}_{\rm beam}$ is a unit vector along the electron beam
direction. The design of $D_{2-\rm jet}$ is meant to capture the
behaviour of $\delta\mathcal{F}_{\rm correl}(\lambda)$ for the actual
$D$-parameter. This observable does not possess a lot of the structure
witnessed in three-jet observables, nevertheless, it allows us to
check our Monte Carlo implementation for obtaining the results in
eq.~\eqref{eq:lnf-average}. We have computed the analytic resummation
of $D_{2-\rm jet}$, although we do not quote the results
here. Fig.~\ref{fig:fakeD} shows the result of subtracting the resummed
{\em differential} distribution from that of EVENT2. Admittedly, the plot in
fig.~\ref{fig:fakeD} does not exhibit a satisfactory asymptotic behaviour,
however, it is quite suggestive. By making use of quadruple precision
one should be able to probe sufficiently small values of
$D_{2-\rm jet}$.
Finally, we have also tried to fully validate our resummation of the
$D$ parameter, at $\mathcal{O}(\alpha_s^3)$, using
NLOJet++. Unfortunately, double precision did not allow us to reach
values of $\ln 1/D$ larger than $10-12$, which are not asymptotic
enough to provide a reliable validation of our NNLL resummation.

\subsection{Matching to fixed order at NLO}
\label{sec:matching}

In order to provide a suitable cumulative distribution that paves
the way for phenomenological studies, one has to match the resummation
to fixed order. Matching is required to provide results across all
values of the observable. The basic idea is to combine the results of
both the resummation and fixed order, while making sure to get rid of
contributions that are double counted. There are two generic
conditions that the matching procedure must satisfy. First, based on
physical grounds the matched total cross section should go to zero as
$v \to 0$. Second, the matched distribution, or likewise the total
cross section, must reproduce the fixed order at the kinematic
endpoint $v \to v_{\rm max}$.

The two most popular matching schemes for $e^+e^-$ annihilation are
the R and log-R schemes~\cite{Catani:1991kz,Catani:1992ua}. In other
contexts, multiplicative matching schemes are
used~\cite{Antonelli:1999kx,Dasgupta:2001eq}. Adopting one over the
other is a choice that depends on the problem at hand. In our case, we
could not achieve a stable matched distribution using the R and log-R
schemes, given that the available data sets for the $D$ parameter
forces us to use low values of $y_{\rm cut}$. The problem with both
schemes is that, given that the various components of the resummed
cross section contain powers of $\ln y_{\rm cut}$, the resummation
does not switch off quickly enough and ends up substantially
contributing to the tail of the matched distribution. This situation
might be expected given that the K-factor NLO/LO is very large,
approximately 100 \%. This is similar to the case of resuming the
distribution in the Higgs transverse momentum $p_{t,H}$ where the K
factor is also known to be large \cite{Bizon:2017rah}. We therefore
need to supplement our matching scheme with a factor that effectively
damps the resummation at large values of $D$.

Based on the above discussion, we use the multiplicative matching
scheme designed in ref.~\cite{Bizon:2017rah}. The goal of that scheme
is precisely to suppress the large terms, present in the resummation,
which emerge outside the resummation region. This enables us to
control the tail of the distribution and achieve a stable matching. In
this scheme, matching is performed on the level of the total cross
section. Given that NLOJet++ simulates the {\em inclusive} cross
section, i.e.\ integrated over Born kinematics with the three-jet
selection cut, we have to match on the same level. Explicitly, we have
\begin{align}
\label{eq:mastermatch}
\Sigma_{\mathcal{H}}^{\rm Mat.}(v) = \left(\Sigma_{\mathcal{H}}^{\rm Res.}(v)\right)^Z \frac{\Sigma^{\rm{FO.}}_{\mathcal{H}}(v)}{\left(\Sigma_{\mathcal{H}}^{\rm Exp.}(v)\right)^Z}\,,
\end{align}
where 
\begin{align}\label{eq:Z}
Z = \left( 1 - \left( \frac{v}{v_0}\right)^u \right)^h \Theta(v-v_0) \,,
\end{align}
controls how quickly the logarithms are shut down outside the resummation
region. In eq.~\eqref{eq:mastermatch}, $\Sigma_{\mathcal{H}}^{\rm Res.}$ is the
resummed cross section, $\Sigma^{\rm{FO.}}_{\mathcal{H}}$ is the corresponding
fixed-order quantity and $\Sigma_{\mathcal{H}}^{\rm Exp.}$ denotes the expansion
of the resummation to NLO. The details of the matching can be found in
appendix~\ref{app:matching} and the expanded version of eq.~\eqref{eq:mastermatch}
is given in full in eq.~\eqref{eq:mastermatch-explicit}. The presence of the step
function in eq.~\eqref{eq:Z} suggests that the transition region, between the
resummation and fixed order, will not be smooth enough. In fact, we verified
that even if the step function is removed from the definition of $Z$, the
resummation still shuts down smoothly well before reaching the kinematic
endpoint $v_{\rm max}$. We carry out the matching using the values $u=1$, $h=3$
and $v_0 = 1/2$.

As is customary in resummed calculations, we need to probe the size of
subleading logarithmic terms. This is done using two simultaneous
variations. The first introduces a rescaling $x_V$ as follows
\begin{equation}\label{eq:xv}
  \ln \frac{1}{v} = \ln \frac{x_V}{v} - \ln x_V \, , \quad x_V \equiv X \cdot X_V \ \ .
\end{equation}
In the above, $X$ is a variable choice to define the resummation
scale, i.e.\ the logarithms being practically resummed, while $X_V$
controls the scale variation. We expand the total cross section around
$\ln x_V/v$ neglecting subleading terms.  Furthermore, the resummed
logarithm, $\ln x_V/v$, must be modified in order to impose that the
total cross section is reproduced at the kinematical endpoint
$v_{\rm max}$ \cite{Dasgupta:2001eq}
\begin{equation}
\label{eq:Ltilde}
  \ln \frac{x_V}{v} \to \tilde{L} \equiv \frac{1}{p} \ln \left( \left( \frac{x_V}{v} \right)^p
    - \left( \frac{x_V}{v_{\rm max}} \right)^p + 1 \right)   \ \ ,
\end{equation}
where $p$ denotes a positive number that controls how quickly the
logarithms are switched off close to the endpoint. The parameter $p$
is free, but is only constrained by the behaviour of the fixed order
distribution near the endpoint \cite{Dasgupta:2001eq}. In our case, we
set $p=1$.

Another estimate for the uncertainty in our matched distribution comes
from varying the renormalisation scale, $\muR$, around a central scale
that we take to be the centre-of-mass energy of the hard scattering,
$Q$. For LEP1 energies, $Q=M_Z$ corresponding to $\as(M_Z) = 0.118$
while for FCCee energies, $Q=500\, \rm GeV$ corresponding to
$\as(500\, \rm GeV) = 0.094$.

We implement two different choices for $X$ in eq.~\eqref{eq:xv}. The
first is referred to as the $X_{\rm const}$ scheme which corresponds
to setting $X=1$ in eq.~\eqref{eq:xv}, while the second is the
$X_{\rm prod}$ scheme which corresponds to setting
\begin{align}
X = \frac{3}{2C_F + C_A} \ln \frac{27 \lambda_1 \lambda_2}{2} \ \ ,
\end{align}
which is a function of Born kinematics.  Finally, we construct the
uncertainty bands by varying $\muR$ by a factor of two in either
direction and $X_V$ by a factor of three-halves in either direction.

\begin{figure}[htbp]
  \begin{minipage}[l]{0.5\linewidth}
    \begin{center}
      \includegraphics[width=\textwidth]{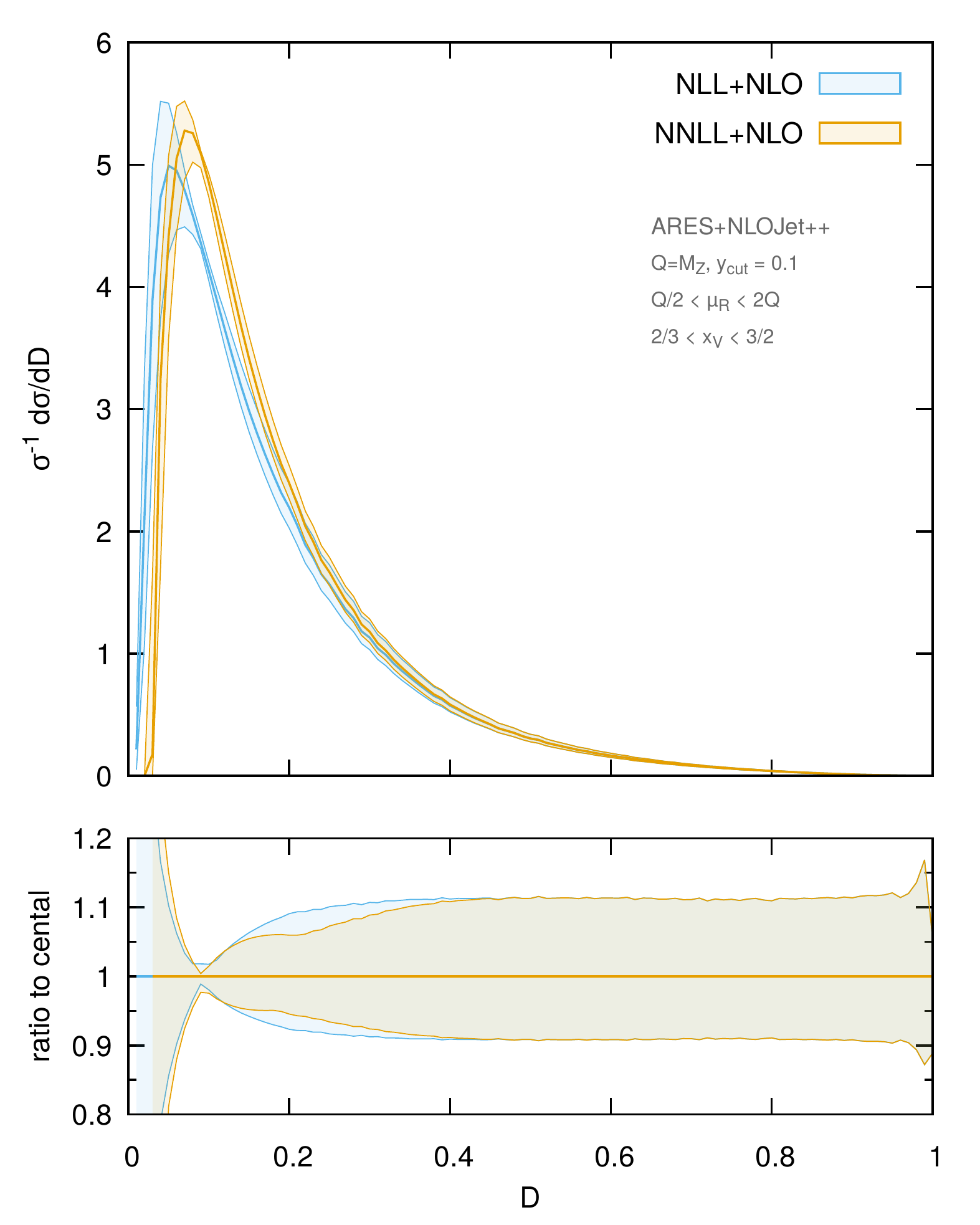}
    \end{center}
  \end{minipage}
  \begin{minipage}[r]{0.5\linewidth}
    \begin{center}
      \includegraphics[width=\textwidth]{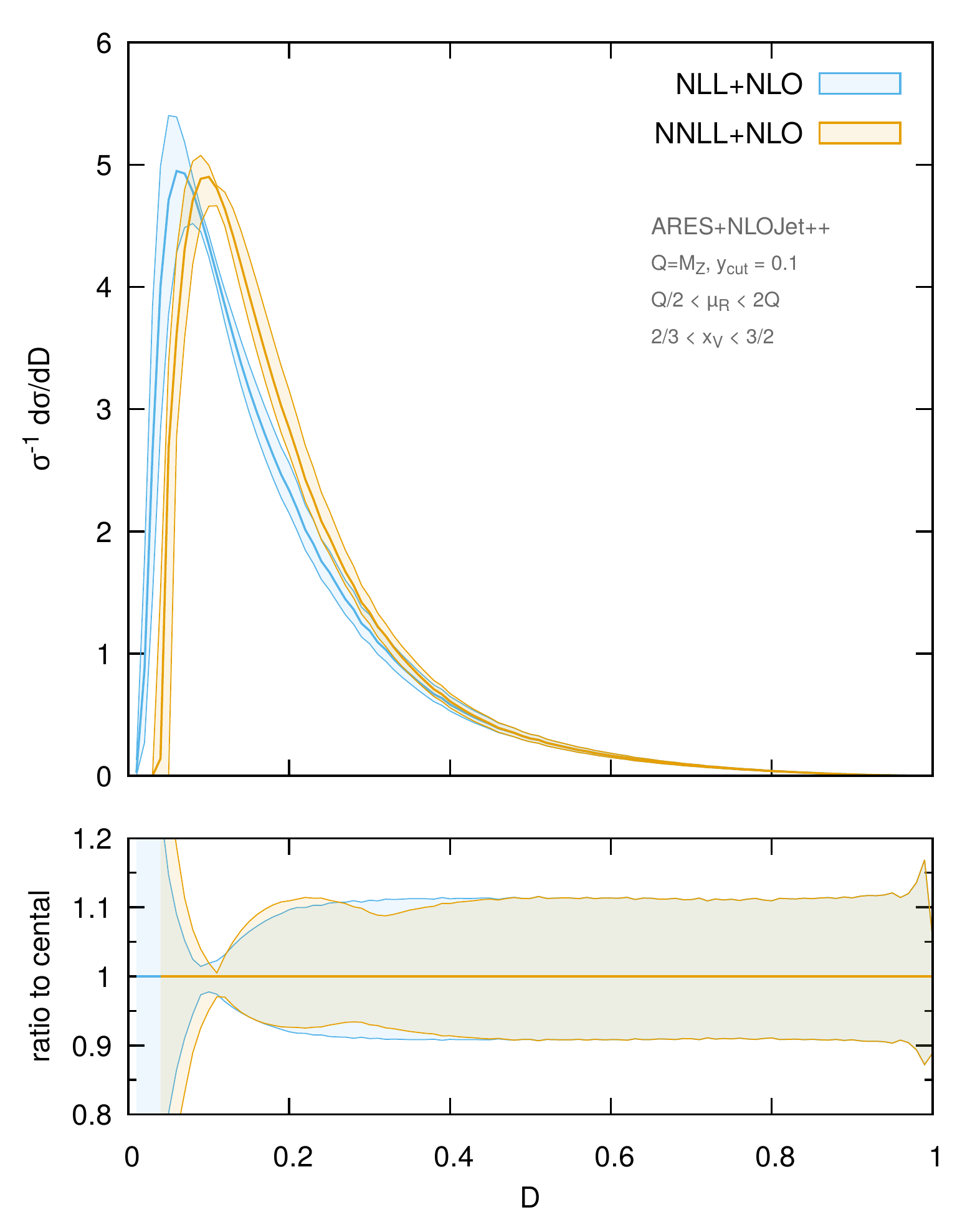}
    \end{center}
  \end{minipage}
  \caption{The matched distribution for $y_{\rm cut}=0.1$ and $Q = M_Z$. The left plot using the
    $X_{\rm const}$ scheme and the right using the $X_{\rm prod}$ scheme.}
  \label{fig:dpar-y0100-MZ-PT}
\end{figure}

\begin{figure}[htbp]
  \begin{minipage}[l]{0.5\linewidth}
    \begin{center}
      \includegraphics[width=\textwidth]{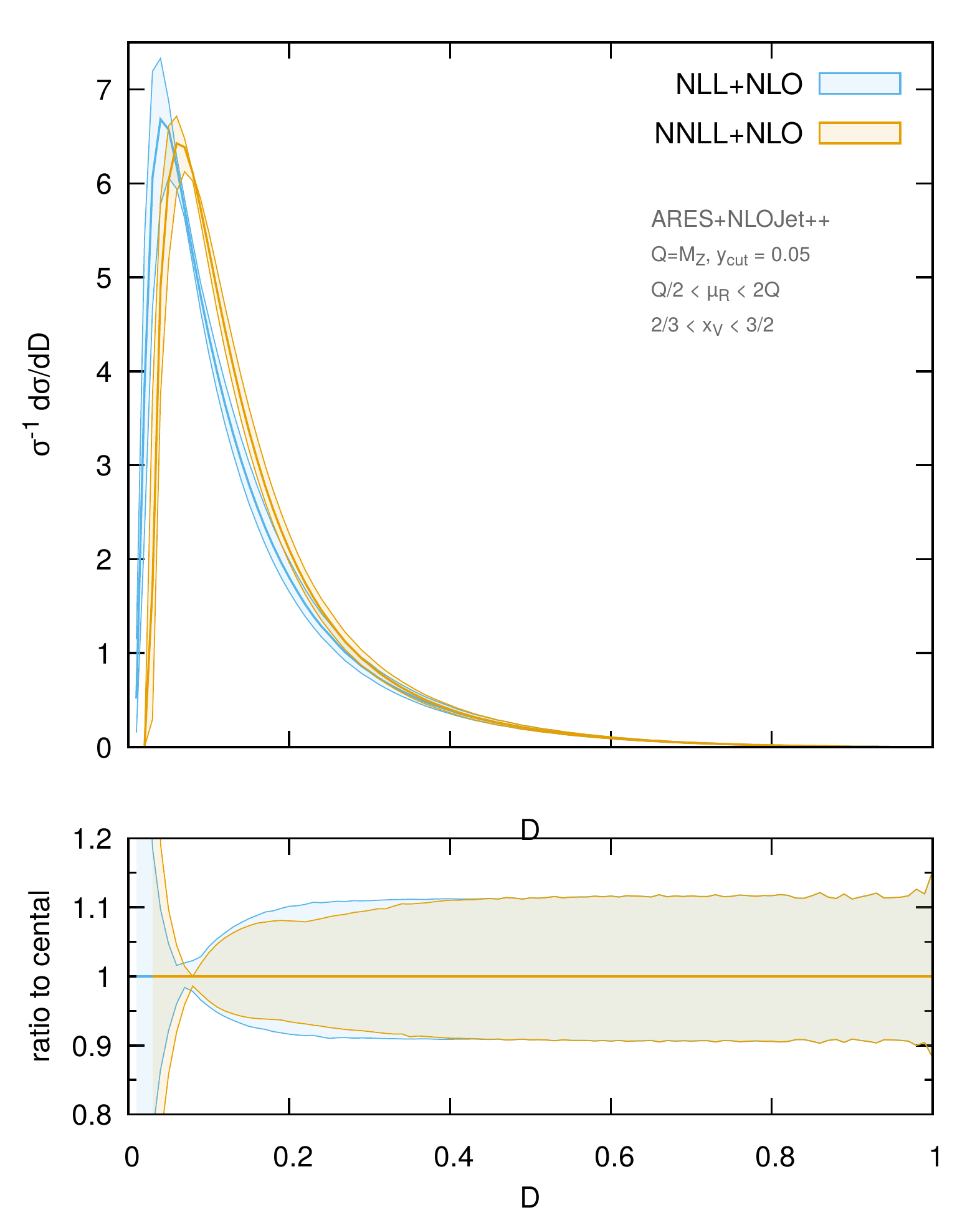}
    \end{center}
  \end{minipage}
  \begin{minipage}[r]{0.5\linewidth}
    \begin{center}
      \includegraphics[width=\textwidth]{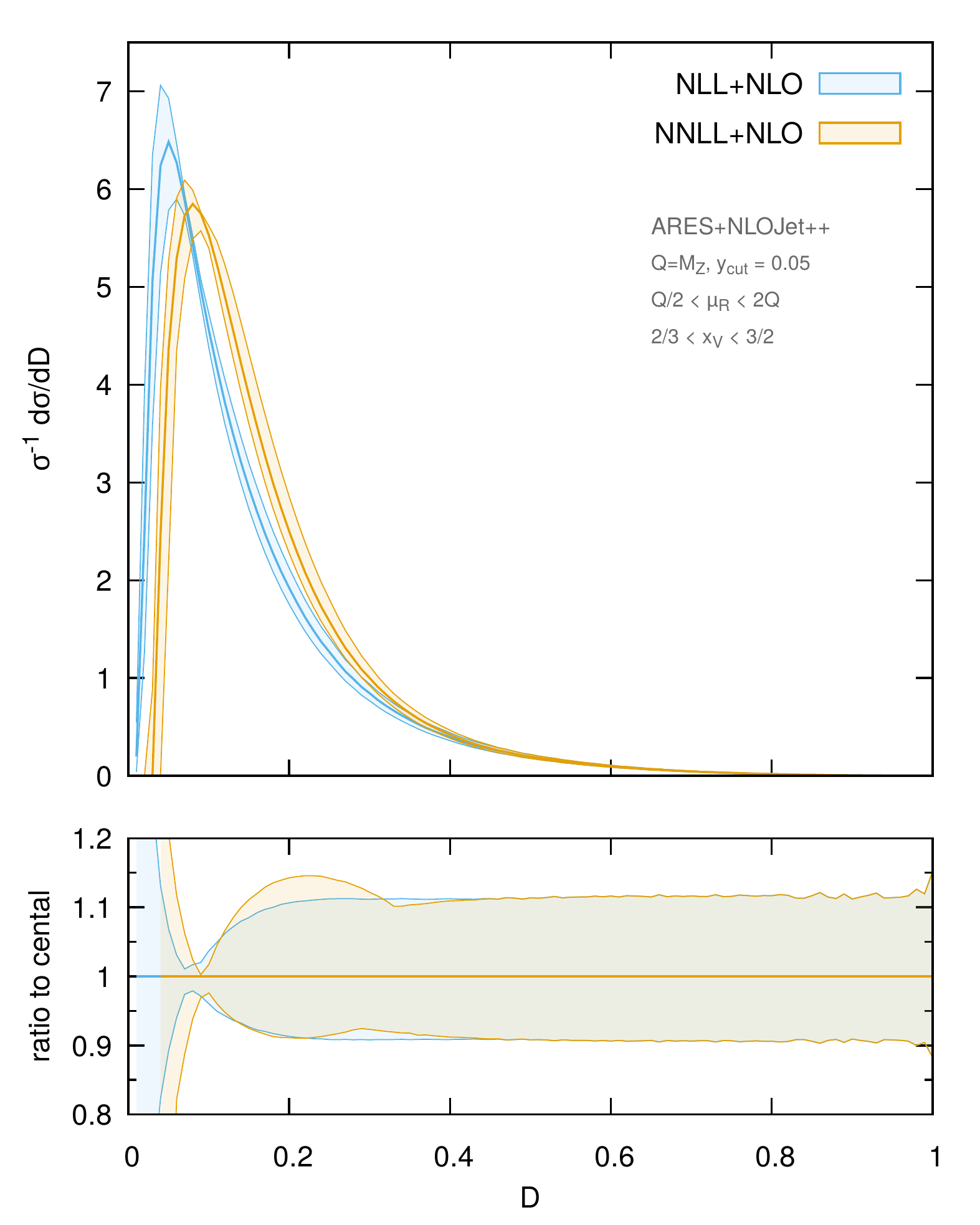}
    \end{center}
  \end{minipage}
  \caption{The matched distribution for $y_{\rm cut}=0.05$ and $Q = M_Z$. The left plot using the
    $X_{\rm const}$ scheme and the right using the $X_{\rm prod}$ scheme.}
  \label{fig:dpar-y0050-MZ-PT}
\end{figure}

In figs.~\ref{fig:dpar-y0100-MZ-PT} and \ref{fig:dpar-y0050-MZ-PT} we
plot the matched distribution for $Q = M_Z$ using the two resummation
schemes and for two different values of $y_{\rm cut}$, namely
$y_{\rm cut}=0.1$ and $y_{\rm cut}=0.05$. We immediately notice the
following features:
\begin{itemize}
\item The uncertainty bands are not drastically reduced when
  increasing the logarithmic accuracy of the resummation, at least
  when compared to the typical situation with two-jet observables.
\item The position of the peak is stable under varying $y_{\rm cut}$.
\item For NLL, the uncertainty bands remain almost unchanged with
  decreasing $y_{\rm cut}$. In contrast, the uncertainty bands for
  NNLL are noticeably enhanced as we increase $y_{\rm cut}$.
\end{itemize}
The above features can be traced to the fact that jet selection
generates terms that go as $\ln ^2y_{\rm cut}$, for each power of
$\alpha_s$ relative to the Born cross section. The largest {\em
  transverse} momentum of soft-collinear emissions, at fixed value of
$D$, is of the order of $\sqrt{D} Q$. Our resummation is strictly
defined when the largest momentum is much smaller than the largest
transverse momentum available, the latter being of the order of
$\sqrt{y_{\rm cut}} Q$. Essentially, our resummation is formally
correct, as $D \ll 1$, but phenomenologically viable only in the limit
$D \ll y_{\rm cut} \ll 1$.  Inspection of
figs.~\ref{fig:dpar-y0100-MZ-PT} and~\ref{fig:dpar-y0050-MZ-PT} shows
that the most probable value of $D$, which corresponds to the position
of the peak of differential distributions, is of the same order as
$y_{\rm cut}$ and that is why we see the features described
above. This is also reflected in the sensitivity of the uncertainty
bands of the NNLL distribution to the variation of $y_{\rm cut}$ in
comparison to NLL. Simply, the NNLL pieces in the cross section,
e.g.\ $\Delta \mathcal{F}_{\rm wa}$, contain extra powers of
$\ln y_{\rm cut}$ compared to NLL. These logarithms are large, for
$y_{\rm cut} = 0.05$-$0.1$, and thus we observe this behaviour of the
uncertainty bands.

\begin{figure}[htbp]
  \begin{minipage}[l]{0.5\linewidth}
    \begin{center}
      \includegraphics[width=\textwidth]{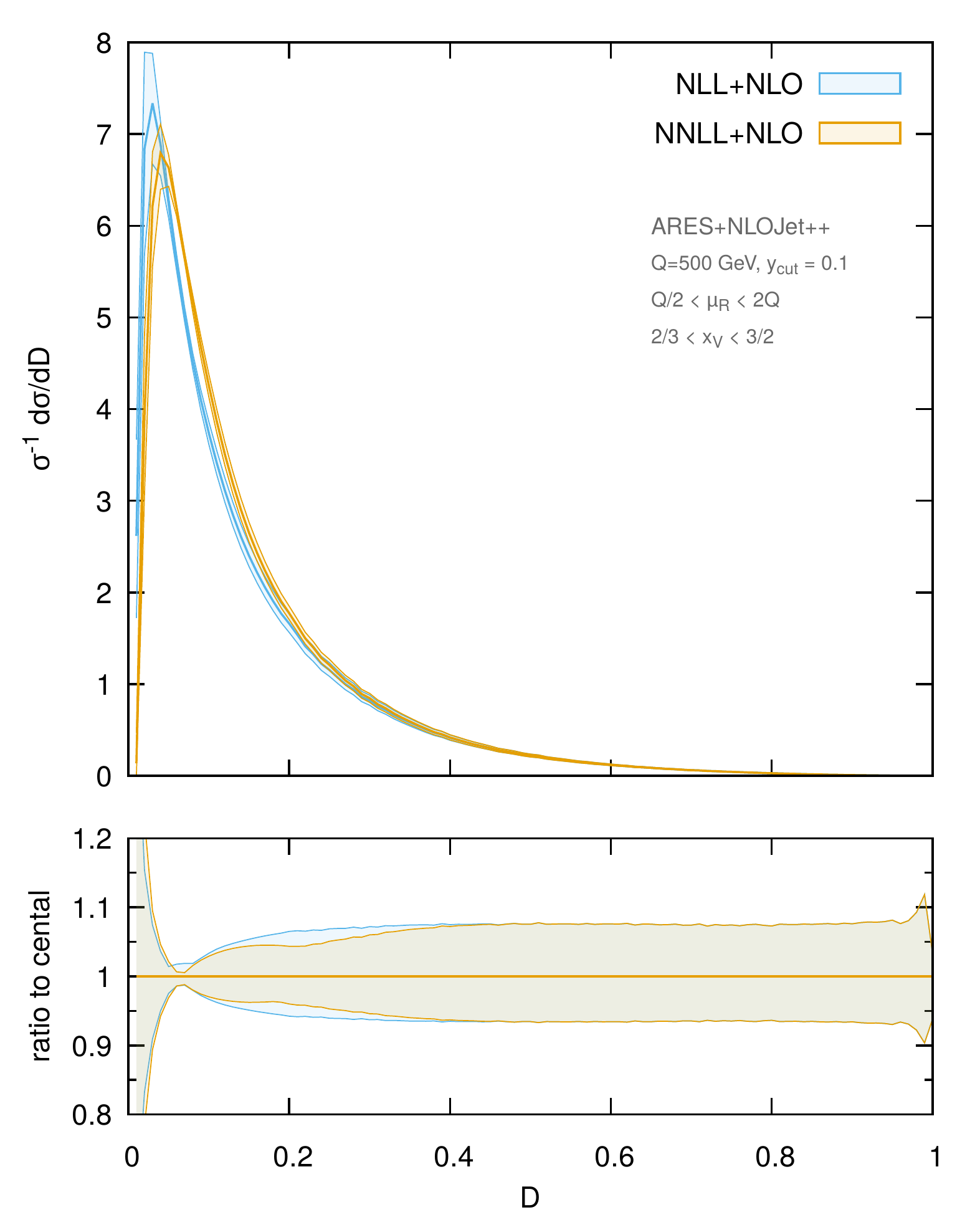}
    \end{center}
  \end{minipage}
  \begin{minipage}[r]{0.5\linewidth}
    \begin{center}
      \includegraphics[width=\textwidth]{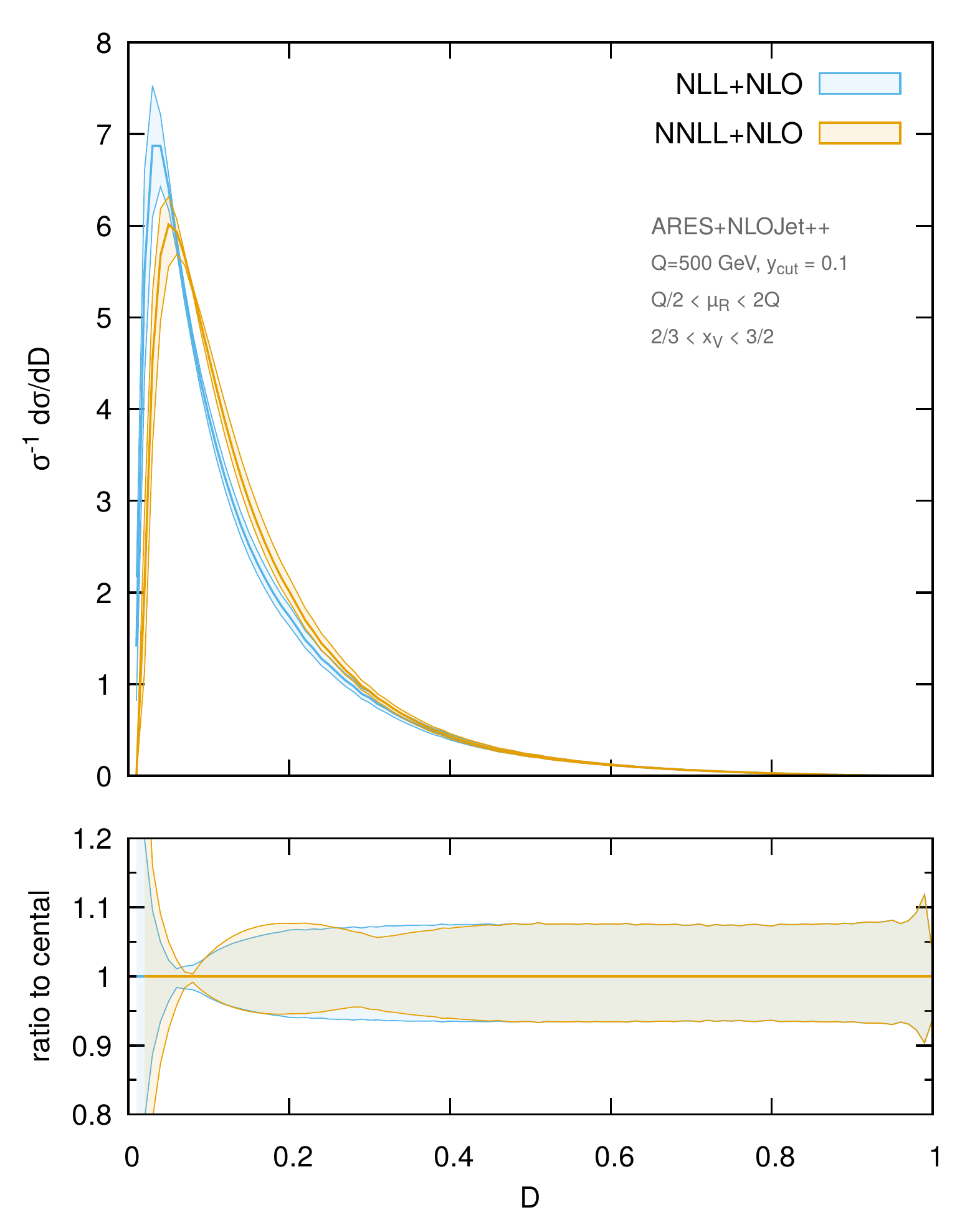}
    \end{center}
  \end{minipage}
  \caption{The matched distribution for $y_{\rm cut}=0.1$ and $Q = 500 \rm GeV$. The left plot using the
    $X_{\rm const}$ scheme and the right using the $X_{\rm prod}$ scheme.}
  \label{fig:dpar-y0100-Q500-PT}
\end{figure}

\begin{figure}[htbp]
  \begin{minipage}[l]{0.5\linewidth}
    \begin{center}
      \includegraphics[width=\textwidth]{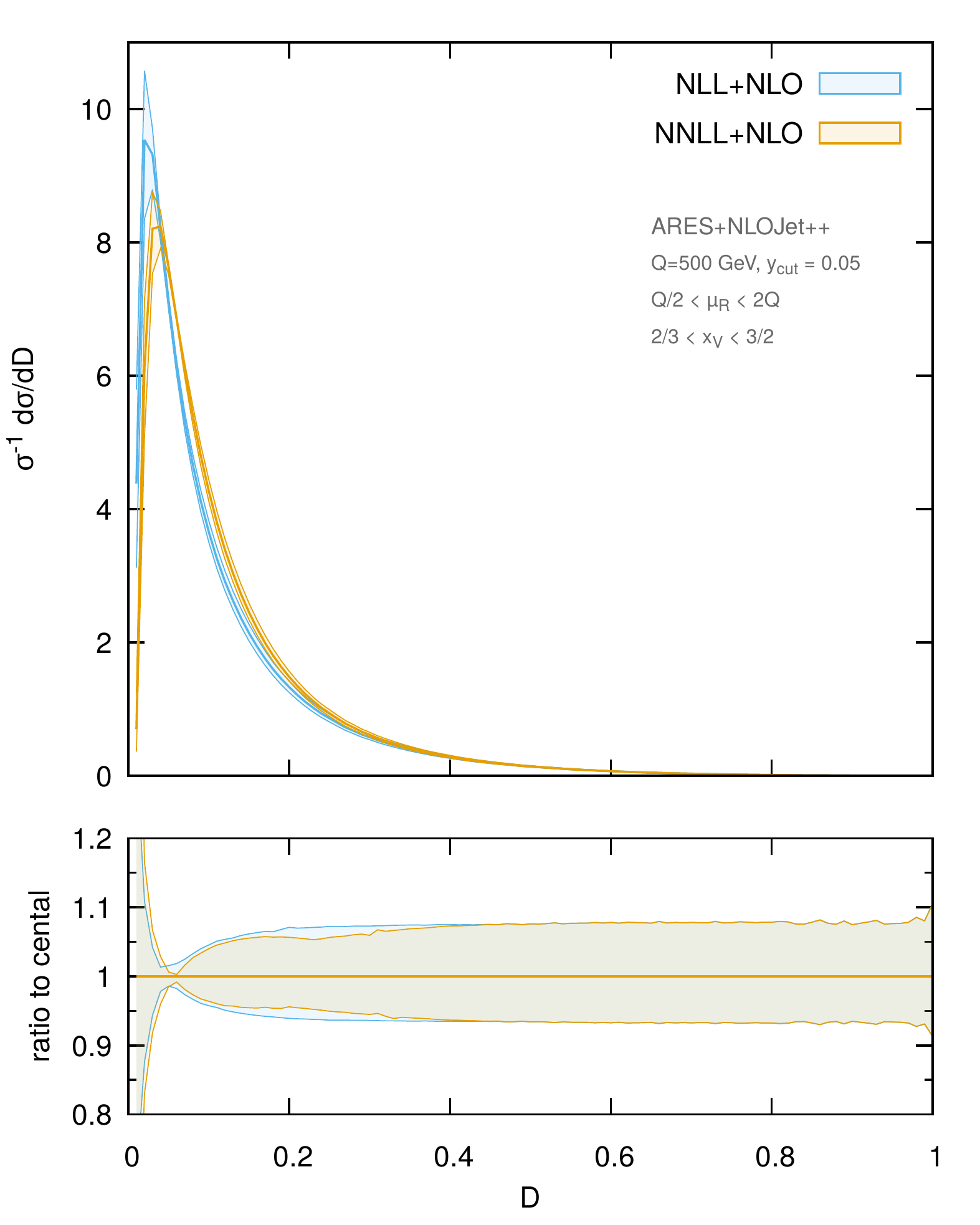}
    \end{center}
  \end{minipage}
  \begin{minipage}[r]{0.5\linewidth}
    \begin{center}
      \includegraphics[width=\textwidth]{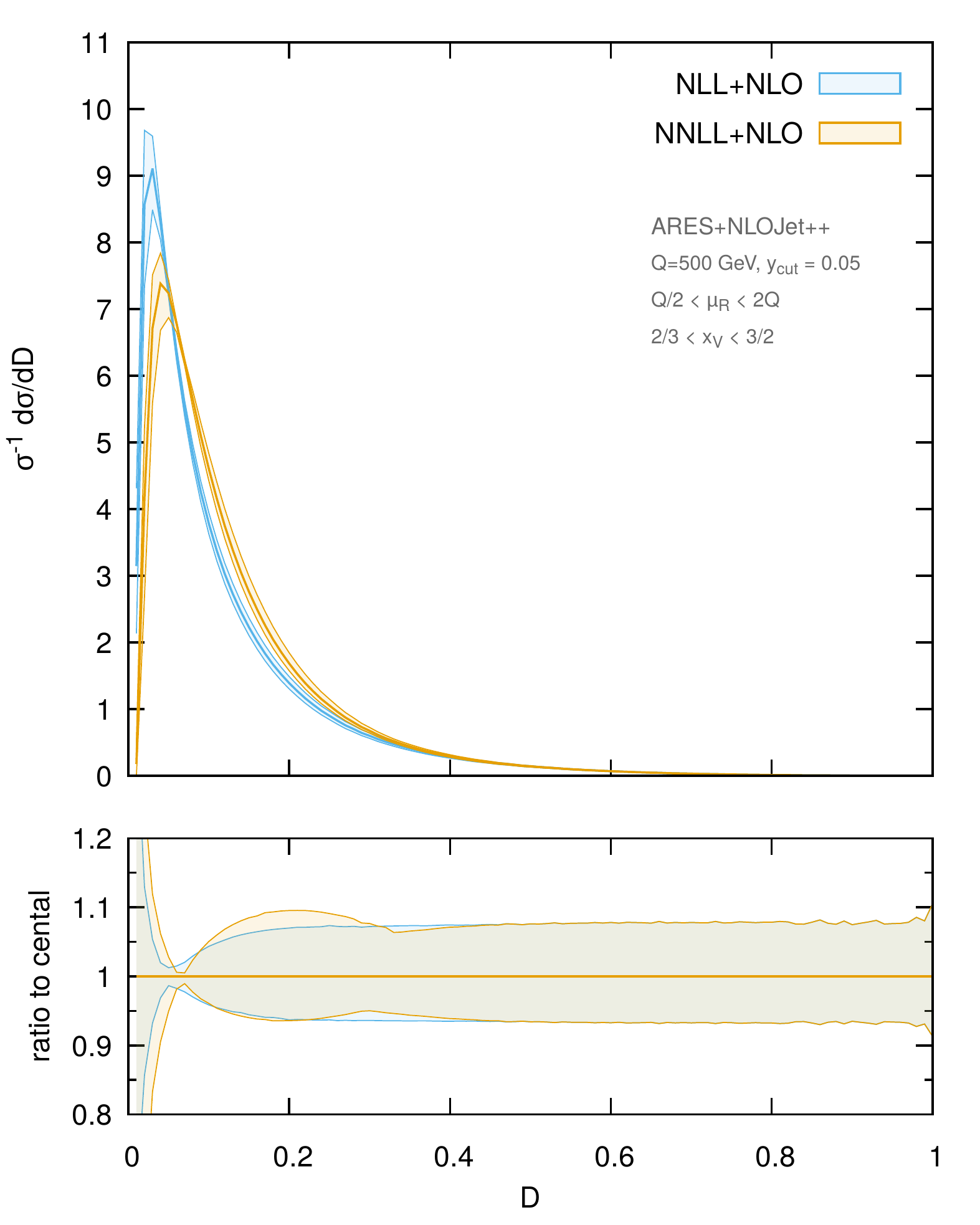}
    \end{center}
  \end{minipage}
  \ \caption{The matched distribution for $y_{\rm cut}=0.05$ and $Q = 500 \rm GeV$. The left plot using the
    $X_{\rm const}$ scheme and the right using the $X_{\rm prod}$ scheme.}
  \label{fig:dpar-y0050-Q500-PT}
\end{figure}

The situation becomes better at FCC-ee energies, as we see clearly in
figs.~\ref{fig:dpar-y0100-Q500-PT} and
\ref{fig:dpar-y0050-Q500-PT}. Noticeably the position of the peak
tends towards smaller values of $D$, and we start approaching the
strict resummation regime $D \ll y_{\rm cut} \ll 1$. Simultaneously we
see a reduction in the uncertainty by almost $50 \%$. To conclude, for
this observable, and depending on the value of $y_{\rm cut}$, we
expect large subleading corrections that are not under control in any
resummation formalism. This calls for a joint resummation of both
types of logarithms, the observable and $y_{\rm cut}$, along the line
of the presented resummation of both $p_{\rm t,H}$ and the transverse
momentum of the leading jet~\cite{Monni:2019yyr}.

Leaving these caveats aside, we note that NNLL corrections generically
yield harder $D$-parameter distributions. The effect is larger using
the $X_{\rm prod}$ scheme, because the resummed logarithms in the
latter scheme are typically larger than the $X_{\rm const}$
scheme. Indeed, this is one of the reasons why the NNLL uncertainty
bands get larger when we use the $X_{\rm prod}$ scheme, while their
counterparts at NLL remain virtually the same. Note that in the
$X_{\rm prod}$ scheme the resummation scale is effectively of the
order $\sqrt{y_{\rm cut}} Q$ which is the appropriate upper bound for
transverse momenta. Therefore, this scheme automatically captures some
of the terms which are enhanced by logarithms of $y_{\rm cut}$. 

Last, we compare our predictions to existing LEP1
data~\cite{alephqcd}. In order to do so, we need to supplement our
perturbative resummation with some estimate of non-perturbative
hadronisation corrections. Before we do this, we need to choose
whether to use $X_{\rm const}$ or $X_{\rm prod}$ as our default choice
for the resummation scale. We have observed that NNLL distributions
obtained with $X_{\rm prod}$ are not very stable with respect to the
choice of the matching parameter $v_0$, which points to the fact that
such a choice brings in numerically large subleading corrections,
which we cannot control within our framework. Therefore, we decide to
present non-perturbative plots using $X_{\rm const}$ as our
resummation scale, and $v_0=1/2$. We have checked that using other
values of $v_0$ does not change considerably our findings.  We include
hadronisation corrections in the dispersive approach of
ref.~\cite{Dokshitzer:1995qm}, where leading hadronisation corrections
result in a shift of the corresponding perturbative distributions. In
our case, we use the non-perturbative shift computed in
ref.~\cite{Banfi:2001pb}, and define
\begin{equation}
  \label{eq:SigmaH-NP}
  \Sigma_{\mathcal{H}}^{\rm NP}(D) = \frac{1}{\sigma_{\mathcal{H}}}
  \int d\Phi_3 \frac{d\sigma_3}{d\Phi_3} \Sigma_{\mathcal{B}}\left( \{p_1, p_2, p_3\},
  D - Z_{\rm NP}\, \delta D \left( \{p_1, p_2, p_3\} \right) \right)\, \mathcal{H}(p_1,p_2,p_3)\, ,
\end{equation}
where
\begin{equation}
  \label{eq:delta-D}
  \delta D(\{p_1, p_2, p_3\}) = \frac{a_{\rm NP}}{Q} 27 \lambda_1 \lambda_2 \sum_{(ij)} C_{(ij)}\,
  g_{ij} \left( \theta_{ij} \right)\,.
\end{equation}
In the above equation, the geometry dependent functions $g_{ij}$ are
the ones of ref.~\cite{Banfi:2001pb}, which we rewrite using our own notation and conventions as follows:
\begin{equation}
  \label{eq:gij}
  g_{ij}(\theta_{ij}) = \sin \frac{\theta_{ij}}{2} \int_0^{2\pi} \frac{d\phi}{2\pi}
  \int_{-\infty}^{\infty} d\eta \frac{\sin^2 \phi}{\cosh \eta + \cos (\theta_{ij}/2) \cos \phi}\,.
\end{equation}
The non-perturbative parameter $a_{\rm NP}$ is given by
\begin{equation}
  \label{eq:aNP}
  a_{\rm NP} = \frac{4\mu_I}{\pi^2}\mathcal{M}\left(\alpha_0(\mu_I)-\alpha_s(Q)-2\beta_0 \alpha^2_s(Q)\left(\ln\frac{Q}{\mu_I}+\frac{K^{(1)}}{4\pi\beta_0}+1\right)\right)\,,
\end{equation}
where
\begin{equation}
  \label{eq:alpha0}
  \alpha_0(\mu_I) = \int_0^{\mu_I}\frac{dk}{\mu_I} \alpha_s(k)\,,
\end{equation}
and $\alpha_s(k)$ is the dispersive coupling defined in
ref.~\cite{Dokshitzer:1995qm}. As in previous non-perturbative
studies, we set $\mu_I=2\,$GeV. In eq.~\eqref{eq:SigmaH-NP}, we have also introduced the factor
\begin{equation}
  \label{eq:Z-NP}
  Z_{\rm NP} = 1 - \left( \frac{D}{D_{\rm max}} \right)^q\,,
\end{equation}
that ensures that the shift vanishes at the endpoint of the
distribution. Also, to ensure that the distribution vanishes at its
endpoint, we replace $\tilde L$ defined in eq.~\eqref{eq:Ltilde} with
\begin{equation}
  \label{eq:Ltilde-NP}
  \tilde{L}_{\rm NP} \equiv \frac{1}{p} \ln \left( \left( \frac{x_D}{D - \delta D} \right)^p
    - \left( \frac{x_D}{D_{\rm max} - \delta D} \right)^p + 1 \right)\,.
\end{equation}
Specifically, we have set $q=2$ and $p=1$.  Last, in order to produce
matched non-perturbative distributions, we compute
$\delta D_{\mathcal{H}}$ defined by
\begin{equation}
  \label{eq:deltaDH}
  \Sigma_{\cal H}(D-\delta D_{\mathcal{H}})=\Sigma_{\mathcal{H}}^{\rm NP}(D)\,,
\end{equation}
and define our matched non-perturbative distribution as $\Sigma^{\rm Mat.}_{\cal H}(D-\delta D_{\mathcal{H}})$.
\begin{figure}[htbp]
  \begin{minipage}[l]{0.5\linewidth}
    \begin{center}
      \includegraphics[width=\textwidth]{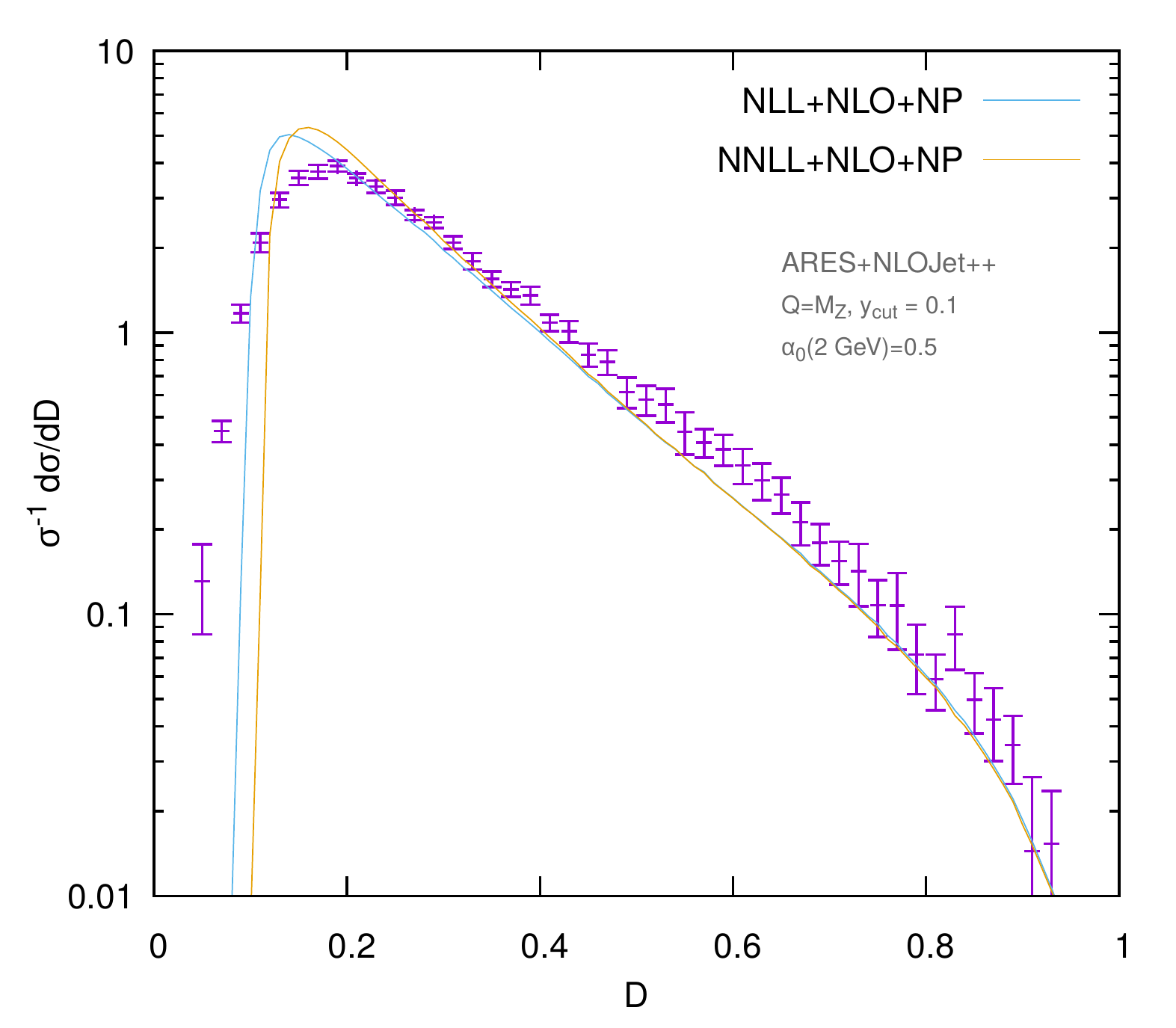}
    \end{center}
  \end{minipage}
  \begin{minipage}[r]{0.5\linewidth}
    \begin{center}
      \includegraphics[width=\textwidth]{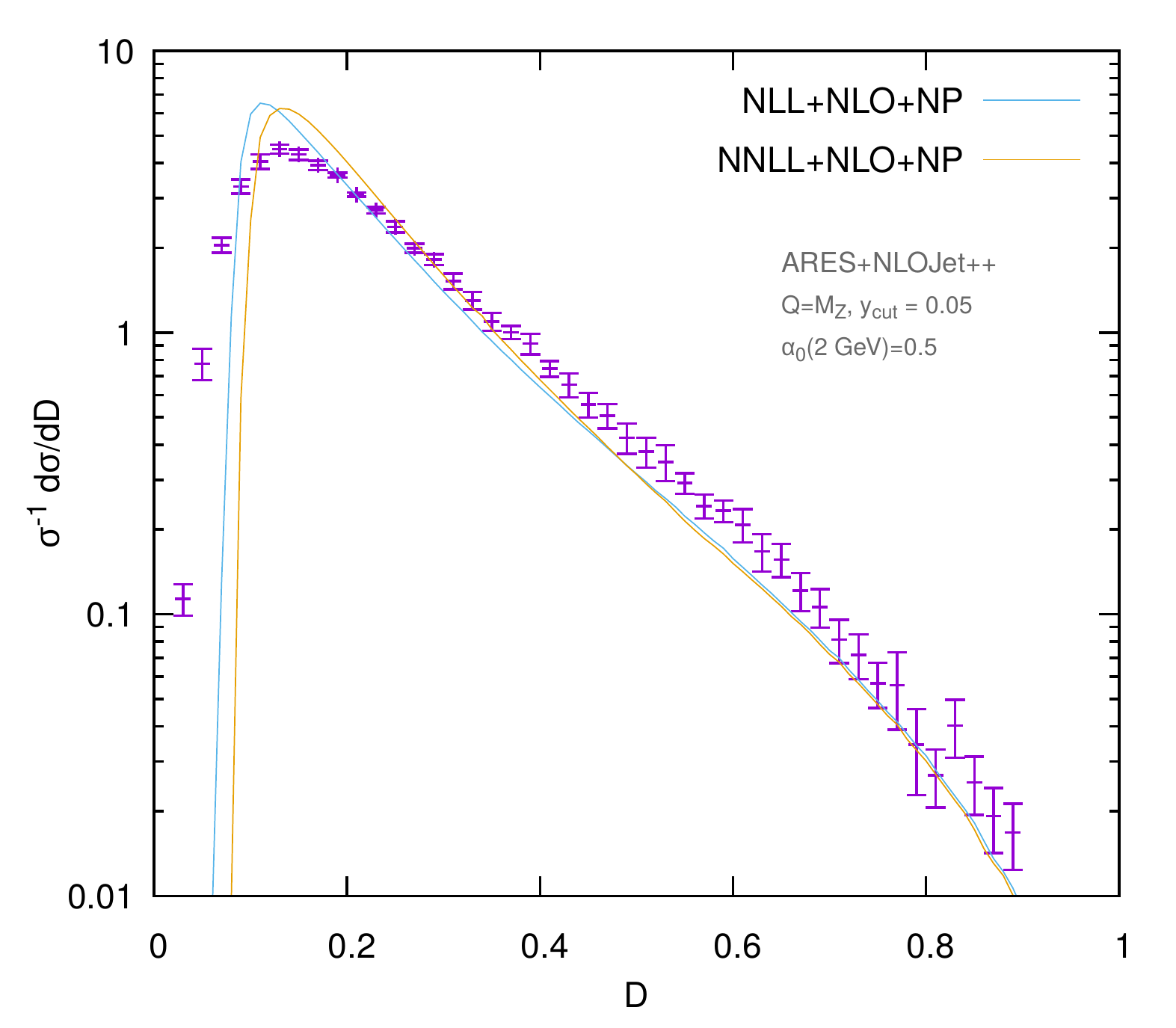}
    \end{center}
  \end{minipage}
  \caption{The matched distribution, including the non-perturbative corrections, is compared to data from LEP1 for the two values of $y_{\rm cut}$ we adopt in this article.}
  \label{fig:dpar-MZ-NP}
\end{figure}
In Fig.~\ref{fig:dpar-MZ-NP} we produce plots for non-perturbative
matched distributions, with central scales, corresponding to NLL and
NNLL accuracy. The non-perturbative shift corresponds to a value of
$\alpha_0(2\,\mathrm{GeV})$ that give us a good agreement with data,
and is within the range favoured by existing fits to event-shape
data~\cite{Salam:2001bd}. We see that NNLL resummation has a shape
that resembles data more closely than NLL resummation, whilst
favouring a similar value of $\alpha_0$, namely $\alpha_0=0.5$. This
trend persists irrespective of the value of $y_{\rm cut}$. This value
of $\alpha_0$ is similar to the central value of a fit obtained with
the NNLL thrust distribution~\cite{Gehrmann:2012sc}.  Finally,
increasing the value of $v_0$ up to $D_{\max}$ does not change the
distributions close to the peak, but gives a better agreement with
data in the tails.

\section{Conclusions}
\label{sec:conclusions}

This article presents a general method to compute the NNLL resummation
of rIRC safe observables for processes characterised by the presence
of three hard emitters. The method is a generalisation of the ARES
approach to NNLL resummations, and paves the way to a general NNLL
resummation with an arbitrary number of hard emitters. Although we concentrate
on three-jet events in $e^+e^-$ annihilation, our treatment
of NNLL contributions induced by final-state radiation is completely
general.

Similar to the two-jet case, we are able to combine unresolved real
radiation and virtual corrections to the Born process into an
analytically computable NNLL radiator. The remaining corrections are
all induced by real radiation, and can be computed for a general
observable using suitable Monte-Carlo procedures. 

Two new functions appear in the three-jet case. First, a new NNLL
correction of soft wide-angle origin appears, due to the fact that now
we have three-hard emitters with non-trivial colour
correlations. Second, since we have a hard gluon initiating a
three-jet event, we need to take into account non-trivial spin
correlations in hard-collinear splittings. These are embedded in a new
NNLL correction that adds to those of hard-collinear origin.

As an example, we have applied our method to the $D$-parameter. Since
this is an additive observable, we are able to compute most NNLL
functions analytically, with a couple of integrals to be computed
numerically. Then, we have performed phenomenological studies
by matching our resummation to exact fixed-order and presenting
predictions for LEP1 and future colliders. Both validation of
resummation and phenomenology is tricky for three-jet
observables. First, while it is possible to easily check NLL
contributions against exact fixed-order, it is impossible to check
NNLL ones without resorting to quadruple precision. With the
aid of a fake two-jet observable that resembles the $D$-parameter, we
have been able to check some NNLL contributions using the NLO code
EVENT2. For what concerns the actual phenomenology, current cuts to
select three-jet events give rise to large subleading effects at LEP1
energies. The situation is a bit better at FCC-ee. Nevertheless, we
envisage that, to improve phenomenological studies of the
$D$-parameter, one should attempt a joint resummation of logarithms of
the $D$-parameter and of the variable determining the three-jet
selection, with a similar procedure to that for angularites, or for
the transverse momentum of a colour singlet and an accompanying
leading jet. 

A comparison with LEP1 data requires the inclusion of non-perturbative
hadronisation corrections. We have added to our NNLL resummation the
leading hadronisation corrections evaluated in the dispersive
model. In general, the NNLL distribution has a shape that is similar
to data, so hadronisation corrections are compatible with a rigid
shift of perturbative distributions. We find that the value of the
non-perturbative parameter $\alpha_0$ determining the size of
hadronisation corrections is comparable to the one obtained with NLL
distributions. It might be very interesting at this stage to perform a
comprehensive simultaneous fit of $\alpha_s$ and $\alpha_0$ using NNLL
resummations for different event shapes.

In conclusion, our study sets the main building blocks for a general NNLL
resummation of rIRC safe final-state observables with an arbitrary number of
hard emitting legs. The only missing ingredient is a general treatment of both
initial-state radiation and soft wide-angle corrections for a system with more
than three hard emitting legs. Despite the technical difficulties, the
philosophy of our method stays unchanged. In particular, ARES does not depend on
the specific factorisation properties of an observable, and gives promise to
achieve a fully general solution to the problem of NNLL resummation in the near
future.

\section*{Acknowledgements}

The work of A.B. and B.K.E. is supported by the Science Technology and
Facilities Council (STFC) under grant number ST/P000819/1. During the final stages of this project, B.K.E has also been supported by the European Research Council (ERC) under the European Union's Horizon 2020 research and innovation programme (grant agreement No. 788223, PanScales).

\appendix

\section{Correlated two-parton emission}
\label{sec:correl-kinematics}
The double-emission function $\mathcal{A}^2$ in eq.~(\ref{eq:doublesoft}) reads
\begin{align}
\mathcal{A}^2 = C_A (2\mathcal{S} + \mathcal{H}_g) + n_f \mathcal{H}_q \ \ ,
\end{align}
where
\begin{subequations}
\begin{align}
  \label{eq:2S}
  2\mathcal{S} & = \frac{1}{z(1-z)}\left[\frac{1-(1-z)\mu^2/z}{u_a^2}+\frac{1-z\mu^2/(1-z)}{u_b^2}\right] \\
  \label{eq:Hg}
  \mathcal{H}_g & =- 4 + \frac{z(1-z)}{1+\mu^2}\left(2\cos  \phi+\frac{(1-2z)\mu}{\sqrt{z(1-z)}}\right)^2\nonumber \\ & +\frac{1}{2(1-z)}\left[1-\frac{1-(1-z)\mu^2/z}{u_a^2}\right] 
+\frac{1}{2z}\left[1-\frac{1-z\mu^2/(1-z)}{u_b^2}\right]
\\
  \label{eq:Hq}
  \mathcal{H}_q& =1-\frac{z(1-z)}{1+\mu^2}\left(2\cos  \phi+\frac{(1-2z)\mu}{\sqrt{z(1-z)}}\right)^2\,.
\end{align}
 \end{subequations}
 In the above equations, we also defined the following quantities
 \begin{equation}
   \label{eq:xqi}
   u_a^2 = 1+2\sqrt{\frac{1-z}{z}}\mu\cos \phi + \frac{1-z}{z}\mu^2\,,\qquad 
   u_b^2 = 1-2\sqrt{\frac{z}{1-z}}\mu\cos \phi + \frac{z}{1-z}\mu^2\,.
 \end{equation}
 Note, in particular, that the quark function $\mathcal{H}_q$ is defined with a factor of 2 to compensate for the symmetry $1/2!$ in the phase space. Now it should be useful to demonstrate explicitly the variables transformations we implemented in the matrix element. Indeed, the form of eq.~(\ref{eq:doublesoft}) does not depend on the specific dipole to which the correlated soft pair belongs to.
Nevertheless, we introduce the Sudakov decomposition of each momentum in a certain dipole
 \begin{align}
 k_a^{(ij)} &= z_a^{(i)} \,p_i + z_a^{(j)} \,p_j + \kappa_a^{(ij)} \cos\phi_a^{(ij)} \,n^{(ij)}_{\rm in} + \kappa^{(ij)} \sin\phi_a^{(ij)} \, n^{(ij)}_{\rm out} \ \ , \\
  k_b^{(ij)} &= z_b^{(i)} \,p_i + z_b^{(j)} \,p_j + \kappa_b^{(ij)} \cos\phi_b^{(ij)} \,n^{(ij)}_{\rm in} + \kappa_b^{(ij)} \sin\phi_b^{(ij)} \, n^{(ij)}_{\rm out}  \ \ .
 \end{align}
 Our task is to express the emission's Sudakov variables in terms of
 the Sudakov variables of the pseudo-parent momentum, defined as
 $k = k_a + k_b$. Now in the Euclidean two-dimensional plane spanned
 by the pair $(\vec{n}^{(ij)}_{\rm in},\vec{n}^{(ij)}_{\rm out})$, we
 define two vectors
\begin{align}
\vec{k}^{(ij)}_{a} \equiv \kappa_a^{(ij)} \cos\phi_a^{(ij)} \,\vec{n}^{(ij)}_{\rm in} + \kappa^{(ij)} \sin\phi_a^{(ij)} \, \vec{n}^{(ij)}_{\rm out}, \quad \vec{k}^{(ij)}_{b} \equiv \kappa_b^{(ij)} \cos\phi_b^{(ij)} \,\vec{n}^{(ij)}_{\rm in} + \kappa_b^{(ij)} \sin\phi_b^{(ij)} \, \vec{n}^{(ij)}_{\rm out} \ \ ,
\end{align}
which play the role of transverse momenta and allows us to directly utilise the results of ref.~\cite{Banfi:2018mcq}. We henceforth list the variables appearing in eq.~(\ref{eq:doublesoft})
\begin{align}
 \frac{z}{1-z}  = \frac{z_a^{(i)}}{z_b^{(i)}}, \quad \vec{q}^{~(ij)} = z \vec{k}^{(ij)}_a + (1-z) \vec{k}^{(ij)}_b, \quad \mu^2 = \frac{(k_a + k_b)^2}{(\vec{k}^{(ij)}_{a} + \vec{k}^{(ij)}_{b})^2}, \quad \cos \phi = \frac{\vec{q}^{~(ij)} \cdot \vec{k}^{(ij)}}{(\vec{k}^{(ij)})^2(\vec{q}^{~(ij)})^2} \ \ .
\end{align}

\section{Three-parton kinematics}
\label{sec:kinematics}

We consider three momenta $p_1,p_2,p_3$, with
$p_1+p_2+p_3=q=(Q,0,0,0)$. Using a flavour-based labelling, $p_1$ is a
quark, $p_2$ an antiquark and $p_3$ a gluon. We define the
dimensionless variables $x_i=2(p_iq)/Q^2<1$, satisfying 
$x_1+x_2+x_3=2$. In terms of these variables,
\begin{equation}
  \label{eq:E-masses}
  E_i=x_i \frac{Q}{2}\,,\qquad 2(p_ip_j)=(x_i+x_j-1)Q^2\,.
\end{equation}
This makes it possible to write the angles between pairs of momenta in terms of the $x_i$'s as follows
\begin{equation}
  \label{eq:angles}
  \sin^2\frac{\theta_{ij}}{2} = \frac{x_i+x_j-1}{x_i\,x_j}\,.
\end{equation}
The three-parton cross section, differential in $x_1$ and $x_2$, in four dimensions reads
\begin{equation}
  \label{eq:3jet-dx12}
  \frac{d\sigma}{dx_1 dx_2} = \sigma_0 C_F\frac{\alpha_s}{2\pi}\frac{x_1^2+x_2^2}{(1\!-\!x_1)(1\!-\!x_2)}
\end{equation}
with $\sigma_0$ the Born cross section for producing a quark-antiquark
pair in $e^+e^-$ annihilation. 

\begin{figure}[h]
  \centering
  \includegraphics[width=.5\textwidth]{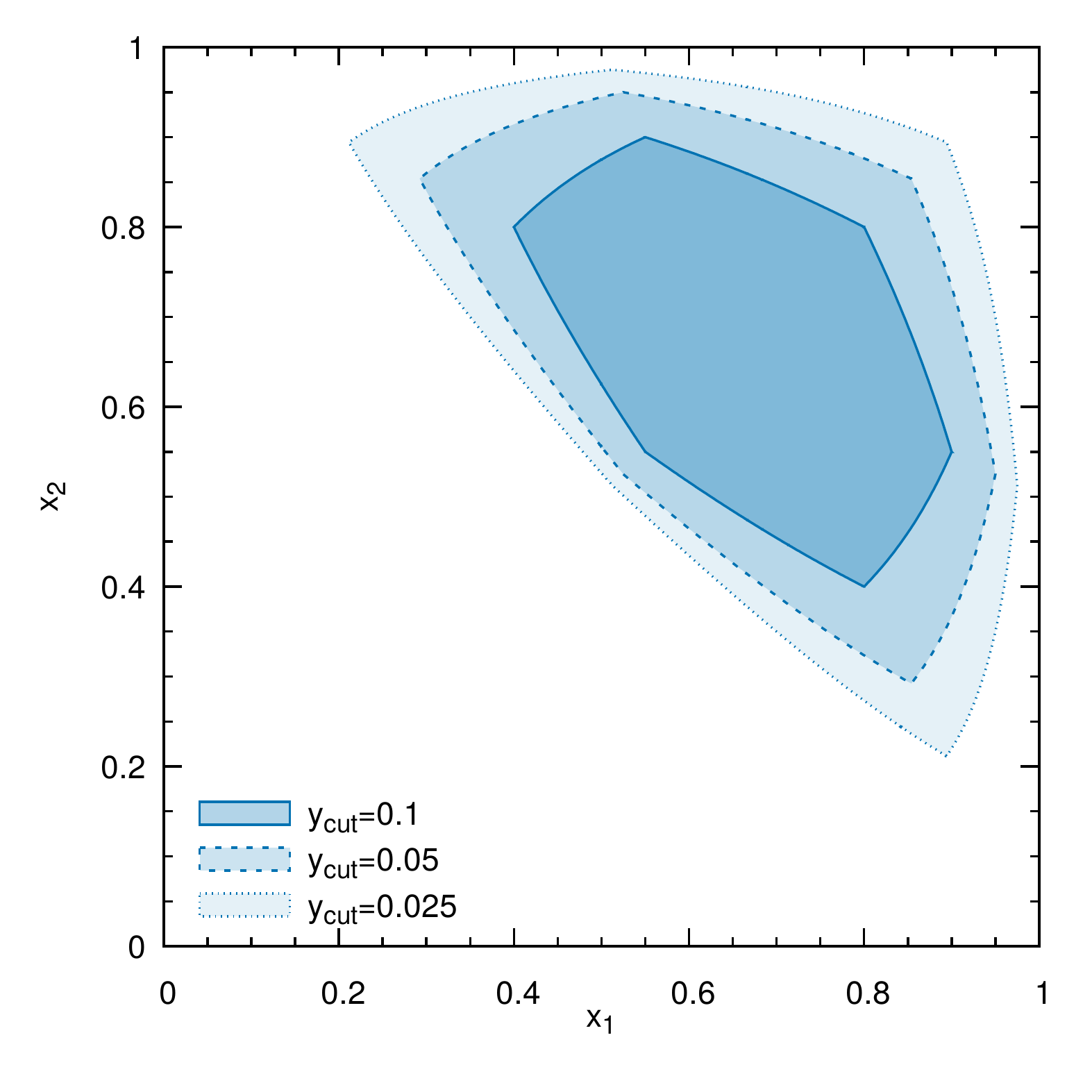}

\vspace{-.5cm}

  \caption{The Durham algorithm three-jet region for three different
    values of $\ycut$.}
  \label{fig:duralg-3j-region}
\end{figure}

To obtain the Born three-jet cross section $\sigma_0(\ycut)$ with the
Durham algorithm~\cite{Catani:1991hj} we need to integrate the
differential cross section in eq.~\eqref{eq:3jet-dx12} with the
constraint $y_3(p_1,p_2,p_3)=\min\{y_{12},y_{13},y_{23}\}>\ycut$,
where $y_{ij}$ is the ``distance'' between pairs of partons defined by
\begin{equation}
  \label{eq:yij}
  y_{ij}\equiv 2\frac{\min\{E_i^2,E_j^2\}}{Q^2} (1-\cos\theta_{ij})=\min\left\{\frac{x_i}{x_j},\frac{x_j}{x_i}\right\}(x_i+x_j-1)\,.
\end{equation}
The Durham algorithm defines a six-sided region in the $(x_1, x_2)$
plane, as shown in Fig.~\ref{fig:duralg-3j-region} for the three different values of $\ycut$ we consider here.
The corresponding Born cross section
$\sigma^{(0)}_{\cal H}$ is
\begin{equation}
  \label{eq:sigma0-y3}
  \begin{split}
  \sigma^{(0)}_{\cal H}&=\sigma_0 C_F \frac{\alpha_s}{2\pi}\int_0^1 dx_1\,\int_0^1 dx_2 \frac{x_1^2+x_2^2}{(1\!-\!x_1)(1\!-\!x_2)}\,\Theta(x_1+x_2\!-\!1)\,\Theta\left(\min\{y_{12},y_{13},y_{23}\}\!-\!\ycut\right)\,.
  \end{split}
\end{equation}
This cross section can be computed analytically. Its expression, not
particularly illuminating, can be found in~\cite{Brown:1991hx}.

\section{Full matching formulae}
\label{app:matching}

In our matching formulae $\Sigma^{\rm Mat.}_{\mathcal{H}}(v)$ we normalise all
of the distributions to the total cross section $\sigma_{\mathcal{H}}$. However
this is not what is provided by NLOjet++, instead it provides the un-normalised
differential distribution for the $D$-parameter. We can transform the output of
NLOjet++ into our conventions as follows. First we compute the un-normalised,
\emph{barred}, total cross section
\begin{equation}
  \label{eq:nlojet-bar}
  \bar{\Sigma}^{(i)}_{\rm NLOJet} = - \int_v^{v_{\rm max}} dv^\prime \frac{d\Sigma^{(i)}_{\rm NLOJet}(v^\prime)}{dv^\prime} \, ,
\end{equation}
where $i$ refers to the power of $\as$ in perturbation theory. To transform this
result into our conventions we perform the following manipulations
\begin{equation}
  \begin{split}
    \bar{\Sigma}^{(1)}_{\rm FO.}(v) &= \frac{\bar{\Sigma}^{(1)}_{\rm NLOJet}(v)}{\sigma^{(0)}_{\mathcal{H}}} \, , \\
    \bar{\Sigma}^{(2)}_{\rm FO.}(v) &= \frac{\bar{\Sigma}^{(2)}_{\rm NLOJet}(v)}{\sigma^{(0)}_{\mathcal{H}}}
    - \frac{\sigma^{(1)}_{\mathcal{H}}}{\sigma^{(0)}_{\mathcal{H}}} \bar{\Sigma}^{(1)}_{\rm FO.}(v) \, .
  \end{split}
\end{equation}
In terms of the barred variables we have
\begin{equation}
  \begin{split}
    \Sigma_{\rm FO.}(v) &= \sum_{i=0}^2 \Sigma^{(i)}_{\rm FO.}(v) \\
    &= 1 + \sum_{i=1}^2 \bar{\Sigma}^{(i)}_{\rm FO.}(v) \, ,
  \end{split}
\end{equation}
and analogously for the expansion of the resummation
\begin{equation}
  \Sigma_{\rm Exp.}(v) = \sum_{i=0}^2 \Sigma^{(i)}_{\rm Exp.}(v) \, .
\end{equation}
Finally we can present the explicit form of our matched distribution in
eq.~\eqref{eq:mastermatch}
\begin{equation}
  \label{eq:mastermatch-explicit}
  \begin{split}
    \Sigma_{\rm Mat.}(v) &= \left( \Sigma_{\rm Res.}(v) \right)^Z
    \left[ 1 + \bar{\Sigma}^{(1)}_{\rm FO.}(v) - Z \Sigma^{(1)}_{\rm Exp.}(v) + \right. \\
      & \left. + \bar{\Sigma}^{(2)}_{\rm FO.}(v) - Z \Sigma^{(2)}_{\rm Exp.}(v) - Z
      \Sigma^{(1)}_{\rm Exp.}(v) \left( \Sigma^{(1)}_{\rm FO.}(v) - \frac{Z+1}{2}
        \Sigma^{(1)}_{\rm Exp.}(v) \right) \right] \, .
  \end{split}
\end{equation}

\bibliographystyle{JHEP}
\bibliography{nnll3jets.bib}

\providecommand{\href}[2]{#2}\begingroup\raggedright\begin{thebibliography}{10}

\bibitem{Patrignani:2016xqp}
{\scshape Particle Data Group} collaboration, \emph{{Review of Particle
  Physics}}, \href{https://doi.org/10.1088/1674-1137/40/10/100001}{\emph{Chin.
  Phys.} {\bfseries C40} (2016) 100001}.

\bibitem{Dasgupta:2001sh}
M.~Dasgupta and G.~P. Salam, \emph{{Resummation of nonglobal QCD observables}},
  \href{https://doi.org/10.1016/S0370-2693(01)00725-0}{\emph{Phys. Lett.}
  {\bfseries B512} (2001) 323}
  [\href{https://arxiv.org/abs/hep-ph/0104277}{{\ttfamily hep-ph/0104277}}].

\bibitem{Dasgupta:2002bw}
M.~Dasgupta and G.~P. Salam, \emph{{Accounting for coherence in interjet E(t)
  flow: A Case study}},
  \href{https://doi.org/10.1088/1126-6708/2002/03/017}{\emph{JHEP} {\bfseries
  03} (2002) 017} [\href{https://arxiv.org/abs/hep-ph/0203009}{{\ttfamily
  hep-ph/0203009}}].

\bibitem{Banfi:2002hw}
A.~Banfi, G.~Marchesini and G.~Smye, \emph{{Away from jet energy flow}},
  \href{https://doi.org/10.1088/1126-6708/2002/08/006}{\emph{JHEP} {\bfseries
  08} (2002) 006} [\href{https://arxiv.org/abs/hep-ph/0206076}{{\ttfamily
  hep-ph/0206076}}].

\bibitem{Ellis:1980wv}
R.~K. Ellis, D.~A. Ross and A.~E. Terrano, \emph{{The Perturbative Calculation
  of Jet Structure in e+ e- Annihilation}},
  \href{https://doi.org/10.1016/0550-3213(81)90165-6}{\emph{Nucl. Phys.}
  {\bfseries B178} (1981) 421}.

\bibitem{GehrmannDeRidder:2007hr}
A.~Gehrmann-De~Ridder, T.~Gehrmann, E.~W.~N. Glover and G.~Heinrich,
  \emph{{NNLO corrections to event shapes in e+ e- annihilation}},
  \href{https://doi.org/10.1088/1126-6708/2007/12/094}{\emph{JHEP} {\bfseries
  12} (2007) 094} [\href{https://arxiv.org/abs/0711.4711}{{\ttfamily
  0711.4711}}].

\bibitem{GehrmannDeRidder:2008ug}
A.~Gehrmann-De~Ridder, T.~Gehrmann, E.~W.~N. Glover and G.~Heinrich, \emph{{Jet
  rates in electron-positron annihilation at O(alpha(s)**3) in QCD}},
  \href{https://doi.org/10.1103/PhysRevLett.100.172001}{\emph{Phys. Rev. Lett.}
  {\bfseries 100} (2008) 172001}
  [\href{https://arxiv.org/abs/0802.0813}{{\ttfamily 0802.0813}}].

\bibitem{Weinzierl:2008iv}
S.~Weinzierl, \emph{{NNLO corrections to 3-jet observables in electron-positron
  annihilation}},
  \href{https://doi.org/10.1103/PhysRevLett.101.162001}{\emph{Phys. Rev. Lett.}
  {\bfseries 101} (2008) 162001}
  [\href{https://arxiv.org/abs/0807.3241}{{\ttfamily 0807.3241}}].

\bibitem{Weinzierl:2009ms}
S.~Weinzierl, \emph{{Event shapes and jet rates in electron-positron
  annihilation at NNLO}},
  \href{https://doi.org/10.1088/1126-6708/2009/06/041}{\emph{JHEP} {\bfseries
  06} (2009) 041} [\href{https://arxiv.org/abs/0904.1077}{{\ttfamily
  0904.1077}}].

\bibitem{Collins:1984kg}
J.~C. Collins, D.~E. Soper and G.~F. Sterman, \emph{{Transverse Momentum
  Distribution in Drell-Yan Pair and W and Z Boson Production}},
  \href{https://doi.org/10.1016/0550-3213(85)90479-1}{\emph{Nucl. Phys.}
  {\bfseries B250} (1985) 199}.

\bibitem{Catani:1991bd}
S.~Catani, G.~Turnock and B.~R. Webber, \emph{{Heavy jet mass distribution in
  e+ e- annihilation}},
  \href{https://doi.org/10.1016/0370-2693(91)91845-M}{\emph{Phys. Lett.}
  {\bfseries B272} (1991) 368}.

\bibitem{Catani:1991kz}
S.~Catani, G.~Turnock, B.~R. Webber and L.~Trentadue, \emph{{Thrust
  distribution in e+ e- annihilation}},
  \href{https://doi.org/10.1016/0370-2693(91)90494-B}{\emph{Phys. Lett.}
  {\bfseries B263} (1991) 491}.

\bibitem{Catani:1992ua}
S.~Catani, L.~Trentadue, G.~Turnock and B.~R. Webber, \emph{{Resummation of
  large logarithms in e+ e- event shape distributions}},
  \href{https://doi.org/10.1016/0550-3213(93)90271-P}{\emph{Nucl. Phys.}
  {\bfseries B407} (1993) 3}.

\bibitem{Dokshitzer:1998kz}
Y.~L. Dokshitzer, A.~Lucenti, G.~Marchesini and G.~P. Salam, \emph{{On the QCD
  analysis of jet broadening}},
  \href{https://doi.org/10.1088/1126-6708/1998/01/011}{\emph{JHEP} {\bfseries
  01} (1998) 011} [\href{https://arxiv.org/abs/hep-ph/9801324}{{\ttfamily
  hep-ph/9801324}}].

\bibitem{Bonciani:2003nt}
R.~Bonciani, S.~Catani, M.~L. Mangano and P.~Nason, \emph{{Sudakov resummation
  of multiparton QCD cross-sections}},
  \href{https://doi.org/10.1016/j.physletb.2003.09.068}{\emph{Phys. Lett.}
  {\bfseries B575} (2003) 268}
  [\href{https://arxiv.org/abs/hep-ph/0307035}{{\ttfamily hep-ph/0307035}}].

\bibitem{Banfi:2001bz}
A.~Banfi, G.~P. Salam and G.~Zanderighi, \emph{{Semi-numerical resummation of
  event shapes}},
  \href{https://doi.org/10.1088/1126-6708/2002/01/018}{\emph{JHEP} {\bfseries
  01} (2002) 018} [\href{https://arxiv.org/abs/hep-ph/0112156}{{\ttfamily
  hep-ph/0112156}}].

\bibitem{Banfi:2003je}
A.~Banfi, G.~P. Salam and G.~Zanderighi, \emph{{Generalized resummation of QCD
  final state observables}},
  \href{https://doi.org/10.1016/j.physletb.2004.01.048}{\emph{Phys. Lett.}
  {\bfseries B584} (2004) 298}
  [\href{https://arxiv.org/abs/hep-ph/0304148}{{\ttfamily hep-ph/0304148}}].

\bibitem{Banfi:2004yd}
A.~Banfi, G.~P. Salam and G.~Zanderighi, \emph{{Principles of general
  final-state resummation and automated implementation}},
  \href{https://doi.org/10.1088/1126-6708/2005/03/073}{\emph{JHEP} {\bfseries
  03} (2005) 073} [\href{https://arxiv.org/abs/hep-ph/0407286}{{\ttfamily
  hep-ph/0407286}}].

\bibitem{Banfi:2004nk}
A.~Banfi, G.~P. Salam and G.~Zanderighi, \emph{Resummed event shapes at hadron
  - hadron colliders},
  \href{https://doi.org/10.1088/1126-6708/2004/08/062}{\emph{JHEP} {\bfseries
  08} (2004) 062} [\href{https://arxiv.org/abs/hep-ph/0407287}{{\ttfamily
  hep-ph/0407287}}].

\bibitem{Banfi:2010xy}
A.~Banfi, G.~P. Salam and G.~Zanderighi, \emph{{Phenomenology of event shapes
  at hadron colliders}},
  \href{https://doi.org/10.1007/JHEP06(2010)038}{\emph{JHEP} {\bfseries 06}
  (2010) 038} [\href{https://arxiv.org/abs/1001.4082}{{\ttfamily 1001.4082}}].

\bibitem{Dasgupta:2018nvj}
M.~Dasgupta, F.~A. Dreyer, K.~Hamilton, P.~F. Monni and G.~P. Salam,
  \emph{{Logarithmic accuracy of parton showers: a fixed-order study}},
  \href{https://doi.org/10.1007/JHEP09(2018)033}{\emph{JHEP} {\bfseries 09}
  (2018) 033} [\href{https://arxiv.org/abs/1805.09327}{{\ttfamily
  1805.09327}}].

\bibitem{Dokshitzer:1998qp}
Y.~L. Dokshitzer, G.~Marchesini and G.~P. Salam, \emph{{Revisiting
  nonperturbative effects in the jet broadenings}},
  \href{https://doi.org/10.1007/s1010599c0003}{\emph{Eur. Phys. J.direct}
  {\bfseries 1} (1999) 3}
  [\href{https://arxiv.org/abs/hep-ph/9812487}{{\ttfamily hep-ph/9812487}}].

\bibitem{Salam:2001bd}
G.~P. Salam and D.~Wicke, \emph{{Hadron masses and power corrections to event
  shapes}}, \href{https://doi.org/10.1088/1126-6708/2001/05/061}{\emph{JHEP}
  {\bfseries 05} (2001) 061}
  [\href{https://arxiv.org/abs/hep-ph/0102343}{{\ttfamily hep-ph/0102343}}].

\bibitem{Gehrmann:2012sc}
T.~Gehrmann, G.~Luisoni and P.~F. Monni, \emph{{Power corrections in the
  dispersive model for a determination of the strong coupling constant from the
  thrust distribution}},
  \href{https://doi.org/10.1140/epjc/s10052-012-2265-x}{\emph{Eur. Phys. J.}
  {\bfseries C73} (2013) 2265}
  [\href{https://arxiv.org/abs/1210.6945}{{\ttfamily 1210.6945}}].

\bibitem{Abbate:2010xh}
R.~Abbate, M.~Fickinger, A.~H. Hoang, V.~Mateu and I.~W. Stewart, \emph{{Thrust
  at N$^3$LL with Power Corrections and a Precision Global Fit for
  $\alpha_s(M_Z)$}},
  \href{https://doi.org/10.1103/PhysRevD.83.074021}{\emph{Phys. Rev.}
  {\bfseries D83} (2011) 074021}
  [\href{https://arxiv.org/abs/1006.3080}{{\ttfamily 1006.3080}}].

\bibitem{Hoang:2014wka}
A.~H. Hoang, D.~W. Kolodrubetz, V.~Mateu and I.~W. Stewart,
  \emph{{$C$-parameter distribution at N$^3$LL? including power corrections}},
  \href{https://doi.org/10.1103/PhysRevD.91.094017}{\emph{Phys. Rev.}
  {\bfseries D91} (2015) 094017}
  [\href{https://arxiv.org/abs/1411.6633}{{\ttfamily 1411.6633}}].

\bibitem{Hoang:2015hka}
A.~H. Hoang, D.~W. Kolodrubetz, V.~Mateu and I.~W. Stewart, \emph{{Precise
  determination of $\alpha_s$ from the $C$-parameter distribution}},
  \href{https://doi.org/10.1103/PhysRevD.91.094018}{\emph{Phys. Rev.}
  {\bfseries D91} (2015) 094018}
  [\href{https://arxiv.org/abs/1501.04111}{{\ttfamily 1501.04111}}].

\bibitem{Becher:2008cf}
T.~Becher and M.~D. Schwartz, \emph{{A precise determination of $\alpha_s$ from
  LEP thrust data using effective field theory}},
  \href{https://doi.org/10.1088/1126-6708/2008/07/034}{\emph{JHEP} {\bfseries
  07} (2008) 034} [\href{https://arxiv.org/abs/0803.0342}{{\ttfamily
  0803.0342}}].

\bibitem{Monni:2011gb}
P.~F. Monni, T.~Gehrmann and G.~Luisoni, \emph{{Two-Loop Soft Corrections and
  Resummation of the Thrust Distribution in the Dijet Region}},
  \href{https://doi.org/10.1007/JHEP08(2011)010}{\emph{JHEP} {\bfseries 08}
  (2011) 010} [\href{https://arxiv.org/abs/1105.4560}{{\ttfamily 1105.4560}}].

\bibitem{Chien:2010kc}
Y.-T. Chien and M.~D. Schwartz, \emph{{Resummation of heavy jet mass and
  comparison to LEP data}},
  \href{https://doi.org/10.1007/JHEP08(2010)058}{\emph{JHEP} {\bfseries 08}
  (2010) 058} [\href{https://arxiv.org/abs/1005.1644}{{\ttfamily 1005.1644}}].

\bibitem{Becher:2012qa}
T.~Becher and M.~Neubert, \emph{{Factorization and NNLL Resummation for Higgs
  Production with a Jet Veto}},
  \href{https://doi.org/10.1007/JHEP07(2012)108}{\emph{JHEP} {\bfseries 07}
  (2012) 108} [\href{https://arxiv.org/abs/1205.3806}{{\ttfamily 1205.3806}}].

\bibitem{deFlorian:2004mp}
D.~de~Florian and M.~Grazzini, \emph{{The Back-to-back region in e+ e-
  energy-energy correlation}},
  \href{https://doi.org/10.1016/j.nuclphysb.2004.10.051}{\emph{Nucl. Phys.}
  {\bfseries B704} (2005) 387}
  [\href{https://arxiv.org/abs/hep-ph/0407241}{{\ttfamily hep-ph/0407241}}].

\bibitem{Tulipant:2017ybb}
Z.~Tulipánt, A.~Kardos and G.~Somogyi, \emph{{Energyenergy correlation in
  electronpositron annihilation at NNLL + NNLO accuracy}},
  \href{https://doi.org/10.1140/epjc/s10052-017-5320-9}{\emph{Eur. Phys. J.}
  {\bfseries C77} (2017) 749}
  [\href{https://arxiv.org/abs/1708.04093}{{\ttfamily 1708.04093}}].

\bibitem{Moult:2018jzp}
I.~Moult and H.~X. Zhu, \emph{{Simplicity from Recoil: The Three-Loop Soft
  Function and Factorization for the Energy-Energy Correlation}},
  \href{https://doi.org/10.1007/JHEP08(2018)160}{\emph{JHEP} {\bfseries 08}
  (2018) 160} [\href{https://arxiv.org/abs/1801.02627}{{\ttfamily
  1801.02627}}].

\bibitem{Frye:2016aiz}
C.~Frye, A.~J. Larkoski, M.~D. Schwartz and K.~Yan, \emph{{Factorization for
  groomed jet substructure beyond the next-to-leading logarithm}},
  \href{https://doi.org/10.1007/JHEP07(2016)064}{\emph{JHEP} {\bfseries 07}
  (2016) 064} [\href{https://arxiv.org/abs/1603.09338}{{\ttfamily
  1603.09338}}].

\bibitem{Procura:2018zpn}
M.~Procura, W.~J. Waalewijn and L.~Zeune, \emph{{Joint resummation of two
  angularities at next-to-next-to-leading logarithmic order}},
  \href{https://doi.org/10.1007/JHEP10(2018)098}{\emph{JHEP} {\bfseries 10}
  (2018) 098} [\href{https://arxiv.org/abs/1806.10622}{{\ttfamily
  1806.10622}}].

\bibitem{Bell:2018gce}
G.~Bell, A.~Hornig, C.~Lee and J.~Talbert, \emph{{$e^+ e^-$ angularity
  distributions at NNLL$^\prime$ accuracy}},
  \href{https://doi.org/10.1007/JHEP01(2019)147}{\emph{JHEP} {\bfseries 01}
  (2019) 147} [\href{https://arxiv.org/abs/1808.07867}{{\ttfamily
  1808.07867}}].

\bibitem{Kang:2013nha}
D.~Kang, C.~Lee and I.~W. Stewart, \emph{{Using 1-Jettiness to Measure 2 Jets
  in DIS 3 Ways}},
  \href{https://doi.org/10.1103/PhysRevD.88.054004}{\emph{Phys. Rev.}
  {\bfseries D88} (2013) 054004}
  [\href{https://arxiv.org/abs/1303.6952}{{\ttfamily 1303.6952}}].

\bibitem{Kang:2013wca}
Z.-B. Kang, X.~Liu, S.~Mantry and J.-W. Qiu, \emph{{Probing nuclear dynamics in
  jet production with a global event shape}},
  \href{https://doi.org/10.1103/PhysRevD.88.074020}{\emph{Phys. Rev.}
  {\bfseries D88} (2013) 074020}
  [\href{https://arxiv.org/abs/1303.3063}{{\ttfamily 1303.3063}}].

\bibitem{Kang:2013lga}
Z.-B. Kang, X.~Liu and S.~Mantry, \emph{{1-jettiness DIS event shape: NNLL+NLO
  results}}, \href{https://doi.org/10.1103/PhysRevD.90.014041}{\emph{Phys.
  Rev.} {\bfseries D90} (2014) 014041}
  [\href{https://arxiv.org/abs/1312.0301}{{\ttfamily 1312.0301}}].

\bibitem{Bozzi:2005wk}
G.~Bozzi, S.~Catani, D.~de~Florian and M.~Grazzini, \emph{{Transverse-momentum
  resummation and the spectrum of the Higgs boson at the LHC}},
  \href{https://doi.org/10.1016/j.nuclphysb.2005.12.022}{\emph{Nucl. Phys.}
  {\bfseries B737} (2006) 73}
  [\href{https://arxiv.org/abs/hep-ph/0508068}{{\ttfamily hep-ph/0508068}}].

\bibitem{Becher:2010tm}
T.~Becher and M.~Neubert, \emph{{{Drell-Yan} Production at Small $q_T$,
  Transverse Parton Distributions and the Collinear Anomaly}},
  \href{https://doi.org/10.1140/epjc/s10052-011-1665-7}{\emph{Eur. Phys. J.}
  {\bfseries C71} (2011) 1665}
  [\href{https://arxiv.org/abs/1007.4005}{{\ttfamily 1007.4005}}].

\bibitem{Banfi:2011dx}
A.~Banfi, M.~Dasgupta and S.~Marzani, \emph{{QCD predictions for new variables
  to study dilepton transverse momenta at hadron colliders}},
  \href{https://doi.org/10.1016/j.physletb.2011.05.028}{\emph{Phys. Lett.}
  {\bfseries B701} (2011) 75}
  [\href{https://arxiv.org/abs/1102.3594}{{\ttfamily 1102.3594}}].

\bibitem{Stewart:2010pd}
I.~W. Stewart, F.~J. Tackmann and W.~J. Waalewijn, \emph{{The Beam Thrust Cross
  Section for Drell-Yan at NNLL Order}},
  \href{https://doi.org/10.1103/PhysRevLett.106.032001}{\emph{Phys. Rev. Lett.}
  {\bfseries 106} (2011) 032001}
  [\href{https://arxiv.org/abs/1005.4060}{{\ttfamily 1005.4060}}].

\bibitem{Berger:2010xi}
C.~F. Berger, C.~Marcantonini, I.~W. Stewart, F.~J. Tackmann and W.~J.
  Waalewijn, \emph{{Higgs Production with a Central Jet Veto at NNLL+NNLO}},
  \href{https://doi.org/10.1007/JHEP04(2011)092}{\emph{JHEP} {\bfseries 04}
  (2011) 092} [\href{https://arxiv.org/abs/1012.4480}{{\ttfamily 1012.4480}}].

\bibitem{Becher:2015lmy}
T.~Becher, X.~Garcia~i Tormo and J.~Piclum, \emph{{Next-to-next-to-leading
  logarithmic resummation for transverse thrust}},
  \href{https://doi.org/10.1103/PhysRevD.93.054038,
  10.1103/PhysRevD.93.079905}{\emph{Phys. Rev.} {\bfseries D93} (2016) 054038}
  [\href{https://arxiv.org/abs/1512.00022}{{\ttfamily 1512.00022}}].

\bibitem{Becher:2013xia}
T.~Becher, M.~Neubert and L.~Rothen, \emph{{Factorization and
  $N^{3}LL_{p}$+NNLO predictions for the Higgs cross section with a jet veto}},
  \href{https://doi.org/10.1007/JHEP10(2013)125}{\emph{JHEP} {\bfseries 10}
  (2013) 125} [\href{https://arxiv.org/abs/1307.0025}{{\ttfamily 1307.0025}}].

\bibitem{Banfi:2012jm}
A.~Banfi, P.~F. Monni, G.~P. Salam and G.~Zanderighi, \emph{{Higgs and Z-boson
  production with a jet veto}},
  \href{https://doi.org/10.1103/PhysRevLett.109.202001}{\emph{Phys. Rev. Lett.}
  {\bfseries 109} (2012) 202001}
  [\href{https://arxiv.org/abs/1206.4998}{{\ttfamily 1206.4998}}].

\bibitem{Stewart:2013faa}
I.~W. Stewart, F.~J. Tackmann, J.~R. Walsh and S.~Zuberi, \emph{{Jet $p_T$
  resummation in Higgs production at $NNLL'+NNLO$}},
  \href{https://doi.org/10.1103/PhysRevD.89.054001}{\emph{Phys. Rev.}
  {\bfseries D89} (2014) 054001}
  [\href{https://arxiv.org/abs/1307.1808}{{\ttfamily 1307.1808}}].

\bibitem{Catani:2014qha}
S.~Catani, M.~Grazzini and A.~Torre, \emph{{Transverse-momentum resummation for
  heavy-quark hadroproduction}},
  \href{https://doi.org/10.1016/j.nuclphysb.2014.11.019}{\emph{Nucl. Phys.}
  {\bfseries B890} (2014) 518}
  [\href{https://arxiv.org/abs/1408.4564}{{\ttfamily 1408.4564}}].

\bibitem{Zhu:2012ts}
H.~X. Zhu, C.~S. Li, H.~T. Li, D.~Y. Shao and L.~L. Yang,
  \emph{{Transverse-momentum resummation for top-quark pairs at hadron
  colliders}},
  \href{https://doi.org/10.1103/PhysRevLett.110.082001}{\emph{Phys. Rev. Lett.}
  {\bfseries 110} (2013) 082001}
  [\href{https://arxiv.org/abs/1208.5774}{{\ttfamily 1208.5774}}].

\bibitem{Stewart:2010tn}
I.~W. Stewart, F.~J. Tackmann and W.~J. Waalewijn, \emph{{N-Jettiness: An
  Inclusive Event Shape to Veto Jets}},
  \href{https://doi.org/10.1103/PhysRevLett.105.092002}{\emph{Phys. Rev. Lett.}
  {\bfseries 105} (2010) 092002}
  [\href{https://arxiv.org/abs/1004.2489}{{\ttfamily 1004.2489}}].

\bibitem{Jouttenus:2011wh}
T.~T. Jouttenus, I.~W. Stewart, F.~J. Tackmann and W.~J. Waalewijn, \emph{{The
  Soft Function for Exclusive N-Jet Production at Hadron Colliders}},
  \href{https://doi.org/10.1103/PhysRevD.83.114030}{\emph{Phys. Rev.}
  {\bfseries D83} (2011) 114030}
  [\href{https://arxiv.org/abs/1102.4344}{{\ttfamily 1102.4344}}].

\bibitem{Bizon:2017rah}
W.~Bizon, P.~F. Monni, E.~Re, L.~Rottoli and P.~Torrielli,
  \emph{{Momentum-space resummation for transverse observables and the Higgs
  p$_{\perp}$ at N$^{3}$LL+NNLO}},
  \href{https://doi.org/10.1007/JHEP02(2018)108}{\emph{JHEP} {\bfseries 02}
  (2018) 108} [\href{https://arxiv.org/abs/1705.09127}{{\ttfamily
  1705.09127}}].

\bibitem{Bizon:2018foh}
W.~Bizo?, X.~Chen, A.~Gehrmann-De~Ridder, T.~Gehrmann, N.~Glover, A.~Huss
  et~al., \emph{{Fiducial distributions in Higgs and Drell-Yan production at
  N$^{3}$LL+NNLO}}, \href{https://doi.org/10.1007/JHEP12(2018)132}{\emph{JHEP}
  {\bfseries 12} (2018) 132}
  [\href{https://arxiv.org/abs/1805.05916}{{\ttfamily 1805.05916}}].

\bibitem{Bauer:2000yr}
C.~W. Bauer, S.~Fleming, D.~Pirjol and I.~W. Stewart, \emph{{An Effective field
  theory for collinear and soft gluons: Heavy to light decays}},
  \href{https://doi.org/10.1103/PhysRevD.63.114020}{\emph{Phys. Rev.}
  {\bfseries D63} (2001) 114020}
  [\href{https://arxiv.org/abs/hep-ph/0011336}{{\ttfamily hep-ph/0011336}}].

\bibitem{Banfi:2016zlc}
A.~Banfi, H.~McAslan, P.~F. Monni and G.~Zanderighi, \emph{{The two-jet rate in
  $e^+ e^-$ at next-to-next-to-leading-logarithmic order}},
  \href{https://doi.org/10.1103/PhysRevLett.117.172001}{\emph{Phys. Rev. Lett.}
  {\bfseries 117} (2016) 172001}
  [\href{https://arxiv.org/abs/1607.03111}{{\ttfamily 1607.03111}}].

\bibitem{Banfi:2000si}
A.~Banfi, G.~Marchesini, Y.~L. Dokshitzer and G.~Zanderighi, \emph{{QCD
  analysis of near-to-planar three jet events}},
  \href{https://doi.org/10.1088/1126-6708/2000/07/002}{\emph{JHEP} {\bfseries
  07} (2000) 002} [\href{https://arxiv.org/abs/hep-ph/0004027}{{\ttfamily
  hep-ph/0004027}}].

\bibitem{Catani:2011st}
S.~Catani, D.~de~Florian and G.~Rodrigo, \emph{{Space-like (versus time-like)
  collinear limits in QCD: Is factorization violated?}},
  \href{https://doi.org/10.1007/JHEP07(2012)026}{\emph{JHEP} {\bfseries 07}
  (2012) 026} [\href{https://arxiv.org/abs/1112.4405}{{\ttfamily 1112.4405}}].

\bibitem{Catani:1991hj}
S.~Catani, Y.~L. Dokshitzer, M.~Olsson, G.~Turnock and B.~R. Webber, \emph{{New
  clustering algorithm for multi - jet cross-sections in e+ e- annihilation}},
  \href{https://doi.org/10.1016/0370-2693(91)90196-W}{\emph{Phys. Lett.}
  {\bfseries B269} (1991) 432}.

\bibitem{Banfi:2018mcq}
A.~Banfi, B.~K. El-Menoufi and P.~F. Monni, \emph{{The Sudakov radiator for jet
  observables and the soft physical coupling}},
  \href{https://doi.org/10.1007/JHEP01(2019)083}{\emph{JHEP} {\bfseries 01}
  (2019) 083} [\href{https://arxiv.org/abs/1807.11487}{{\ttfamily
  1807.11487}}].

\bibitem{Falcioni:2014pka}
G.~Falcioni, E.~Gardi, M.~Harley, L.~Magnea and C.~D. White, \emph{{Multiple
  Gluon Exchange Webs}},
  \href{https://doi.org/10.1007/JHEP10(2014)010}{\emph{JHEP} {\bfseries 10}
  (2014) 10} [\href{https://arxiv.org/abs/1407.3477}{{\ttfamily 1407.3477}}].

\bibitem{Gardi:2009qi}
E.~Gardi and L.~Magnea, \emph{{Factorization constraints for soft anomalous
  dimensions in QCD scattering amplitudes}},
  \href{https://doi.org/10.1088/1126-6708/2009/03/079}{\emph{JHEP} {\bfseries
  03} (2009) 079} [\href{https://arxiv.org/abs/0901.1091}{{\ttfamily
  0901.1091}}].

\bibitem{Banfi:2014sua}
A.~Banfi, H.~McAslan, P.~F. Monni and G.~Zanderighi, \emph{{A general method
  for the resummation of event-shape distributions in $e^{+} e^{−}$
  annihilation}}, \href{https://doi.org/10.1007/JHEP05(2015)102}{\emph{JHEP}
  {\bfseries 05} (2015) 102} [\href{https://arxiv.org/abs/1412.2126}{{\ttfamily
  1412.2126}}].

\bibitem{deFlorian:2001zd}
D.~de~Florian and M.~Grazzini, \emph{{The Structure of large logarithmic
  corrections at small transverse momentum in hadronic collisions}},
  \href{https://doi.org/10.1016/S0550-3213(01)00460-6}{\emph{Nucl. Phys.}
  {\bfseries B616} (2001) 247}
  [\href{https://arxiv.org/abs/hep-ph/0108273}{{\ttfamily hep-ph/0108273}}].

\bibitem{Larkoski:2018cke}
A.~J. Larkoski and A.~Procita, \emph{{New Insights on an Old Problem:
  Resummation of the D-parameter}},
  \href{https://doi.org/10.1007/JHEP02(2019)104}{\emph{JHEP} {\bfseries 02}
  (2019) 104} [\href{https://arxiv.org/abs/1810.06563}{{\ttfamily
  1810.06563}}].

\bibitem{Antonelli:1999kx}
V.~Antonelli, M.~Dasgupta and G.~P. Salam, \emph{{Resummation of thrust
  distributions in DIS}},
  \href{https://doi.org/10.1088/1126-6708/2000/02/001}{\emph{JHEP} {\bfseries
  02} (2000) 001} [\href{https://arxiv.org/abs/hep-ph/9912488}{{\ttfamily
  hep-ph/9912488}}].

\bibitem{Dasgupta:2001eq}
M.~Dasgupta and G.~P. Salam, \emph{{Resummation of the jet broadening in DIS}},
  \href{https://doi.org/10.1007/s100520200915}{\emph{Eur. Phys. J.} {\bfseries
  C24} (2002) 213} [\href{https://arxiv.org/abs/hep-ph/0110213}{{\ttfamily
  hep-ph/0110213}}].

\bibitem{Monni:2019yyr}
P.~F. Monni, L.~Rottoli and P.~Torrielli, \emph{{Higgs transverse momentum with
  a jet veto: a double-differential resummation}},
  \href{https://arxiv.org/abs/1909.04704}{{\ttfamily 1909.04704}}.

\bibitem{alephqcd}
\url{http://aleph.web.cern.ch/aleph/QCD/dat/91/lep1.html}.

\bibitem{Dokshitzer:1995qm}
Y.~L. Dokshitzer, G.~Marchesini and B.~R. Webber, \emph{{Dispersive approach to
  power behaved contributions in QCD hard processes}},
  \href{https://doi.org/10.1016/0550-3213(96)00155-1}{\emph{Nucl. Phys.}
  {\bfseries B469} (1996) 93}
  [\href{https://arxiv.org/abs/hep-ph/9512336}{{\ttfamily hep-ph/9512336}}].

\bibitem{Banfi:2001pb}
A.~Banfi, Y.~L. Dokshitzer, G.~Marchesini and G.~Zanderighi, \emph{{QCD
  analysis of $D$-parameter in near to planar three jet events}},
  \href{https://doi.org/10.1088/1126-6708/2001/05/040}{\emph{JHEP} {\bfseries
  05} (2001) 040} [\href{https://arxiv.org/abs/hep-ph/0104162}{{\ttfamily
  hep-ph/0104162}}].

\bibitem{Brown:1991hx}
N.~Brown and W.~J. Stirling, \emph{{Finding jets and summing soft gluons: A New
  algorithm}}, \href{https://doi.org/10.1007/BF01559740}{\emph{Z. Phys.}
  {\bfseries C53} (1992) 629}.

\end{thebibliography}\endgroup

\end{document}